\documentclass[11pt, a4paper]{article}
\pdfoutput=1

\usepackage{jcappub}

\usepackage{comment}
\usepackage{amsmath}
\usepackage{amsfonts}
\usepackage{amssymb}
\usepackage{graphicx}
\usepackage{mathrsfs}
\usepackage{latexsym}
\usepackage{color}
\usepackage{slashed, cancel}
\usepackage{hyperref}
\usepackage{cleveref}
\usepackage{float}
\usepackage{tikz}
\usepackage{bbold}
\usepackage{soul} 
\usepackage{mathtools}
\usepackage[utf8]{inputenc} 
\usepackage{csquotes}
\usepackage{support-caption} 
\usepackage{subcaption}
\usepackage{enumitem}
\usepackage{journals}
\allowdisplaybreaks
\usepackage{multirow}
\usepackage{pifont}

\usepackage[version=3]{mhchem}
\hypersetup{urlcolor=blue, colorlinks=true} 

\usepackage{fancyhdr}
\usepackage{pdfpages} 

\def\supplementfilename{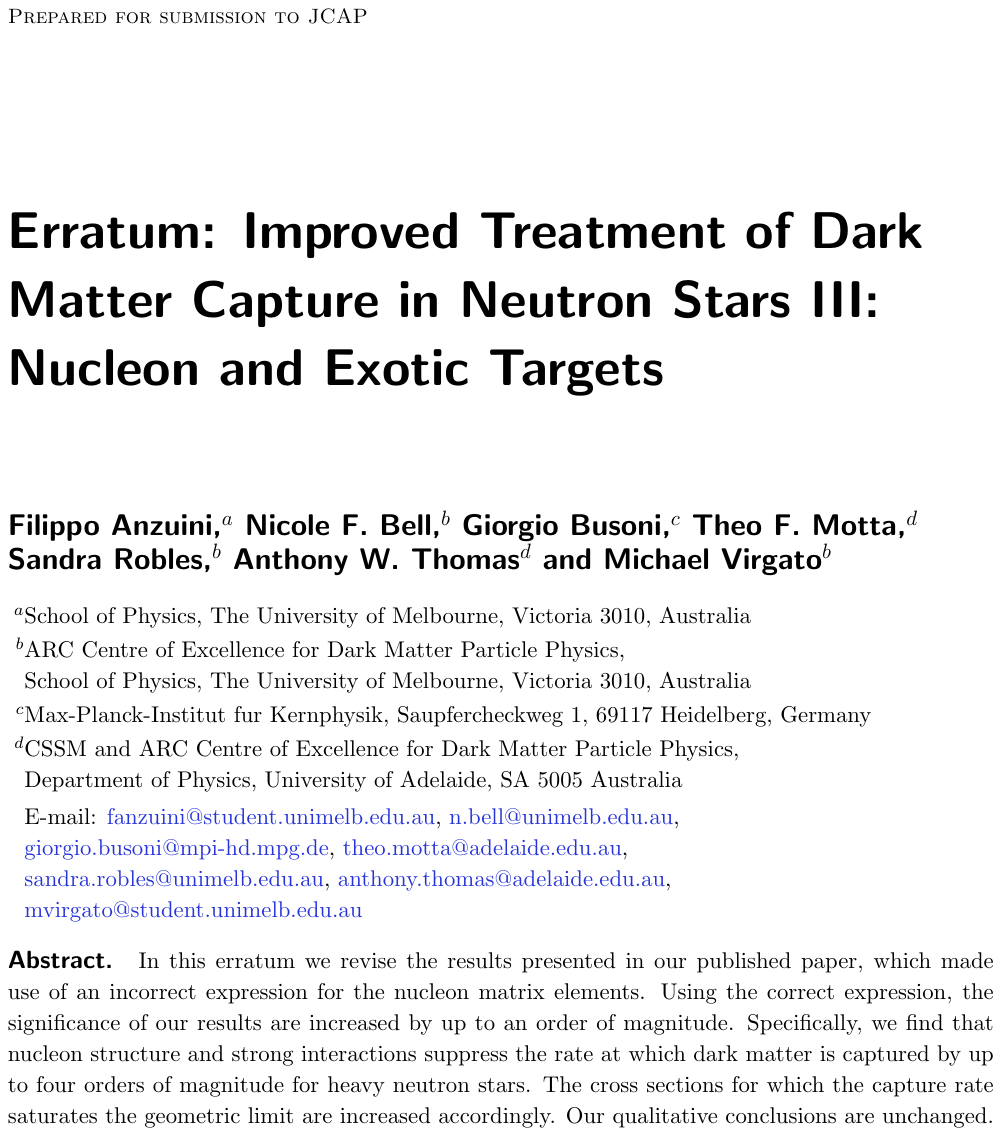}

\pdfximage{\supplementfilename}
\def\numbersupplementpages{\the\pdflastximagepages}

\renewcommand{\arraystretch}{1}
\numberwithin{equation}{section}

\definecolor{orange}{rgb}{1,0.4,0}
\definecolor{green}{rgb}{0,0.65,0}
\definecolor{rossos}{rgb}{0.8,0.2,0.3}
\definecolor{bluscuro}{rgb}{0.15, 0.2, .85}
\definecolor{bluchiaro}{cmyk}{1,.3,0.,0.1}
\hypersetup{colorlinks, citecolor=bluscuro, linkcolor=bluscuro, urlcolor=bluscuro}

\makeatother   
\newcommand{\GeV}{{\rm \,GeV}}
\newcommand{\TeV}{{\rm \,TeV}}
\newcommand{\MeV}{{\rm \,MeV}}
\newcommand{\keV}{{\rm \,keV}}

\newcommand{\m}{{\rm \,m}}
\newcommand{\cm}{{\rm \,cm}}
\newcommand{\fm}{{\rm \,fm}}
\newcommand{\km}{{\rm \,km}}
\newcommand{\s}{{\rm \,s}}
\newcommand{\g}{{\rm \,g}}

\newcommand{\Msun}{M_\odot}
\newcommand{\Mstar}{M_\star}
\newcommand{\Rstar}{R_\star}
\newcommand{\vstar}{v_\star}

\newcommand{\cut}{\nonumber\\}

 \newcommand{\muFi}{\mu_{F,i}}
\newcommand{\kinFn}{\varepsilon_{F,n}}
\newcommand{\kinFp}{\varepsilon_{F,p}}
\newcommand{\kinFi}{\varepsilon_{F,i}}

\newcommand{\fMB}{f_{\rm MB}}
\newcommand{\fFD}{f_{\rm FD}}

\newcommand{\sigmathi}{\sigma^{th}_{i\chi}}
\newcommand{\sigmathn}{\sigma^{th}_{n\chi}}
\newcommand{\sigmathp}{\sigma^{th}_{p\chi}}
\newcommand{\sigmath}{\sigma_{th}}
\newcommand{\mstar}{m_i^*}
\newcommand{\mnstar}{m_n^*}

\newcommand{\mneff}{m_n^{\rm eff}}
\newcommand{\mbeff}{m_i^{\rm eff}}
\newcommand{\mBeff}{m_{\cal B}^{\rm eff}}

\newcommand{\erf}{{\rm \,Erf}}
\newcommand{\Msq}{|\overline{M}|^2}

 \def\be   {\begin{equation}}   \def\ee   {\end{equation}}
 \def\ba   {\begin{array}}      \def\ea   {\end{array}}
 \def\bea  {\begin{eqnarray}}   \def\eea  {\end{eqnarray}}
 \def\bean {\begin{eqnarray*}}  \def\eean {\end{eqnarray*}}
 \def\nn{\nonumber}

\begin{document}

\hfill ADP-21-11/T1158

\title{Improved Treatment of Dark Matter Capture in Neutron Stars III: Nucleon and Exotic Targets}

\author[a]{Filippo Anzuini,}
\author[b]{Nicole F.\ Bell,}
\author[c]{Giorgio Busoni,}
\author[d]{Theo F.\ Motta,}
\author[b]{Sandra Robles,}
\author[d]{Anthony W. Thomas}
\author[b]{and Michael Virgato}
\affiliation[a]{School of Physics, The University of Melbourne, Victoria 3010, Australia}
\affiliation[b]{ARC Centre of Excellence for Dark Matter Particle Physics, \\
School of Physics, The University of Melbourne, Victoria 3010, Australia}
\affiliation[c]{Max-Planck-Institut fur Kernphysik, Saupfercheckweg 1, 69117 Heidelberg, Germany}
\affiliation[d]{
CSSM and ARC Centre of Excellence for Dark Matter Particle Physics, \\
Department of Physics, University of Adelaide, SA 5005 Australia
}
\emailAdd{fanzuini@student.unimelb.edu.au}
\emailAdd{n.bell@unimelb.edu.au}
\emailAdd{giorgio.busoni@mpi-hd.mpg.de}
\emailAdd{theo.motta@adelaide.edu.au}
\emailAdd{sandra.robles@unimelb.edu.au}
\emailAdd{anthony.thomas@adelaide.edu.au}
\emailAdd{mvirgato@student.unimelb.edu.au}

\abstract{
We consider the capture of dark matter (DM) in neutron stars via scattering on hadronic targets, including neutrons, protons and hyperons. We extend previous analyses by including momentum dependent form factors, which account for hadronic structure, and incorporating the effect of baryon strong interactions in the dense neutron star interior, rather than modelling the baryons as a free Fermi gas. The combination of these effects suppresses the DM capture rate over a wide mass range, thus increasing the cross section for which the capture rate saturates the geometric limit. In addition, variation in the capture rate associated with the choice of neutron star equation of state is reduced.
For proton targets, the use of the interacting baryon approach to obtain the correct Fermi energy is essential for an accurate evaluation of the capture rate in the Pauli-blocked regime. For heavy neutron stars, which are expected to contain exotic matter, we identify cases where DM scattering on hyperons contributes significantly to the total capture rate. Despite smaller neutron star capture rates, compared to existing analyses, we find that the projected DM-nucleon scattering sensitivity greatly exceeds that of nuclear recoil experiments for a wide DM mass range. 
}

\maketitle

\section{Introduction}

\bigskip

Uncovering the nature of dark matter (DM) remains a major goal of modern physics, with direct detection experiments at the forefront of our attempts. Though such searches have seen an impressive increase in sensitivity over recent years, their reach is limited by practicalities such as the achievable mass of the detector target material and the recoil energy resolution threshold. It is then natural to look to alternative systems in which DM interactions lead to observable consequences. A prominent such alternative is to look at the effect of DM capture in stars and compact objects such as white dwarfs and neutron stars (NSs). The latter is the focus of this work. 

The basic framework for the capture of DM in stars is well established, built on the original formalism of Gould~\cite{Gould:1987ir, Gould:1987ju} and others~\cite{Press:1985ug,Griest:1986yu,Silk:1985ax,Krauss:1985ks}. It has primarily been applied in the context of the Sun~\cite{Jungman:1995df,Busoni:2013kaa,Garani:2017jcj,Busoni:2017mhe} and stars~\cite{Ilie:2020nzp,Ilie:2020iup,Ilie:2021iyh}. Observable consequences of DM accumulation in the Sun include modifications to thermal transport in the solar interior~\cite{Gould:1989ez,Gould:1989hm, Vincent:2013lua, Geytenbeek:2016nfg}, or signals arising from the annihilation of the accumulated DM, in the form of high energy neutrinos~\cite{Tanaka:2011uf, Choi:2015ara,Bell:2021esh, Adrian-Martinez:2016gti, Adrian-Martinez:2016ujo, Aartsen:2016zhm} or cosmic and gamma ray fluxes~\cite{Batell:2009zp, Schuster:2009au, Bell:2011sn, Feng:2016ijc, Leane:2017vag, HAWC:2018szf, Bell:2021pyy}.

Another prominent application is the capture of DM in compact objects such as white dwarfs~\cite{ McCullough:2010ai, Hooper:2010es, Amaro-Seoane:2015uny, Cermeno:2018qgu, Dasgupta:2019juq, Panotopoulos:2020kuo, Ilie:2020vec, Bell:2021fye} and NSs~\cite{Goldman:1989nd,Kouvaris:2007ay,Kouvaris:2010vv,deLavallaz:2010wp,McDermott:2011jp,Bell:2013xk,Baryakhtar:2017dbj, Raj:2017wrv, Bell:2018pkk, Bell:2019pyc, Garani:2018kkd,  Acevedo:2019agu, Joglekar:2019vzy, Joglekar:2020liw, Bell:2020jou, Bell:2020lmm, Bell:2020obw,Leane:2021ihh}. 
Much of this attention is generated by the possibility that DM will thermalise~\cite{Kouvaris:2010vv, Bertoni:2013bsa, Garani:2020wge} and annihilate within these objects, leading to a potentially observable level of heating~\cite{Baryakhtar:2017dbj, Raj:2017wrv, Bell:2018pkk, Bell:2019pyc, Dasgupta:2020dik, Bell:2021fye}. For NSs, additional consequences include the possibility of  induced collapse to black holes~\cite{Kouvaris:2010jy,McDermott:2011jp, Kouvaris:2011fi, Capela:2013yf, Guver:2012ba, Bell:2013xk,Bramante:2013nma, Ellis:2017jgp, Ellis:2018bkr} in the case of asymmetric or non-annihilating DM, or modifications to the rate and gravitational wave signatures of mergers~\cite{Bramante:2017ulk, Ellis:2017jgp, Ellis:2018bkr, Nelson:2018xtr}.
It is important to note that the DM capture process in NSs differs significantly from that in the Sun, because of the extreme conditions present. Therefore, Gould's original formalism requires substantial modification to properly account for the physics of this extreme environment. Such corrections have gradually been introduced. Ref.~\cite{Bell:2020jou} presented a new formalism that consistently incorporated many of the relevant physical effects. These include: the NS internal structure~\cite{Garani:2018kkd, Bell:2020jou, Bell:2020lmm}, through solving the Tolman–Oppenheimer–Volkoff (TOV) equations~\cite{Tolman:1939jz, Oppenheimer:1939ne} coupled to an appropriate equation of state, 
 correct treatment of Pauli blocking for degenerate targets~\cite{Garani:2018kkd, Bell:2020jou, Bell:2020lmm}, relativistic kinematics~\cite{Joglekar:2019vzy, Joglekar:2020liw, Bell:2020jou, Bell:2020lmm}, gravitational focusing~\cite{Goldman:1989nd, Kouvaris:2010jy, Bell:2020jou, Bell:2020lmm}, and multiple scattering effects for heavy DM~\cite{Bramante:2017xlb, Joglekar:2020liw,Bell:2020jou,Ilie:2020vec, Bell:2020lmm}.

In this work, we build on these improvements by focusing on the physics pertinent to baryonic targets in the extremely dense NS interior. The two main physical effects are (i) baryons experience strong interactions in the NS interior and (ii) the momentum transfer in DM-baryon collisions is sufficiently large that baryons cannot be treated as point particles. These features were first introduced in ref.~\cite{Bell:2020obw}, in the context of scattering from neutrons, where we demonstrated an impact on the capture rate of up to three orders of magnitude for heavy NSs. We generalize that analysis here to consider scattering from other baryonic species in the neutron star, namely protons and hyperons. We also extend this treatment to the multiple scattering regime relevant for heavy DM, and project sensitivity limits across a wide DM mass range.

Accounting for the momentum-dependent form factors and strong interactions leads to several non-trivial consequences that are missing from previous treatments.
Incorporating the hadronic structure via momentum dependent form factors acts to suppress collisions with large energy transfer, resulting in a considerable reduction of the capture rate over a wide mass range. 
In addition, because the average DM energy loss per interaction is reduced, multiple scattering effects become relevant at lower DM masses than previously expected, and the suppression of the capture rate in this regime becomes larger.
The effects of strong interactions on the capture rates are primarily observed in the intermediate and large DM mass range. However, in the case of DM-proton scattering, we find that the usual approximations can lead to highly misleading conclusions for the low DM mass range. 
Furthermore, the presence of exotic matter can have a somewhat surprising impact on the capture rate for particular DM-nucleon interactions, despite their lower abundance.

This work is organised as follows: In section~\ref{sec:ns} we discuss relevant properties of NSs, including the internal structure and the equation of state adopted. In section~\ref{sec:captureintrates} we outline the DM capture rate formalism for NSs, and incorporate the effects of the momentum dependent form factors and strong interactions. Our results are presented in section~\ref{sec:results}, with a discussion of their implications in section~\ref{sec:discussion}. Our conclusions are presented in section~\ref{sec:conclusion}.

\section{Neutron Stars}
\label{sec:ns}

Neutron stars, the remnants of core collapse supernova explosions of massive stars, are the densest known stellar objects. This extreme density provides a unique cosmic laboratory in which to probe DM interactions with ordinary matter. 
Despite recent major theoretical and observational breakthroughs, the exact NS composition and internal structure remain unknown. In this section, we outline the typical NS structure and the general properties of matter in the different NS layers. We also introduce the equations of state and benchmark models adopted in this work.

\subsection{Internal structure}
\label{sec:NSintstruct}

The internal structure of neutron stars can be broadly divided into a core region and a crust, with the latter having a typical thickness of approximately $1\km$. The outer crust extends from the stellar surface down to a layer with baryonic density $\rho$ equal to the neutron drip density, $\rho_{\rm ND}\sim 4.3 \times 10^{11} {\g \cm^{-3}}$~\cite{Ruester:2005fm,RocaMaza:2008ja,Pearson:2011,Kreim:2013rqa,Chamel:2015oqa,Antic:2020zuk}. In particular, the region from the surface down to the layer where $\rho \sim 10^{10} $ g cm$^{-3}$ is commonly referred to as the \enquote{blanketing envelope} \cite{Gudmundsson:1983, Potekhin:1997mn}, which insulates the stellar surface from the hotter, internal layers of the star. 
The outer crust is composed of a Coulomb lattice of heavy nuclei surrounded by a gas of electrons, which become degenerate for $\rho\gg 10^6 \g\cm^{-3}$. As the density increases, electron capture by the nuclei leads to the formation of heavier chemical elements, until the baryonic density reaches $\rho = \rho_{\rm ND}$ and neutrons leak out of the nuclei. The inner crust extends from $\rho_{\rm ND}$ up to $\sim 0.5 \rho_0$ (with $\rho_0=2.8 \times 10^{14} {\g \cm^{-3}}$ being the nuclear saturation density) and is characterized by the presence of free neutrons and electrons surrounding neutron-proton clusters  \cite{Chamel:2008Lr,Chamel:2015oqa}. The latter can assume non-spherical shapes at the bottom of the inner crust~\cite{Chamel:2008Lr}. For $\rho \gtrsim 0.5\rho_0$ the clusters dissolve, and nuclear matter becomes homogeneous.

The outer core ($0.5\rho_0 \lesssim \rho \lesssim 2\rho_0$) contains a liquid (or possible superfluid) of  neutrons  with an admixture of protons, as well as a gas of electrons and muons. This is the so-called $npe\mu$ matter in beta equilibrium.
In low mass NSs, with central densities $\rho_c\lesssim2-3\rho_0$, $npe\mu$ matter constitutes the entire NS core. The central density of the most massive NSs is expected to reach $\rho_c\gtrsim 5\rho_0$ and the radius of their inner core is expected to be several kilometers. The lack of understanding of strong interactions at densities above nuclear saturation means that the composition of the inner core is much less certain. 
Many possible models have been put forth which predict the emergence of exotic phases of matter. These include the appearance of hyperons, meson condensates, or a phase transition into quark matter~\cite{Haensel:2007yy,Weber:2006ep,Baym:2017whm,RikovskaStone:2006ta,Whittenbury:2015ziz}. 
In this paper, we focus on NSs in beta equilibrium, composed of an admixture of nucleons, leptons and hyperonic matter. Of the various exotic matter possibilities, the inclusion of hyperons is the most straightforward to study within our formalism, as scattering of dark matter on hyperons can be treated in the same manner as scattering on neutrons and protons, with the relevant form factors provided in Appendix~\ref{apx:hypff}.

\subsection{Equation of state}
\label{sec:EoS}

The key ingredient to solve the structure equations is the equation of state (EoS) of dense matter. For matter in beta equilibrium in a strongly degenerate state, the EoS  depends only on one parameter, frequently taken to be  the baryon number density, $n_b$. 
An EoS which includes hyperonic matter is required in order to obtain the structure
of NSs with masses beyond about $1.7\Msun$.
Once the $\Lambda^0$ chemical potential equals the neutron chemical potential (which equals the electron plus proton chemical potentials) hyperons appear. 
The $\Xi^-$ appears when the neutron plus electron chemical potential equals the $\Xi^-$ chemical potential.
In other words, high momentum neutrons are replaced by low momentum hyperons, thereby lowering the matter pressure.
Given that the time scales for weak interactions leading to hyperon formation are nano-seconds, far shorter than the times associated with NS formation, it is natural that they should be included in the EoS. 
As large densities are required for significant hyperon abundance to occur, it is expected that the resulting NS will be quite massive, and so many EoS are ruled out from observations~\cite{Fortin:2014mya}. 

As shown in ref.~\cite{Motta:2019tjc}, within the QMC model, assuming beta equilibrium and hence including hyperons, the energy density is the sum of Hartree and Fock terms, $\epsilon = \epsilon_{\rm Hartree} + \epsilon_\text{Fock}$, where
	\begin{eqnarray}
	\epsilon_{\rm Hartree}&=& \frac{m_\sigma^2\sigma^2}{2} + \frac{m_\omega^2\omega^2}{2}+ \frac{m_\rho^2b^2}{2} + \frac{m_\delta^2\delta^2}{2} \cut 
	&+& \frac{1}{\pi^2}\sum_{\cal B}\int_{0}^{k_F^{\cal B}}{k^2}{\sqrt{k^2+\mBeff(\sigma,\delta)^2}dk}   
	+ \frac{1}{\pi^2}\sum_L \int_{0}^{k_F^L}{k^2}{\sqrt{k^2+m_L^2}dk} \, . 
	\end{eqnarray}
Here $\sigma, \, \omega, b$ and $\delta$ denote the isoscalar scalar and vector fields and the isovector vector and scalar fields, respectively. The sums run over the full baryon octet ${\cal B}=(n, p, \Lambda^0, \Sigma^{0,\pm},  \Xi^{0,-})$ and leptons 
$L=(e^-, \mu^-)$. 
The meson mean field equations are
\begin{eqnarray}
\label{eq:mean-field}
	m_\sigma^2\sigma &=& \frac{1}{\pi^2} \sum_{\cal B} \big(-\partial_\sigma 
	\mBeff(\sigma,\delta)\big)
	 \int_0^{k_F^{\cal B}} k^2 
	\frac{\mBeff(\sigma,{\delta})}{\sqrt{k^2 + \mBeff(\sigma,{\delta})^2}}dk,  \\
	m_\omega^2 \omega &=& \sum_{\cal B} n_{\cal B} g_\omega \times \left( 1+\frac{s_{\cal B}}{3} \right) =
	\sum_{\cal B}  n_{\cal B} g_\omega^{\cal B} \, , \\
	m_\rho^2 b&=& \sum_{\cal B}  n_{\cal B} g_\rho \times t_{3\cal B} =\sum_{\cal B}  n_{\cal B} g_\rho^{\cal B} \, , \label{eq:meanrho} \\
	m_\delta^2{\delta} &=&  \sum_{\cal B} \big(-\partial_\delta \mBeff(\sigma,\delta)\big)
	\frac{1}{\pi^2}\int_0^{k_F^{\cal B}} k^2 
	\frac{\mBeff(\sigma,{\delta})}{\sqrt{k^2 + \mBeff(\sigma,{\delta})^2}}dk \, .
	\end{eqnarray}
Finally, the Fock terms take the form
	\begin{eqnarray}
		\epsilon_\text{Fock} &=& 
		\frac{1}{(2\pi)^6} \sum_{\cal B} \int_{k_1,k_2} 
		\frac{\partial_\sigma \mBeff(\sigma,\delta)^2}{(\vec{k_1}-\vec{k_2})^2+m_\sigma^2}\times 
		\left[\frac{\mBeff(\sigma,\delta)}{\sqrt{k_1^2+\mBeff(\sigma,\delta)^2}}\right] 
		\left[\frac{m^{\rm eff}_{\cal B'}(\sigma,\delta)}
		{\sqrt{k_2^2+m^{\rm eff}_{\cal B'}(\sigma,\delta)^2}}\right] \nonumber \\
	&+&\frac{1}{(2\pi)^6} \sum_{\cal B,B'} \int_{k_1,k_2}  
		\frac{Z_{t_{3\cal B} t_{3\cal B'}}}{(\vec{k_1}-\vec{k_2})^2+m_\delta^2} \times
		\left[\frac{\mBeff(\sigma,\delta)}{\sqrt{k_1^2+\mBeff(\sigma,\delta)^2}}\right] 
		\left[\frac{m^{\rm eff}_{\cal B'}(\sigma,\delta)}{\sqrt{k_2^2+m^{\rm eff}_{\cal B'}(\sigma,\delta)^2}}\right] \cut
	&-&\frac{1}{(2\pi)^6} \sum_{\cal B} \int_{k_1,k_2}  
	\frac{{g^{\cal B}_\omega}^2}{(\vec{k_1} - \vec{k_2})^2 + m_\omega^2} 
		- \sum_{\cal B,B'} \int_{k_1,k_2}  \frac{g_\rho^2 I_{t_{3\cal B} t_{3\cal B'}}}{(\vec k_1 - \vec{k_2})^2 + m_\rho^2}  \, , 
		\nonumber
		\end{eqnarray}
where
	\begin{equation}
	I_{t_{3\cal B} t_{3\cal B'}}
	=  \delta_{t_{3\cal B} t_{3\cal B'}} + {(\delta_{t_{3\cal B}, t_{3\cal B'}+1}+\delta_{t_{3\cal B'}, t_{3\cal B}+1})}{t_{\cal B}} 
	\end{equation}
and
	\begin{equation}
			Z_{t_{3\cal B} t_{3\cal B'}}
			= \partial_\delta \mBeff(\sigma,\delta)\partial_\delta 
			m^{\rm eff}_{\cal B'}(\sigma,\delta)  \delta_{t_{3\cal B} t_{3\cal B'}} 
			+ g^{\cal B}_\delta(\delta,\sigma)g^{\cal B'}_\delta(\delta,\sigma)  {(\delta_{t_{3\cal B}, t_{3\cal B'}+1}+\delta_{t_{3\cal B'}, t_{3\cal B}+1})}{t_{\cal B}}.
	\end{equation}
The chemical potentials for each species are defined as 
\begin{equation}
    \muFi = \frac{\partial \epsilon}{\partial n_i} \, .
\end{equation}
The relevant coupling constants of the scalar and vector mesons to the baryons, as well as the meson masses and the resulting nuclear matter properties are given in ref.~\cite{Motta:2019tjc}, where it was shown that the model produces neutron stars compatible with all observational constraints.	

The density dependent scalar couplings are calculated by solving the MIT bag model~\cite{Chodos:1974pn} equations of motion in the scalar fields and thus our calculation includes the self-consistent adjustment of the internal structure of the baryons in the medium at the corresponding local density~\cite{Guichon:1987jp,Guichon:1995ue,Guichon:2018uew}. The density dependence of these couplings is equivalent to the inclusion of repulsive three-body forces between all of the baryons, which arise naturally once allowance is made for the modification of baryon structure in the medium, without any new 
parameters~\cite{Guichon:2004xg,Thomas:2021kio}.

\subsection{Benchmark configurations} 
\label{sec:NSmodels}

\begin{table}[tb]
\centering
\begin{tabular}{|l|c|c|c|c|}
\hline
\bf EoS & \bf QMC-1 & \bf QMC-2 & \bf QMC-3 & \bf QMC-4 \\ \hline
$n_{\cal B}^c$ $[\fm^{-3}]$ & 0.325 & 0.447 & 0.540 & 0.872\\
$\Mstar$ $[\Msun]$ & 1.000 & 1.500 & 1.750 & 1.900  \\
$\Rstar$ [km] &  13.044 & 12.847 & 12.611 & 12.109 \\
$B(\Rstar)$ & 0.772 & 0.653 & 0.588 & 0.535\\
\hline
\end{tabular} 
\caption{Benchmark NSs for four different  configurations of the QMC equation of state. 
EoS configurations are determined by the central number density $n_{\cal B}^c$.
}
\label{tab:eos}
\end{table} 

We  assume a non-rotating, non-magnetized, spherically symmetric NS and couple the QMC EoS to the Tolman-Oppenheimer-Volkoff (TOV) equations~\cite{Tolman:1939jz,Oppenheimer:1939ne}, to obtain the NS mass, $\Mstar$, the NS radius, $\Rstar$, and radial profiles of the quantities needed in our analysis, namely particle number fractions, effective masses, chemical potentials and  general relativity corrections encoded in $B(r)$, which is the time part of the Schwarzschild metric (see  Fig.~1 of ref.~\cite{Bell:2020jou} for an example of the $B$ profile).

\begin{figure}[t] 
\centering
\includegraphics[width=0.925\textwidth]{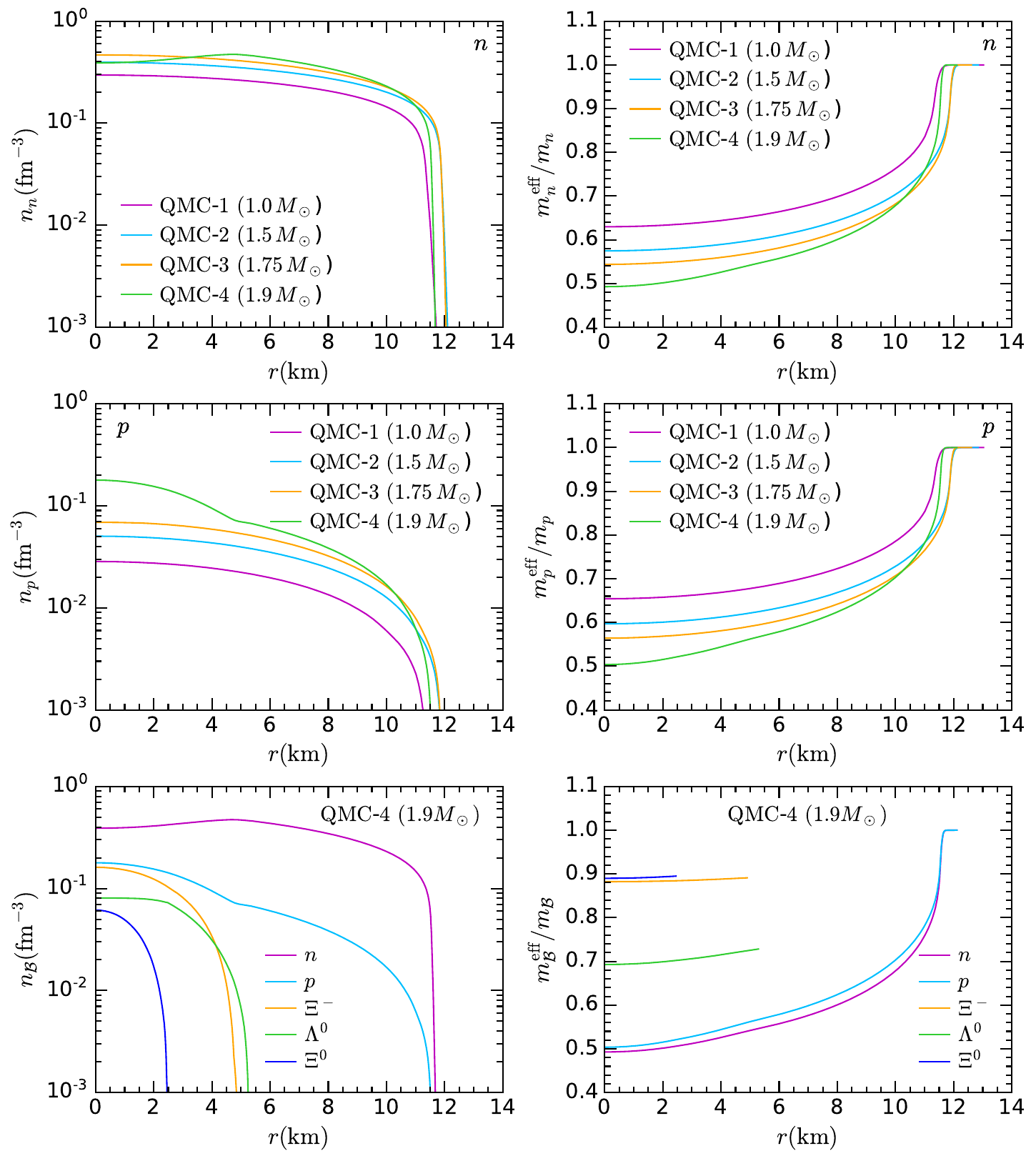}
\caption{Number density profiles (left) and   ratio of the effective mass to the bare mass (right) for neutrons (top) and protons (middle) for the NS benchmark configurations with the QMC EoS given in Table~\ref{tab:eos}.
In the bottom panels, we show the same profiles for all baryonic species in the heaviest NS considered, QMC-4, which contains hyperonic matter.}
\label{fig:NSradprofs1}
\end{figure}

In Fig.~\ref{fig:NSradprofs1}, we show radial profiles for the number densities, $n_{\cal B}$, and effective masses, $\mBeff$, of the baryonic species, for four benchmark NSs obtained using the QMC EoS, given in Table~\ref{tab:eos}.  The choice of the NS mass range is motivated by observations~\cite{Ozel:2016oaf,Antoniadis:2016hxz}. 
Notice that the neutron number density increases with the NS mass.  However in the QMC-4 configuration, corresponding to the heaviest NS we consider, the neutron content is slightly depleted towards the NS centre, because of the appearance of hyperonic matter (see bottom left panel). 
Unlike  the neutron number density, there is more variation in $n_p(r)$ (middle left panel) among the benchmark NS configurations. 
In the right panels, we observe that the nucleon effective mass can reach values as low as $\sim0.5m_N$ at the centre of the massive NS QMC-4. Even in a  NS with $\Mstar=1\Msun$ (QMC-1) the neutron effective mass progressively decreases in the degenerate stellar interior, down to $\mneff\sim0.63 m_n$. The proton effective masses are very similar to those of the neutrons. 
Hyperons start to appear in the NS core as we increase the NS mass beyond about 1.7 $\Msun$, with $\Lambda^0$ present in the QMC-3 configuration, which has a central baryon density of $n_{\cal B }^c\sim0.54\fm^{-3}$. 
In the more massive QMC-4 NS configuration (bottom panels), three hyperonic species are present, $\Lambda^0$, $\Xi^-$ and $\Xi^0$.  Of these, $\Xi^-$ is the most abundant, followed by  $\Lambda^0$, which is populated out to a slightly larger radius within the inner core than  $\Xi^-$. Note that the effective masses of the hyperons do not vary much along the NS radius, in contrast to those of the nucleons.

\section{Capture and Interaction Rates for interacting baryons}
\label{sec:captureintrates}

The capture of dark matter in neutron stars involves dealing with several physical effects. First, the strong gravitational potential of a NS  accelerates DM particles to quasi-relativistic speeds. The targets available in NSs can be relativistic themselves, particularly the degenerate leptons. A fully relativistic treatment which also accounts for gravitational focusing is thus required. Second, the scattering of light DM is limited by the number of free
final states of the target species 
in the outer shell of the Fermi sphere, leading to the capture rate being suppressed by Pauli blocking. Heavy DM, on the other hand, requires more than a single collision to lose enough energy to become gravitationally bound to the star. These effects depend on the internal structure of the NS, as well as the microphysics provided by the EoS. General expressions for the capture and interaction rate incorporating all these effects were derived in  refs.~\cite{Bell:2020jou, Bell:2020lmm}. Those results assume that the targets are point-like particles that behave as an ideal Fermi gas, which is a good approximation for the scattering of DM from leptons. However, there are two additional considerations inherent to the physics of NSs which affect the scattering of DM from baryonic species~\cite{Bell:2020obw}. Specifically: (i) at the extreme densities found in NSs, nucleons (or, more generally, baryons) 
experience strong Lorentz scalar and vector mean fields, with the former giving the baryons an effective mass that differs from its bare mass, 
and (ii) the momentum transfer in the DM-baryon scattering process can be sufficiently large that the targets can no longer be treated as point-like particles. 
These two effects introduce sizeable suppressions of the capture rate, especially in the DM mass range that is not subject to Pauli blocking. 
In the following we expand upon the discussion in ref.~\citep{Bell:2020obw} and evaluate these effects for a wider DM mass range and for all relevant baryonic targets.

\subsection{Interaction Rate}
\label{sec:intrate}

A key component of the capture rate calculation is the DM interaction rate. It is required to construct the probability density function (PDF) of the DM energy transferred in each collision. 
The definition of the capture probability  after $N$ scatterings~\cite{Bell:2020jou}, which is relevant to the capture of heavy DM via multiple scattering, is based on this PDF.

Below we give the key expression for the DM scattering rate, which holds both in the approach in which the target particles are modelled as a free Fermi gas or, with appropriate modifications to the target mass and Fermi energy, in the case where the targets are taken to be strongly interacting.
The DM scattering rate, $\Gamma$, is defined as a function of the target response function, $S(q_0,q)$~\cite{Bertoni:2013bsa,Bell:2020jou}
\begin{eqnarray}
\Gamma &=& \int \frac{d^3k^{'}}{(2\pi)^3} \frac{1}{(2E_\chi)(2E^{'}_\chi)(2m_i)(2m_i)}\Theta(E^{'}_\chi-m_\chi)\Theta(q_0)S(q_0,q), \label{eq:intratedeftext}\\
S(q_0,q) &=& 2\int \frac{d^3p}{(2\pi)^3}\int \frac{d^3p^{'}}{(2\pi)^3} \frac{m_i^2}{E_i E^{'}_i}|\overline{M}(s,t,m_i)|^2 (2\pi)^4\delta^4\left(k_\mu+p_\mu-k_\mu^{'}-p_\mu^{'}\right)\nonumber\\
 &&\times\fFD(E_i)(1-\fFD(E^{'}_i))\Theta(E_i-m_i)\Theta(E^{'}_i-m_i),
 \label{eq:responsefunc}
\end{eqnarray}
where $|\overline{M}|^2$ is the squared matrix element, $k^\mu=(E_\chi$,$\vec{k})$ and $k^{'\mu}=(E^{'}_\chi,\vec{k'})$ are the DM initial and final momenta, $p^\mu=(E_i,\vec{p})$ and $p^{'\mu}=(E^{'}_i,\vec{p'})$ are the target particle initial and final momenta, $m_i$ is the target mass, $m_\chi$ the DM mass, $q_0=E_i^{'}-E_i$ is the DM energy loss, $\vec{q}=\vec{p}-\vec{p}^{'}$ is the three-momentum exchanged with $q$ its magnitude, and $\fFD$ is the Fermi Dirac distribution. 
To calculate the interaction rate,  
we consider the interactions of Dirac fermion DM with SM quarks, described by the dimension 6 effective operators listed in Table~\ref{tab:operatorshe}, where the strength of the coupling is parametrised by the cutoff scale $\Lambda$ and 
\begin{equation}
    \mu=\frac{m_\chi}{m_i}.
\end{equation}

The spin-averaged squared matrix elements $|\overline{M}|^2$ for DM interactions with baryons in Table~\ref{tab:operatorshe} contain effective coefficients $c_i^I(t)$ that depend on the transferred momentum. In the case of  scalar and pseudoscalar interactions, these coefficients also depend on the baryon mass $m_i$. 
In general~\cite{Thomas:2001kw}, 
\begin{eqnarray}
c_i^I(t) &= c_i^I(0) F(t),\quad I\in\{S,P,V,A,T\},\label{eq:tdep}
\end{eqnarray}
where S, P, V, A and T denote scalar, pseudoscalar, vector, axial and tensor interactions, respectively. The coefficients $c_i^I(0)$ are given in appendix~\ref{sec:operators}, and 
\begin{equation}
    F(t) = \frac{1}{(1-t/Q_0^2)^2}, 
    \label{eq:formfactor}
\end{equation}
where the energy scale $Q_0$ depends on the specific hadronic form factor. Here, we assume $Q_0=1\GeV$~\cite{Bell:2020obw} for all operators and baryonic target species. 
This is a conservative choice based on the known values of $Q_0=0.9\GeV\pm0.1\GeV$, which covers the range of nucleon form factors~\cite{Zanotti:2017bte,Alarcon:2017ivh}.

\begin{table}
\centering
{\renewcommand{\arraystretch}{1.3}
\begin{tabular}{ | c | c | c | c | c |}
  \hline                        
  Name & Operator & $g_q$ & $g_i^2$ & $|\overline{M}(s,t,m_i)|^2$   \\   \hline
  D1 & $\bar\chi  \chi\;\bar q  q $ & $\frac{y_q}{\Lambda^2}$ & $\frac{c_i^S(t)}{\Lambda^4}$ & $ g_i^2(t)\frac{\left(4 m_{\chi }^2-t\right) \left(4 m_{\chi }^2-\mu ^2
   t\right)}{\mu ^2}$ \\  \hline
  D2 & $\bar\chi \gamma^5 \chi\;\bar q q $ & $i\frac{y_q}{\Lambda^2}$ & $\frac{c_i^S(t)}{\Lambda^4} $ & $g_i^2(t)\frac{t \left(\mu ^2 t-4 m_{\chi }^2\right)}{\mu ^2}$ \\  \hline
  D3 & $\bar\chi \chi\;\bar q \gamma^5  q $&  $i\frac{y_q}{\Lambda^2}$ & $\frac{c_i^P(t) }{\Lambda^4}$ &  $g_i^2(t) t \left(t-4 m_{\chi }^2\right)$ \\  \hline
  D4 & $\bar\chi \gamma^5 \chi\; \bar q \gamma^5 q $ & $\frac{y_q}{\Lambda^2}$ & $\frac{c_i^P(t)}{\Lambda^4}$ & $g_i^2(t) t^2$ \\  \hline
  D5 & $\bar \chi \gamma_\mu \chi\; \bar q \gamma^\mu q$ & $\frac{1}{\Lambda^2}$ & $\frac{c_i^V(t)}{\Lambda^4}$ &  $2 g_i^2(t) \frac{2 \left(\mu ^2+1\right)^2 m_{\chi }^4-4 \left(\mu ^2+1\right) \mu ^2 s m_{\chi }^2+\mu ^4 \left(2 s^2+2 s t+t^2\right)}{\mu^4}$ \\  \hline
  D6 & $\bar\chi \gamma_\mu \gamma^5 \chi\; \bar  q \gamma^\mu q $ & $\frac{1}{\Lambda^2}$ & $\frac{c_i^V(t)}{\Lambda^4}$ & $2  g_i^2(t)\frac{2 \left(\mu ^2-1\right)^2 m_{\chi }^4-4 \mu ^2 m_{\chi }^2 \left(\mu ^2 s+s+\mu ^2 t\right)+\mu ^4 \left(2 s^2+2 s
   t+t^2\right)}{\mu^4}$  \\  \hline
  D7 & $\bar \chi \gamma_\mu  \chi\; \bar q \gamma^\mu\gamma^5  q$ & $\frac{1}{\Lambda^2}$ & $\frac{c_i^A(t)}{\Lambda^4}$ &  $2  g_i^2(t) \frac{2 \left(\mu ^2-1\right)^2 m_{\chi }^4-4 \mu ^2 m_{\chi }^2 \left(\mu ^2 s+s+t\right)+\mu ^4 \left(2 s^2+2 s t+t^2\right)}{\mu^4}$ \\  \hline
  D8 & $\bar \chi \gamma_\mu \gamma^5 \chi\; \bar q \gamma^\mu \gamma^5 q $ & $\frac{1}{\Lambda^2}$ & $\frac{c_i^A(t)}{\Lambda^4}$ &  $2  g_i^2(t) \frac{2 \left(\mu ^4+10 \mu ^2+1\right) m_{\chi }^4-4 \left(\mu ^2+1\right) \mu ^2
   m_{\chi }^2 (s+t)+\mu ^4 \left(2 s^2+2 s t+t^2\right)}{\mu ^4}$ \\  \hline
  D9 & $\bar \chi \sigma_{\mu\nu} \chi\; \bar q \sigma^{\mu\nu} q $ & $\frac{1}{\Lambda^2}$ & $\frac{c_i^T(t)}{\Lambda^4}$ & $8  g_i^2(t) \frac{4 \left(\mu ^4+4 \mu ^2+1\right) m_{\chi }^4-2 \left(\mu ^2+1\right) \mu ^2 m_{\chi
   }^2 (4 s+t)+\mu ^4 (2 s+t)^2}{\mu ^4}$  \\  \hline
 D10 & $\bar \chi \sigma_{\mu\nu} \gamma^5\chi\; \bar q \sigma^{\mu\nu} q \;$ & $\frac{i}{\Lambda^2}$ & $\frac{c_i^T(t) }{\Lambda^4}$ &  $8  g_i^2(t)\frac{4 \left(\mu ^2-1\right)^2 m_{\chi }^4-2 \left(\mu ^2+1\right) \mu ^2 m_{\chi }^2 (4 s+t)+\mu ^4 (2 s+t)^2}{\mu^4}$ \\  \hline
\end{tabular}}
\caption{Dimension 6 effective operators~\cite{Goodman:2010ku} for the scattering of Dirac DM from quarks. The effective couplings for each operator are given as a function of the quark Yukawa coupling, $y_q$, and the cutoff scale, $\Lambda$. The fourth column shows the squared  effective coefficient $g_i^2$ for DM interactions with baryon species $i$, and the fifth column the squared matrix elements at high energy $|\overline{M}(s,t,m_i)|^2$ as a function of the Mandelstam variables $s$ and $t$ and the mass of the target $m_i$. 
The momentum dependent coefficients $c_i^I(t)$ are defined in Eq.~\ref{eq:tdep}. 
\label{tab:operatorshe} }
\end{table}

As in the case of point-like targets, where $F(t)=1$~\cite{Bell:2020jou,Bell:2020lmm}, when the DM-nucleon differential cross section depends exclusively on powers of the Mandelstam variable $t$ (operators D1-D4 in Table~\ref{tab:operatorshe}), then the scattering rates of Eq.~\ref{eq:intratedeftext} reduce to 
\begin{equation}
\Gamma^{-}(E_\chi) \propto \frac{1}{2^7\pi^3E_\chi k }\int_0^{E_\chi-m_\chi}q_0 dq_0 \int \frac{t_E^n dt_E }{\sqrt{q_0^2+t_E}} F(t) \left[1-g_0\left(\frac{E_i^{\,t^{-}}-\kinFi}{q_0}\right)\right],
\label{eq:gammaFFfinaltext}
\end{equation}
where $t_E=-t=q^2-q_0^2$, 
 $\kinFi$ is the Fermi energy 
 as in Fig.~\ref{fig:muradprofs},  
\begin{equation}
E_i^{\, t^{-}} = -\left(m_i+\frac{q_0}{2}\right) + \sqrt{\left(m_i+\frac{q_0}{2}\right)^2+\left(\frac{\sqrt{q^2-q_0^2}}{2}-\frac{m_i q_0}{\sqrt{q^2-q_0^2}}\right)^2}, 
\end{equation}
is the minimum energy of the target before the collision, obtained from kinematics,  and $g_0(x)$ is a step function with a smooth transition, 
\begin{align}
g_0(x) =  \begin{cases}
\, 1 \quad &x>0, \\
\, 1+x \quad &-1<x<0,\\
\, 0 \quad &x<-1.
\end{cases}
\end{align}
Details of the derivation of the integration domain for Eq.~\ref{eq:gammaFFfinaltext} can be found in ref.~\cite{Bell:2020jou}. 
An expression  similar to Eq.~\ref{eq:gammaFFfinaltext}  holds  for  amplitudes that depend on powers of the kind $t^n s^m$ (see Appendix~A of ref.~\cite{Bell:2020lmm}). It is worth noting that, in all cases, it is  possible to perform the integral over $t_E$ analytically in the zero temperature approximation, for both scattering amplitudes $\propto t^n$ and $\propto t^n s^m$, following the methodology outlined in refs.~\citep{Bell:2020jou,Bell:2020lmm}.

\begin{figure}[t] 
\centering
\includegraphics[width=0.75\textwidth]{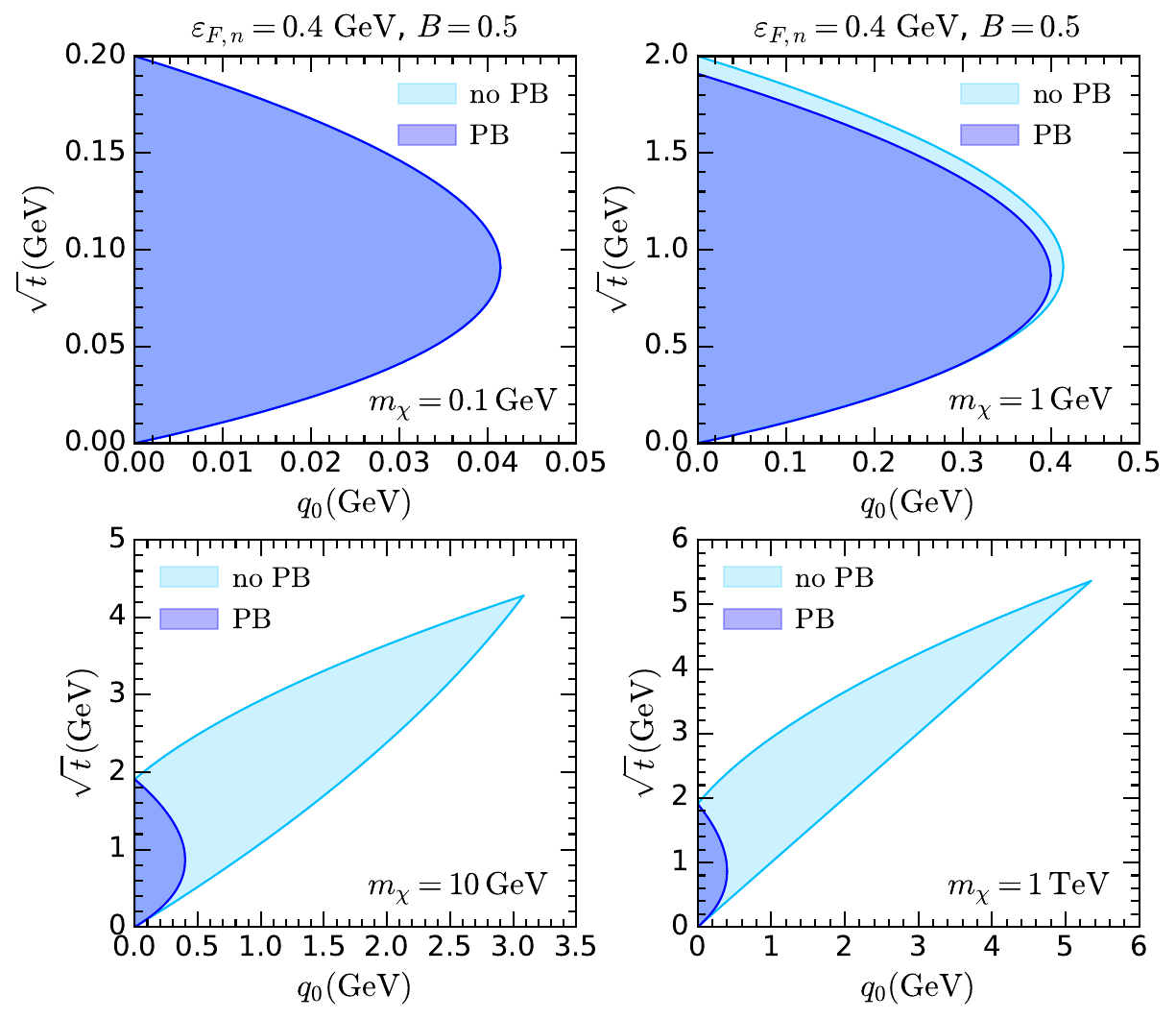}
\caption{Integration domain for the interaction rate, for different choices of DM mass, assuming   
  $B=0.5$ and $\kinFn=0.4\GeV$. The kinematically allowed region is shaded in light blue and the Pauli suppressed region (PB) in dark blue. 
}
\label{fig:intC}
\end{figure}

In Fig.~\ref{fig:intC}, we show how Pauli blocking affects the integration domain of Eq.~\ref{eq:gammaFFfinaltext}, which is controlled by the smoothed step function $g_0(x)$, for some representative choices of DM mass and the benchmark values $\kinFn=0.4\GeV$ and $B=0.5$.  The region where $g_0(x)=1$ is not kinematically allowed and hence not shown in Fig.~\ref{fig:intC}.
The light blue area represents the region that is not affected by Pauli blocking (PB), i.e. $1-g_0(x)=-x$, while the dark blue shaded area indicates the PB region, where $g_0(x)=0$. 
We clearly observe that for light DM masses such as  $m_\chi=0.1\GeV$ (top left panel), the whole domain lies in the PB region. For $m_\chi=1\GeV$ (top right panel), there is a tiny slice of the domain that is not Pauli suppressed. The picture changes dramatically when the DM mass is increased to $m_\chi=10\GeV$ (bottom left panel), where almost the whole integration domain is unaffected by PB. Increasing the DM mass even further, e.g., up to $m_\chi=1\TeV$ (bottom right panel) does not result in much further change to the shape of the domain, demonstrating that the transition between the Pauli blocked and non-PB regimes occurs between $m_\chi\sim1\GeV$ and $m_\chi\sim10\GeV$.

\begin{figure}[t] 
\centering
\includegraphics[width=0.7\textwidth]{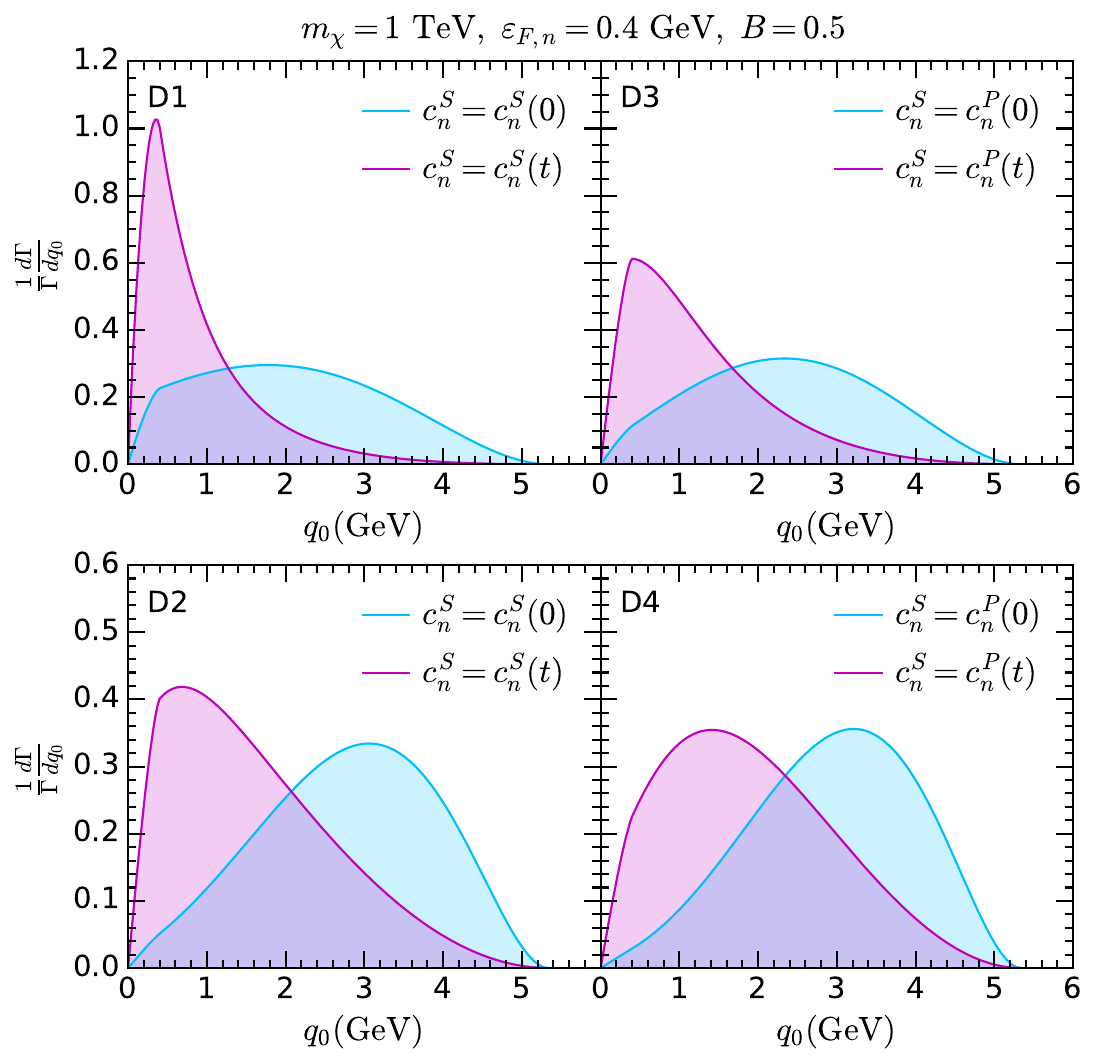} 
\caption{Normalised differential DM-neutron interaction rate as a function of the DM energy loss, $q_0$, for the operators D1 (top left), D3 (top right), D2 (bottom left) and D4 (bottom right). The light blue lines denote the interaction rate calculated using constant neutron form factors $c_n^S(0)$, $c_n^P(0)$, while the magenta lines correspond to that for momentum-dependent couplings $c_n^S(t)$, $c_n^P(t)$. We have set $m_\chi=1\TeV$, $B=0.5$, $\kinFn=0.4\GeV$.
}
\label{fig:intrateqtr1}
\end{figure}

In Fig.~\ref{fig:intrateqtr1} we show the normalised differential interaction rates for operators D1-D4 as a function of the energy loss $q_0$, calculated with (magenta) and without (light blue) momentum dependent couplings, for neutron targets and $m_\chi=1\TeV$. In all cases, we observe that the inclusion of $t$-dependent hadronic matrix elements shifts the peak of the spectrum towards lower energy transfers, $q_0$. 
Thus, the inclusion of $c_i^I(t)$ in Eq.~\ref{eq:gammaFFfinaltext} suppresses DM-neutron scatterings with large momentum transfer. 
Replacing the target mass with the corresponding effective mass $\mbeff\leq m_i$ also shifts  the average energy transfer to lower energies.  When both effects are present, a  lighter target mass reduces the suppression arising from $t$-dependent form factors.

Finally, we comment on deep inelastic scattering (DIS). Given that the momentum transfer in the DM-nucleon scattering process is sufficiently large that we cannot treat the nucleons as point particles, one may wonder if there is a sizeable DIS cross section.
However, despite momentum transfers for which DIS might be expected to become relevant, $Q>1\GeV$\footnote{The precise values is limited by our knowledge of the parton distribution functions.}, it is always a subdominant process.
This is a consequence of the limited phase space available in the scattering of heavy quasi-relativistic DM, as is applicable to the NS capture process.  In contrast, most well-known examples of DIS scattering in the literature involve a very different kinetic regime, such as the DIS scattering of relativistic neutrinos on nucleon targets, or the scattering of boosted DM.
For NS capture, the reduced phase space implies that DIS never dominates the total cross section, even for the heaviest NS considered here. Accounting for baryon effective masses further suppresses the DIS contribution to the total cross section, because of the smaller target mass. In addition, Pauli blocking of the baryonic final state is expected to drastically reduce the already sub-leading DIS contribution~\cite{BEBCWA59:1989ayi,Melnitchouk:1992gd}. Details of the DIS calculation can be found in Appendix~\ref{sec:dis}, together with plots of ratios of the DIS to total cross section (elastic plus DIS) in Fig.~\ref{fig:DISratio}.

In addition to DIS, there may, in principle, be a contribution from the excitation of baryon resonances.  However, the resonance excitation cross section is also expected to be suppressed. In NSs, the mass of the $\Delta$ baryon, which gives the largest contribution to this cross section, increases with the baryon number density by between $\sim100$ and $300 \MeV$~\cite{Motta:2019ywl}. This will suppress the  resonance excitation cross section significantly. Therefore, we consider only elastic collisions.

\subsection{Capture Rate}
\label{sec:caprateB}

DM capture in NSs occurs under different regimes, depending on the DM mass and the DM-target cross section \cite{Bell:2020jou}. 
For cross sections much larger than a threshold value  $\sigmathi$, we are in the optically thick regime, i.e., the DM sees the NS as a rigid sphere. This is the so called geometric limit. In this case, the capture rate  can be estimated using the following expression~\cite{Bell:2018pkk}, 
\begin{equation}
C_{geom} =  \frac{\pi R_\star^2[1-B(R_\star)]}{v_\star B(R_\star)} \frac{\rho_\chi}{m_\chi} \erf\left(\sqrt{\frac{3}{2}}\frac{v_\star}{v_d}\right),
\label{eq:capturegeom}    
\end{equation}
where $\rho_\chi$ is the local DM density, $\vstar$ is the NS velocity and $v_d$ is the DM velocity dispersion.

On the other hand, for cross sections well below $\sigmathi$, 
DM capture occurs in the optically thin limit. In this regime, DM is assumed to have a small probability to interact with the NS constituents.  
DM is free to propagate within the star and capture can, in principle, take place anywhere in the stellar interior. However, only a fraction of the DM flux traversing the star is effectively captured. 
Below, we derive general expressions for the capture rate in the optically thin limit, for various DM mass regimes, correctly incorporating the effects of baryon structure and strong interactions.

\subsubsection{Low and intermediate mass regime}
\label{sec:capsingle}

In the simplest approach, where we consider baryons as point-like particles that form an ideal Fermi gas within a NS, the most general expression for the capture rate that accounts for the stellar structure, Pauli blocking and GR effects is~\cite{Bell:2020jou,Bell:2020lmm} 
\begin{eqnarray}
C &=& \frac{4\pi}{\vstar} \frac{\rho_\chi}{m_\chi} {\rm Erf }\left(\sqrt{\frac{3}{2}}\frac{\vstar}{v_d}\right)\int_0^{\Rstar}  r^2 \frac{\sqrt{1-B(r)}}{B(r)} \Omega^{-}(r)  \, dr, \label{eq:capturefinalM2text} \\ 
\Omega^{-}(r) &=& \frac{\zeta(r)}{32\pi^3}\int dt dE_i ds  \frac{|\overline{M}(s,t,m_i)|^2}{s^2-[m_i^2-m_\chi^2]^2}\frac{E_i}{m_\chi}\sqrt{\frac{B(r)}{1-B(r)}}\frac{s}{\gamma(s,m_i)}\nn\\
&& \times\fFD(E_i,r)(1-\fFD(E_i^{'},r)).
\label{eq:intrate}
\end{eqnarray}
where 
$E_i$ and $E_i^{'}$ are the initial and final energy, respectively, of the target $i$, and  
\begin{eqnarray}
\gamma(s,m_i) &=& \sqrt{(s-m_i^2-m_\chi^2)^2-4m_i^2m_\chi^2}.
\end{eqnarray}
The quantity $\zeta(r)$ is a correction factor~\cite{Garani:2018kkd} that was introduced to allow for the use of realistic profiles for the target number density $n_i(r)$, while treating the targets as a free Fermi gas, 
\begin{eqnarray}
\zeta(r) &=& \frac{n_i(r)}{n_{free}(\kinFi(r),m_i)},\\
n_{free}(\kinFi(r),m_i) &=& \frac{[\kinFi(r)(2m_i+\kinFi(r))]^{3/2}}{3\pi^2}.
\end{eqnarray}
The integration intervals for $s$, $t$ and $E_i$ are given in ref.~\cite{Bell:2020jou}, and $\kinFi=\muFi-m_i$ is the Fermi energy. 
Note that Eq.~\ref{eq:intrate} accounts for Pauli blocking, which is given by the $1-\fFD$ term and is important for DM masses $m_\chi\lesssim 1\GeV$. 

Accounting for strong interactions between baryons, leads to two important modifications to the capture rate calculation~\cite{Bell:2020obw}:
\begin{enumerate}
    \item The baryon bare mass $m_i$ in Eq.~\ref{eq:intrate} is  replaced with its effective mass $\mbeff(r)$, which is no longer a constant but instead exhibits radial dependence. See the right-hand columns of Fig.~\ref{fig:NSradprofs1}.
    \item We calculate the Fermi energy as a function of the target number density and effective mass according to 
    \begin{equation}
        \kinFi(r) = \sqrt{[k_{F,i}(n_i(r))]^2+[\mbeff(r)]^2} - \mbeff(r),
    \end{equation}
    where $k_{F,i}$ is the Fermi momentum of the species $i$. 
    This is now the input of the $\fFD$ distributions. In other words, $\kinFi(r)$ is a function of the baryon effective mass and its number density such that
\begin{equation}
    n_{free}(\kinFi(r),\mbeff(r))=n_i(r). 
\end{equation}    
\end{enumerate}

\begin{figure}[t]
    \centering
\includegraphics[width=0.495\textwidth]{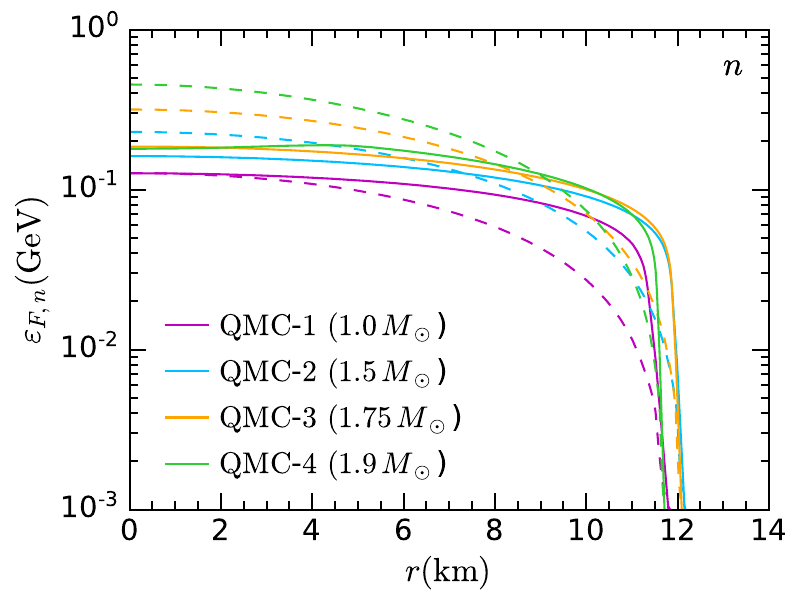}
\includegraphics[width=0.495\textwidth]{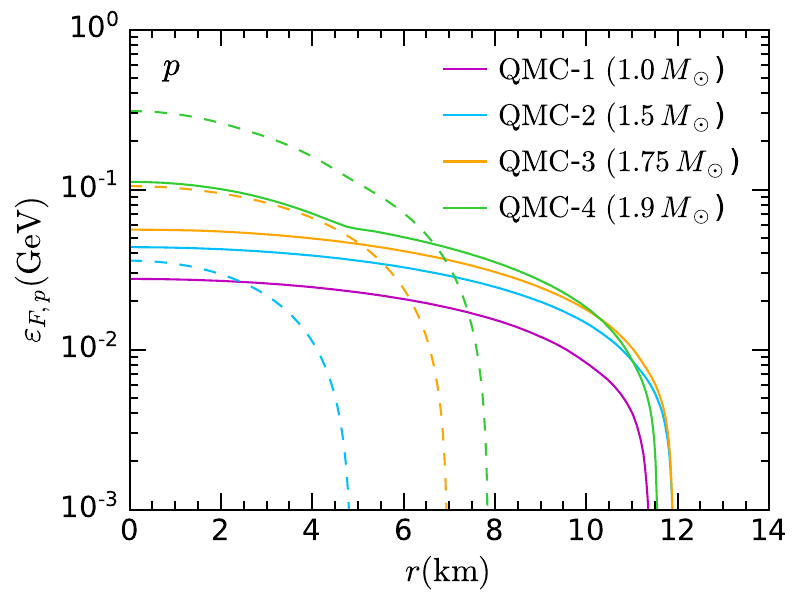}
    \caption{Radial profiles of the Fermi energy for neutrons (left) and protons (right). Results are shown for the free Fermi gas (dashed) and interacting baryon (solid) approaches, for the benchmark NS configurations of Table~\ref{tab:eos}.
    }
    \label{fig:muradprofs}
\end{figure}

Consequently, in the interacting baryon framework, we now set $\zeta(r)=1$ in Eq. \ref{eq:intrate} and the new expression for the scattering rate becomes~\cite{Bell:2020obw}
\begin{eqnarray} 
\Omega^{-}(r) &=& \frac{1}{32\pi^3}\int dt dE_i ds  \frac{|\overline{M}(s,t,\mbeff)|^2}{s^2-[(\mbeff)^2-m_\chi^2]^2}\frac{E_i}{m_\chi}\sqrt{\frac{B(r)}{1-B(r)}}\frac{s}{\gamma(s,\mbeff)}\nn\\
&& \times\fFD(E_i,r)(1-\fFD(E_i^{'},r)).
\label{eq:intratefinal}
\end{eqnarray}
Note that the dependence of the baryonic form factors on the momentum transfer is embedded in the definition of the squared matrix element; see Table~\ref{tab:operatorshe}.

In Fig.~\ref{fig:muradprofs} we show the values of the Fermi energies for neutrons (left panel) and protons (right panel). The dashed lines are the radial profiles used in the free Fermi gas approximation and the solid lines correspond to the  calculation outlined above for  the effective mass  approach. We immediately notice that the profiles for the free Fermi gas approximation are steeper, having larger values in the core and decreasing more rapidly towards the surface, while  those obtained with the effective mass approach are quite flat, especially in the core, and go to zero only very close to the surface as they follow the corresponding nucleon number density. Note that the difference between these two calculations  is more prominent for protons. Specifically, in the free Fermi gas approach, we find that protons are non-degenerate in the outer regions of the core and, for light NS configurations such as QMC-1, are in fact non-degenerate throughout the whole star.\footnote{Not shown in Fig.~\ref{fig:muradprofs} because of the logarithmic scale.}
In contrast, protons are degenerate throughout the star in the interacting baryon treatment. 
As we shall see, these different Fermi energies will have important consequences when calculating DM capture and interaction rates, especially for proton targets.

For DM masses  $m_\chi\gtrsim 1\GeV$, Pauli blocking does not affect the capture rate. In this case,  a simplified expression for the capture rate can be derived 
following the procedure in Appendix~\ref{sec:capratethinnondeg}. 
Thus, in the large mass range,  for $|\overline{M}|^2 = \bar{g}(s) t^n F(t)$, with $n=0,1,2$, we find that Eqs.~\ref{eq:capturefinalM2text} and \ref{eq:intratefinal} reduce to 
\begin{eqnarray}
C &\sim& \frac{1}{4 \vstar} \frac{\rho_\chi}{m_\chi^3}  {\rm Erf}\left(\sqrt{\frac{3}{2}}\frac{\vstar}{v_d}\right)\int_0^{\Rstar} r^2 dr  \, \left[4(\mbeff)^2\right]^n n_i(r)   \frac{\bar{g}(s_0)}{n+1}  \left[\frac{1-B(r)}{B(r)}\right]^{n+1} \mathcal{F}_n(|t_{min}|), 
\label{eq:cnongenerateapproxfinal}
\end{eqnarray}
where $s_0$ is defined in Appendix~\ref{sec:capratethinnondeg}  
and
\begin{eqnarray}
\mathcal{F}_n(|t_{min}|) &=&\frac{\int_0^{|t_{min}|} dt \, t^n F(t)}{\int_0^{|t_{min}|} dt \,  t^n}<1,\\
|t_{min}| &=& \frac{\gamma^2(s)}{s} = \frac{4(1-B(r))m_\chi^2}{B(r)\left(1+\mu^2\right)+2\sqrt{B(r)}\mu} \sim \frac{4(1-B(r))(\mbeff)^2}{B(r)}.
\end{eqnarray}
This simplified equation is valid 
when either $m_\chi\gg m_i$ or $\kinFi\ll m_i$. In this case, we are in the $m_\chi\gg m_i$ regime. 
It is evident from Eq. \ref{eq:cnongenerateapproxfinal} that there is an additional source of suppression in this regime stemming from the factor $\mathcal{F}_n(|t_{min}|)$. The $\mathcal{F}_n$ functions can be calculated analytically, 
\begin{eqnarray}
\mathcal{F}_0(t) &=&\frac{Q_0^2}{Q_0^2+t},\\
\mathcal{F}_1(t) &=&2\frac{Q_0^4}{t^2}\left[\log\left(1+\frac{t}{Q_0^2}\right)-\frac{t}{Q_0^2+t}\right],\\
\mathcal{F}_2(t) &=&6\frac{Q_0^6}{t^3}\left[-\log\left(1+\frac{t}{Q_0^2}\right)+\frac{t(2Q_0^2+t)}{2Q_0^2(Q_0^2+t)}\right]. 
\end{eqnarray}
Note that $\mathcal{F}_n$ are decreasing functions of $n$ and $|t_{min}|$, and therefore of the target mass. 
When accounting for the baryon structure in the free Fermi gas approximation, the capture rate acquires 
  a similar suppression factor $\mathcal{F}_n(|t_{min}|)$, with $\mbeff$ replaced by the rest mass $m_i$ in Eq.~\ref{eq:cnongenerateapproxfinal}.

\subsubsection{Large mass regime}
\label{sec:caprateFFMS}

For DM masses larger than a threshold denoted $\mstar$,  a quantity that depends on the target species $i$ and the nature of the interaction,  
a single collision is not sufficient for a DM particle to lose enough energy to become gravitationally bound to the NS, 
i.e., multiple scatterings are required~\cite{Bell:2020jou}. 
To account for this effect, 
we estimate the value of $\mstar$ by equating the DM initial energy 
(at infinity)
    \begin{equation}
        E_\chi^\infty=\frac{1}{2}m_\chi u_\chi^2,
    \end{equation}
(where $u_\chi^2\sim \vstar^2+v_d^2$), to the average energy loss $\bar{q}_0$, which is computed using the interaction rate. For further details, see ref.~\citep{Bell:2020jou}. 
To account for the transition from the single to the multiple scattering regime ($m_\chi\sim \mstar$), we also rely on the methodology described in ref \citep{Bell:2020jou}. Accordingly, we  define $c_1$ to be the probability that a single scattering interaction will result in capture of the DM particle
\begin{equation}
    c_1=\frac{\int_0^{\infty}du_\chi \frac{\fMB(u_\chi)}{u_\chi}P_1\left(\frac{1}{2} \frac{m_\chi u_\chi^2}{\sqrt{B(r)}}\right)}{\int_0^{\infty}du_\chi \frac{\fMB(u_\chi)}{u_\chi}}, 
    \label{eq:corectc1}
    \end{equation} 
where we have assumed that the DM velocity at infinity follows a Maxwell-Boltzmann (MB) distribution, and $P_1(\delta q_0)$ is the probability that the DM loses an amount of energy of at least $\delta q_0$ after a single scattering~\cite{Bell:2020jou}
\begin{equation}
P_1(\delta q_0)=\int_{\delta q_0}^\infty dx \frac{1}{\Gamma(x,B(r),\kinFi(r))}\dfrac{d\Gamma}{dx}(x,B(r),\kinFi(r)). \label{eq:P1}
\end{equation}
Note that $c_1$ depends on $\kinFi(r)$, $B(r)$ and $m_i$ through the interaction rate $\Gamma$ (see section~\ref{sec:intrate}). In the limit $m_\chi\ll \mstar$ we recover $c_1=1$, i.e., capture via a single collision. For $m_\chi\gtrsim \mstar$ we instead find 
\begin{eqnarray}
   c_1 = \frac{1}{n^*_i} &=& 1-e^{-\mstar/m_\chi},
   \label{eq:mstar}
\end{eqnarray}
where $n^*_i$ can be understood as the average number of collisions with the target species $i$ required to remove DM particles from the incoming flux. In the limit $m_\chi\gg \mstar$, we simply have $c_1 \rightarrow  \frac{\mstar}{m_\chi}$.

A suitable capture rate approximation that accounts for multiple scattering is obtained by inserting $c_1$ in Eq.~\ref{eq:capturefinalM2text}~\cite{Bell:2020jou}
\begin{eqnarray}
C_{approx}^* =\frac{4\pi}{\vstar}\frac{\rho_\chi}{m_\chi}{\rm Erf}\left(\sqrt{\frac{3}{2}}\frac{\vstar}{v_d}\right)  \int  r^2 dr  \frac{\sqrt{1-B(r)}}{B(r)}\Omega^{-}(r) \frac{1}{n^*(r)}. 
\label{eq:chighmass}
\end{eqnarray}
We see that the parameter $\mstar$ quantifies the degree of suppression of the capture rate in the multiple-scattering regime.
In the  free Fermi gas approximation with point-like targets it is calculated using Eqs.~\ref{eq:corectc1}, \ref{eq:P1} and \ref{eq:mstar}. Note that, aside from $B$, $\mstar$ depends only on 2 quantities that have dimension of energy: the  mass of the target $m_i$ and its Fermi energy $\kinFi$. 
Interestingly, if one rescales both $m_i$ and $\kinFi$ by the same factor, the resulting $\mstar$ will just acquire this  factor. 
This means that for baryon couplings at zero momentum transfer, we can calculate $\mstar$ for a generic target mass $\mbeff$ by rescaling the values obtained in the free Fermi gas approximation in the following way: 
\begin{equation}
\mstar(\mbeff, \kinFi)= \frac{\mbeff}{m_i} \mstar(m_i,\kinFi\frac{m_i}{\mbeff}).
\end{equation}
When including the dependence of the baryon couplings on the transferred  momentum, an additional energy scale comes into play, namely $Q_0$, and hence it is no longer possible to rescale the $\mstar$ values obtained with $c_i^I(0)$.

  \begin{figure}[t]
    \centering
\includegraphics[width=0.49\textwidth]{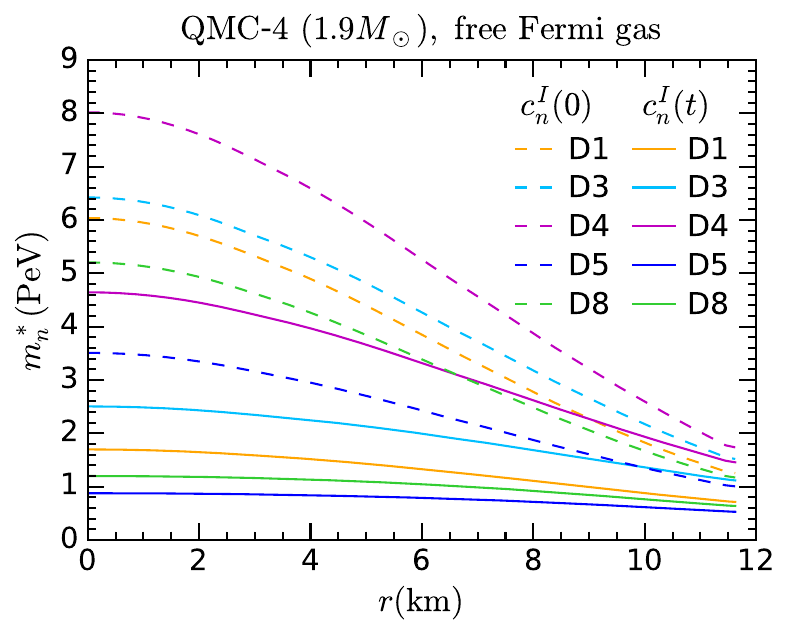}    
\includegraphics[width=0.5025\textwidth]{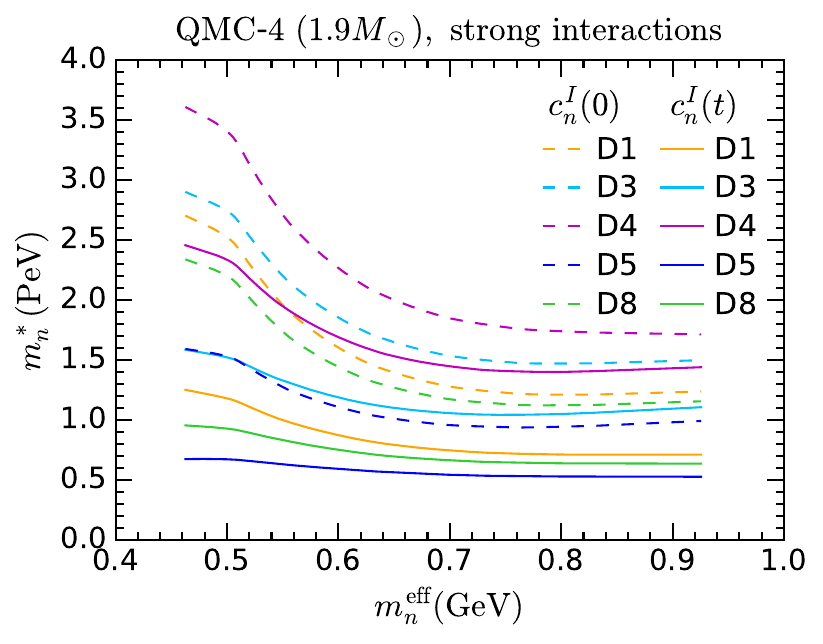}
    \caption{Radial profiles  of $\mnstar$ in the free Fermi gas approach (left) and $\mnstar$ as a function of the  neutron effective mass (right) in the interacting baryon approach, for selected operators, assuming the NS configuration QMC-4 ($1.9\Msun$). In both cases,  $\mnstar$ has been calculated using neutron couplings at zero momentum transfer (dashed lines) and including the dependence on $t$ (solid lines). }
    \label{fig:mstar}
\end{figure}

In the left panel of Fig.~\ref{fig:mstar}, we  show radial profiles of $\mnstar$ for five representative  operators in the case of a QMC-4 NS. These profiles have been calculated in the free Fermi gas approximation  with (solid lines) and without (dashed lines) including momentum dependent neutron form factors $c_i^I(t)$, i.e. the neutron mass remains constant. 
The profiles illustrate the variation of $\mnstar$ with $B(r)$ and $\kinFn(r)$ in the stellar interior. Note that $\mnstar$ also depends on the  DM velocity distribution, which we have taken to be Maxwell-Boltzmann. We can see that the inclusion of $t$-dependent neutron couplings lowers $\mnstar$ by a factor in excess of $\sim 1.2 - 4.5$, with D4 and D3 being less affected. 
This is because of the suppression of large energy transfers when introducing  $c_i^I(t)$ (as illustrated in Fig.~\ref{fig:intrateqtr1}) which results in less energy being lost by the DM particle per collision.  A lower value of $\mnstar$ means that multiple scattering is relevant at smaller DM masses than those expected with $c_i^I(0)$. 
In the interactive baryon approach, $\mnstar$ is also a function of $\mneff(r)$, as plotted in the right hand panel of Fig.~\ref{fig:mstar}. 
 The values of $B(r)$, $\kinFn(r)$ and $\mneff(r)$ used in the right hand panel are those corresponding to the appropriate radial coordinate within the NS, as in the left hand panel. 
In this case, when including the dependence on the transferred momentum (solid lines), $\mnstar$ is reduced by a factor of $\sim 1.2 - 2.5$ with respect to the result obtained with hadronic matrix elements at  zero momentum transfer (dashed lines). Thus, multiple scattering is relevant at an even lower DM mass in the complete approach that accounts both for strong interactions and $t$-dependent nucleon couplings. 
The remaining operators show a similar behaviour to those presented in Fig.~\ref{fig:mstar}.

\section{Results}
\label{sec:results}

\subsection{Capture Rate}
\label{sec:caprateresults}
In this section, we present our results for the capture rate, 
for each EFT operator in Table~\ref{tab:operatorshe} and every baryonic  species in the  QMC family. The rates have been calculated in the optically thin limit using Eq.~\ref{eq:capturefinalM2text} for $m_\chi\lesssim \mstar$ and  Eq.~\ref{eq:chighmass} for $m_\chi\gtrsim \mstar$ wherever the target is degenerate, and Eq.~\ref{eq:csimplenomu} in the non-degenerate regime\footnote{We have numerically solved these equations using the \texttt{CUBA} libraries \citep{Hahn:2004fe,Hahn:2014fua} linked to \texttt{Mathematica}~\cite{Mathematica}.}. We compute the capture rate for each target species individually; these can be summed to obtain the total capture rate. Since we shall always work in the optically thin limit (i.e., where the probability for more than one scattering interaction is very low) this procedure is a good approximation, even in the multi-scattering mass region.
We assume a NS located in the Solar neighbourhood, thus $\rho_\chi=0.4\GeV\cm^{-3}$, $\vstar=230\km\s^{-1}$ and $v_d=270\km\s^{-1}$.

\subsubsection{Nucleons}
\label{sec:capresnucleons}

\begin{figure}[t] 
\centering
\includegraphics[width=\textwidth]{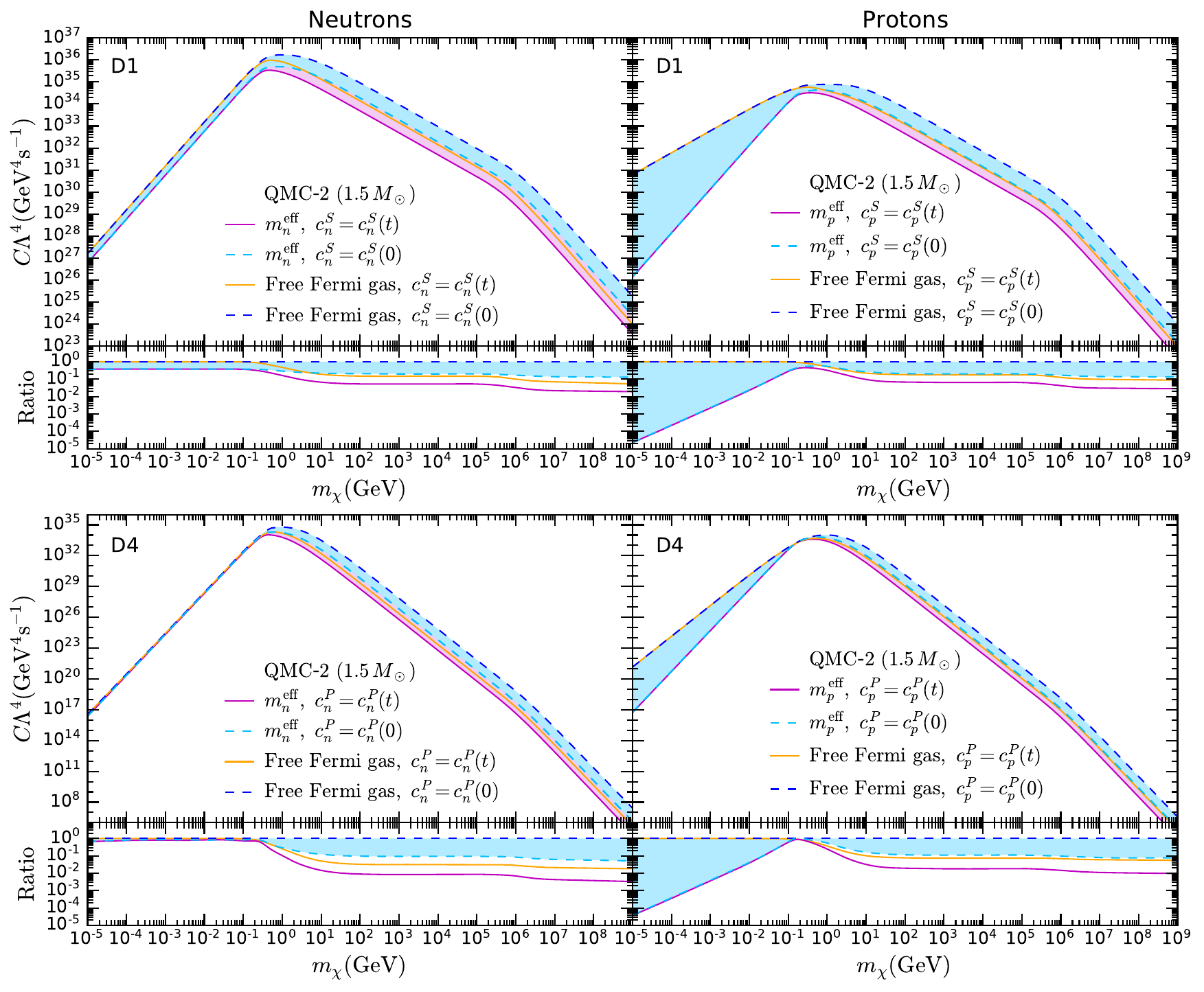}
\caption{
Capture rate in the optically thin limit for the operators D1 (top) and D4 (bottom) as a function of the DM mass $m_\chi$ for neutron (left) and proton (right) targets, using the free Fermi gas approach with constant nucleon couplings (dashed blue) and momentum dependent couplings (solid orange), and the interacting nucleon approach for constant couplings (dashed light blue)  and  momentum dependent couplings (solid magenta), for a QMC-2 NS configuration. Note that these rates scale as $\Lambda^{-4}$. 
The ratio of the capture rate with respect to that for the free Fermi gas approximation for point-like targets (dashed dark blue) is shown in the lower panels. 
}
\label{fig:capratesD1D4}
\end{figure} 

\begin{figure}[t] 
\centering
\includegraphics[width=\textwidth]{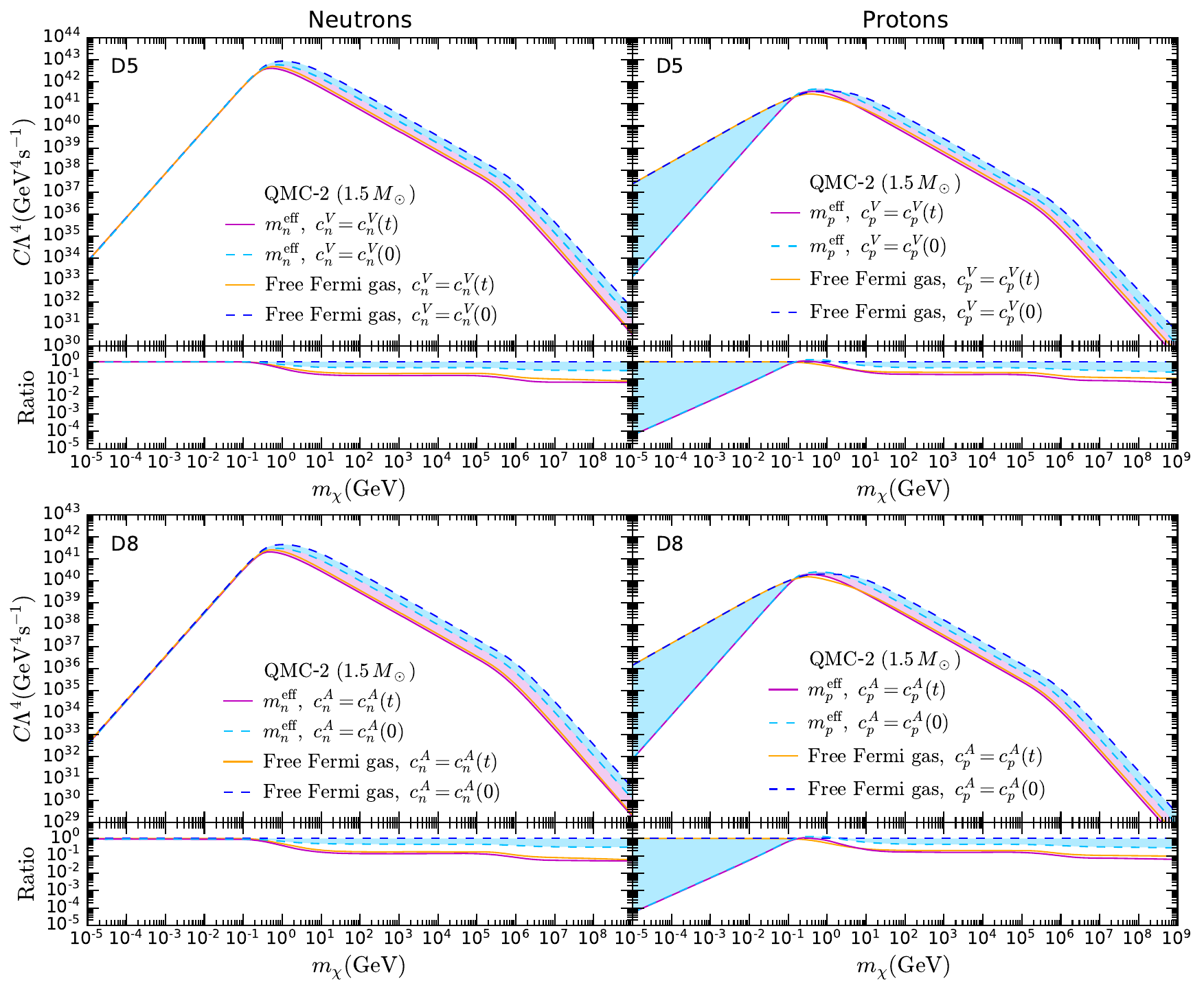}
\caption{
Capture rate in the optically thin limit for the operators D5 (top) and D8 (bottom) as a function of the DM mass $m_\chi$ for neutron (left) and proton (right) targets,
using the free Fermi gas approach with constant nucleon couplings (dashed blue) and momentum dependent couplings (solid orange), and the interacting nucleon approach for constant couplings (dashed light blue)  and  momentum dependent couplings (solid magenta), for a QMC-2 NS configuration. 
Note that these rates scale as $\Lambda^{-4}$. 
The ratio of the capture rate with respect to that for the free Fermi gas approximation for point-like targets (dashed dark blue) is shown in the lower panels. 
}
\label{fig:capratesD5D8}
\end{figure} 

To illustrate the effect of accounting for  strong interactions and nucleon couplings that depend on the momentum transfer, we show in  Figs.~\ref{fig:capratesD1D4} and \ref{fig:capratesD5D8} the capture rate  for four representative operators, D1, D4, D5, and D8, for a QMC-2 NS configuration ($1.5\Msun$). 
Note that in these figures we have not assumed a value of $\Lambda$, i.e. we plot $C \Lambda^4$. 
These figures extend the results of ref.~\cite{Bell:2020obw} for neutrons to the multi-scattering regime, and present our findings for proton targets. Results are shown for the free Fermi gas approximation with constant nucleon couplings (dashed dark blue) and momentum dependent form factors (orange),  and for the interacting baryon framework with (magenta) and without (dashed light blue) $t$-dependent nucleon couplings.  Recall that the interaction rate $\Omega^-(r)$ in Eqs.~\ref{eq:capturefinalM2text} and \ref{eq:chighmass}, is calculated using Eq.~\ref{eq:intrate}  in the ideal Fermi gas approximation and with  Eq.~\ref{eq:intratefinal} in the approach that considers strong interactions.

For neutron targets in the free Fermi gas approach, we find that including  momentum dependent form factors in the scattering cross section does not affect the capture rate for DM masses $m_\chi\lesssim0.2\GeV$, as expected, since the energy scale at which this effect comes into play is $Q_0\sim1\GeV$. For $m_\chi\gtrsim1\GeV$, on the other hand, the ratio of the calculation that accounts for $t$-dependent neutron couplings with respect to that obtained with constant hadronic matrix elements  is ${\cal O}(10^{-2})$ 
for D4 in the large DM mass range; for the remaining operators the suppression exceeds one order of magnitude 
(compare the orange and dashed dark blue lines in the lower panels). 
The fact that this effect is stronger for operators whose matrix elements are a function of larger powers of $t$, such as D4, can be seen from  Eq.~\ref{eq:cnongenerateapproxfinal}, where we note that 
$\mathcal{F}_n$ is a decreasing function of $n$. 
We also note that the suppression caused by the form factors is stronger in the multiple scattering regime, $m_\chi\gtrsim\mnstar$, because the values of $\mnstar$ become lower when $t$-dependent neutron couplings are introduced (see Fig.~\ref{fig:mstar}, left panel). As mentioned above, this is a consequence of the form factors imposing a cutoff on the 
size of the momentum transfer in the capture process (see Fig.~\ref{fig:intrateqtr1}), hence lowering the average DM energy loss per scattering. 
It is worth remarking that the suppression caused by the momentum dependence of the form factors is even more pronounced in heavier NSs~\cite{Bell:2020obw}. 
Similar conclusions are obtained for DM capture associated with scattering from protons (right panels), where the suppression of the capture rate is slightly smaller than that for neutrons. 
As previously stated, we have taken $Q_0=1\GeV$ as a conservative choice; smaller values of $Q_0$ will result in a stronger suppression of the capture rate.

We now turn to the effect of strong interactions on the capture process. For $m_\chi\lesssim0.2\GeV$, the DM mass range where Pauli blocking is in play, the capture rate due to DM-neutron scattering is almost identical for the free Fermi gas (dashed dark blue) and interacting baryon (dashed light blue) approaches, for operators D3-D10. This is because capture occurs very close to the NS surface~\cite{Bell:2020jou}. Operators D1 and D2 suffer a relatively small overall rescaling, because their interaction rates scale with the neutron mass, in this case $(\mneff)^2$.

For protons, however, there is a significant difference in the capture rate in the low DM mass range. In the free Fermi gas approach, protons  are  degenerate only in the innermost region of the NS core, with the exact extent of that region dependent on the NS configuration (see Fig.~\ref{fig:muradprofs}, dashed lines).  For the particular NS model QMC-2, proton targets are affected by Pauli blocking only within a radius of $\sim4\km$ from the NS centre. Consequently, in the free Fermi gas approach (dashed dark blue) the capture rate for scattering on protons is not Pauli suppressed at low DM mass, in contrast to that for neutrons; see the different slope of the capture rates at $m_\chi\lesssim m_p$. (To calculate the capture rate in the non-degenerate region, we use Eq.~\ref{eq:csimplenomu}. For details of the derivation of this expression, see Appendix~\ref{sec:capratethinnondeg}.) 
As a result, capture on proton targets surpasses the contribution of the dominant species, neutrons, in the free Fermi gas approximation and light DM mass regime. 
However, in the interacting baryon approach, protons are degenerate over a much wider region of the stellar interior (see Fig.~\ref{fig:muradprofs}, solid lines). Therefore, the more accurate interacting baryon approach leads to much greater Pauli suppression of the capture rate for scattering on protons. Indeed, the proton contribution to the total capture rate is lower than that of neutrons in most cases. In the case of a QMC-2 NS configuration, the sole exception is for the D4 interaction. Therefore,  the ratio of capture rates for the free Fermi gas and interacting baryon approaches is largest for the scattering of light DM on protons, and exceeds 4 orders of magnitude at $m_\chi=10\keV$, for all operators  (see light blue shaded regions of Fig.~\ref{fig:capratesD1D4} and Fig.~\ref{fig:capratesD5D8}).

For DM masses above $m_\chi\sim m_n$, the capture rate in the interacting baryon framework, with constant nucleon couplings, is lowered by up to one order of magnitude compared to that for the free Fermi gas approach, for both neutron and proton targets, for the case of scalar and pseudoscalar operators. See the light blue shaded regions of Fig.~\ref{fig:capratesD1D4}.
For the remaining operators, the suppression reaches the $\sim30\%$ level in the large DM mass region.  See the light blue shaded region of Fig.~\ref{fig:capratesD5D8}. 
 Note that the capture rate depends on the DM-target reduced mass, which for $m_\chi\gg m_i$ approaches  the target mass. Furthermore, in this mass regime, DM capture can occur deep inside the star, where $\mbeff<m_i$ (see Fig.~\ref{fig:NSradprofs1}), hence the capture rate is suppressed by a lower target mass.
Introducing momentum dependent form factors in the interacting baryon approach (magenta lines) results in a similar reduction of the capture rate to that of the ideal Fermi gas formalism (orange lines), especially for operators D5-D10. For operators D1-D4, the capture rate is lowered by $\sim3$ orders of magnitude in the multiple scattering regime for both neutron and proton targets.

It is worth noting that the combined effect of using the interacting baryon approach and including momentum dependent form factors 
is smaller than the product of the two individual effects. This is because the form factor suppression $\mathcal{F}_n$ in Eq.~\ref{eq:cnongenerateapproxfinal} is a decreasing function of the target mass, which can  reach values as small as $\mneff\sim0.5m_n$ for nucleon targets, thereby resulting in a weaker reduction of the capture rate when compared to the free Fermi gas approach. 
Note also that the DM mass for which multiple scattering effects are relevant is smaller when considering both nucleon effective masses and momentum dependent form factors, being ${\cal O}(10^5\GeV)$.

In Appendix~\ref{sec:uncereos}, we have compared  our results for the QMC EoS with those for the BSk24 functional~\cite{Goriely:2013,Pearson:2018tkr}  (a Skyrme type EoS).  We find that the size of the effects described above are largely independent of the EoS.

\subsubsection{Hyperons}
\label{sec:capresexotic}

\begin{figure}[t] 
\centering
\includegraphics[width=\textwidth]{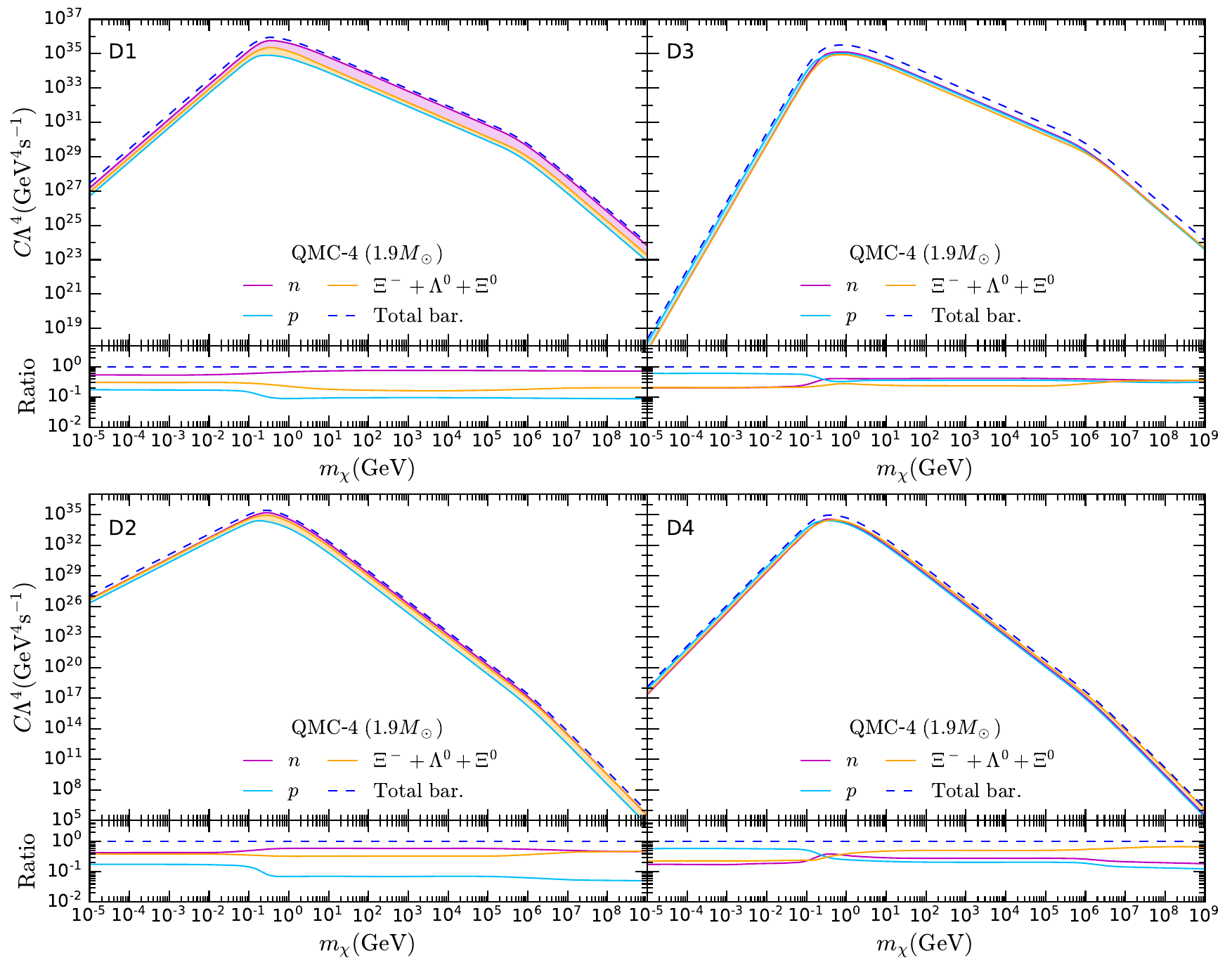}
\caption{Capture rate  in the optically thin limit for operators D1-D4 as a function of the DM mass $m_\chi$ for nucleons and exotic targets in the NS benchmark configuration QMC-4 ($1.9\Msun$). All capture rates were calculated using the complete approach that accounts for strong interactions and momentum dependent form factors for baryons. Note that these rates scale as $\Lambda^{-4}$. The lower panels show the contribution of each baryonic species to the total capture rate associated with DM interactions with baryons (dashed blue line). 
}
\label{fig:capratesD1D4_Hyper}
\end{figure}  

\begin{figure}[t] 
\centering
\includegraphics[width=\textwidth]{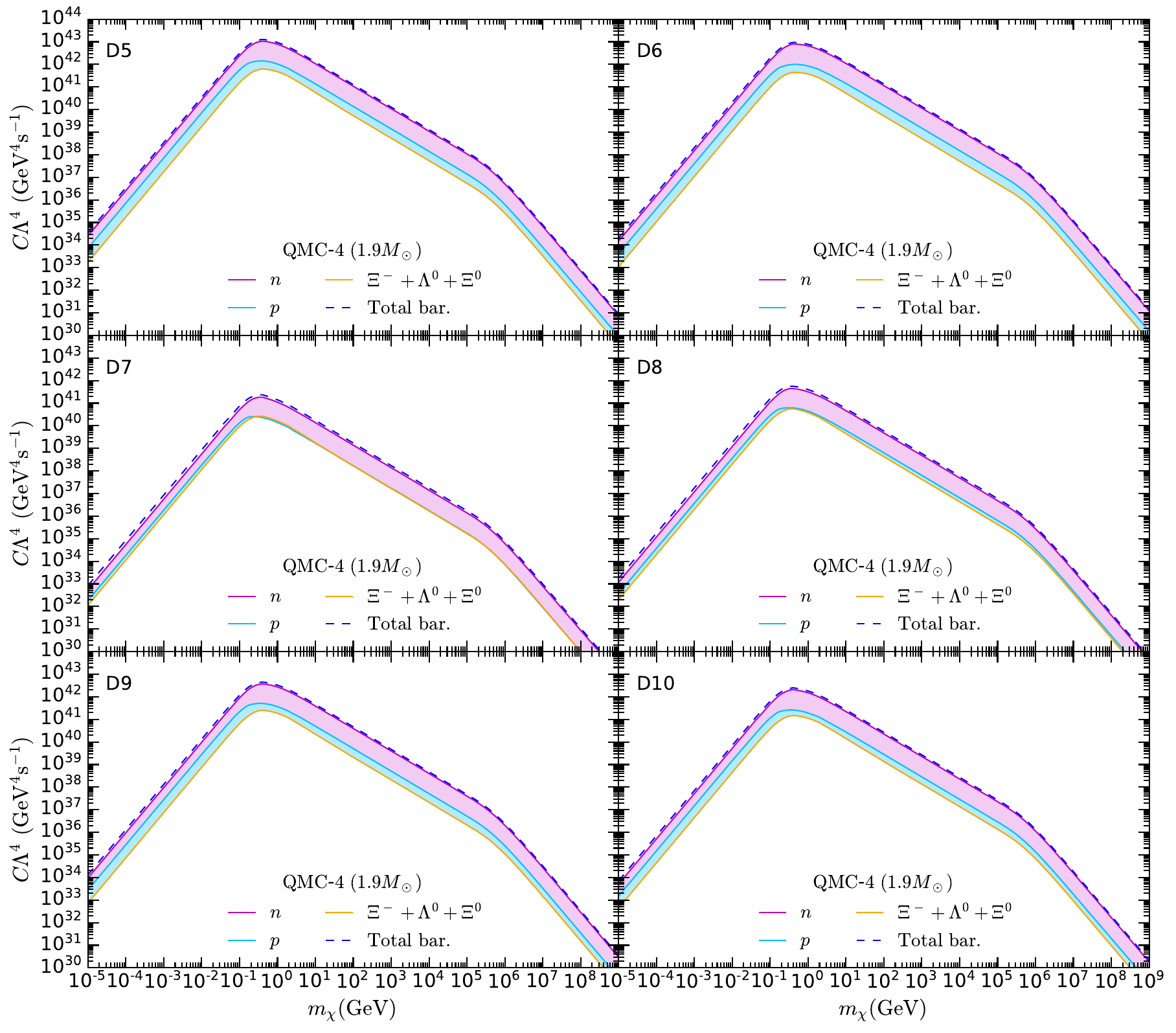}
\caption{Capture rate  in the optically thin limit for operators D5-D10 as a function of the DM mass $m_\chi$ for nucleons and exotic targets in the NS benchmark configuration QMC-4 ($1.9\Msun$). All capture rates were calculated using the complete approach that accounts for strong interactions and momentum dependent form factors for baryons. Note that these rates scale as $\Lambda^{-4}$. 
}
\label{fig:capratesD5D10_Hyper}
\end{figure}  

In Figs.~\ref{fig:capratesD1D4_Hyper} and  \ref{fig:capratesD5D10_Hyper}, we show the capture rates $C \Lambda^4$ for all the  baryon targets in the  benchmark NS QMC-4, calculated using the interactive baryon framework with momentum dependent DM couplings for all the baryonic species. As outlined in section~\ref{sec:NSmodels}, the NS configuration QMC-4 contains $\Lambda^0$, $\Xi^-$ and $\Xi^0$ hyperons in the inner core.  
The orange line represents the sum of the capture rate due to scattering on all of the hyperonic species. Their individual contributions are determined by their abundance in the NS core (see Fig.~\ref{fig:NSradprofs1}, bottom left panel) and hence $\Xi^-$ and $\Lambda^0$ give sizeable contributions to the capture rate while that of $\Xi^0$ is negligible.

For operators D1 and D5-D10, scattering on neutrons (magenta) clearly dominates the total capture rate (dashed dark blue) throughout the whole DM mass range considered here, as expected, followed by protons (light blue) and hyperons. 
For D3 and D4, proton targets provide the largest contribution to the capture rate in the Pauli blocked region $m_\chi\lesssim0.2\GeV$. This  occurs because of the interplay of two facts. First, the  proton Fermi energy is lower than that of neutrons close to the surface (see Fig.~\ref{fig:muradprofs}, solid lines). In fact, the proton contribution to the total capture rate is largest in the light DM mass regime for all operators. 
Second, operators whose matrix elements are a function of larger powers of $t$ (such as D3 and D4) are more greatly affected by Pauli blocking, i.e., the neutron contribution is more suppressed for these operators.

For D3 and D4, the capture rates due to scattering on the three species, neutrons, protons and hyperons, are all of similar magnitude when $m_\chi\gtrsim0.2\GeV$. In fact, hyperons surpass the neutron contribution for D4, especially in the multiple scattering regime.  Recall that hyperon effective masses are much larger than the nucleon masses in the inner core (see Fig.~\ref{fig:NSradprofs1}, bottom right panel). Thus, despite being under-abundant, their contribution to the capture rate is enhanced by a larger DM-target reduced mass and by the D4 matrix element being $\Msq\propto t^2$ ($n=2$ in  Eq.~\ref{eq:cnongenerateapproxfinal}). 
For D2-D4, the capture rates due to scattering on neutrons and hyperons are almost identical for $m_\chi\lesssim0.1\GeV$, with hyperons once again surpassing neutrons in the particular case of D4.  This occurs because the depth of the Fermi sea of hyperons is much lower than that of the neutrons in the NS core, thus leading to less Pauli blocking. 

In addition, we have compared these results with those for a NS configuration of very similar compactness to that of QMC-4, but obtained with a QMC model that does not contain hyperons. We have found that the total capture rate due to DM interactions with baryons for operators D2-D4 is actually enhanced by the presence of hyperonic matter in the NS core. For D3 and D4, this enhancement reaches a factor of $\sim 2$ and $\sim 3$, respectively.

\subsection{Threshold Cross Section}
\label{sec:thxs}

\begin{figure}[t]
    \centering
\includegraphics[width=\textwidth]{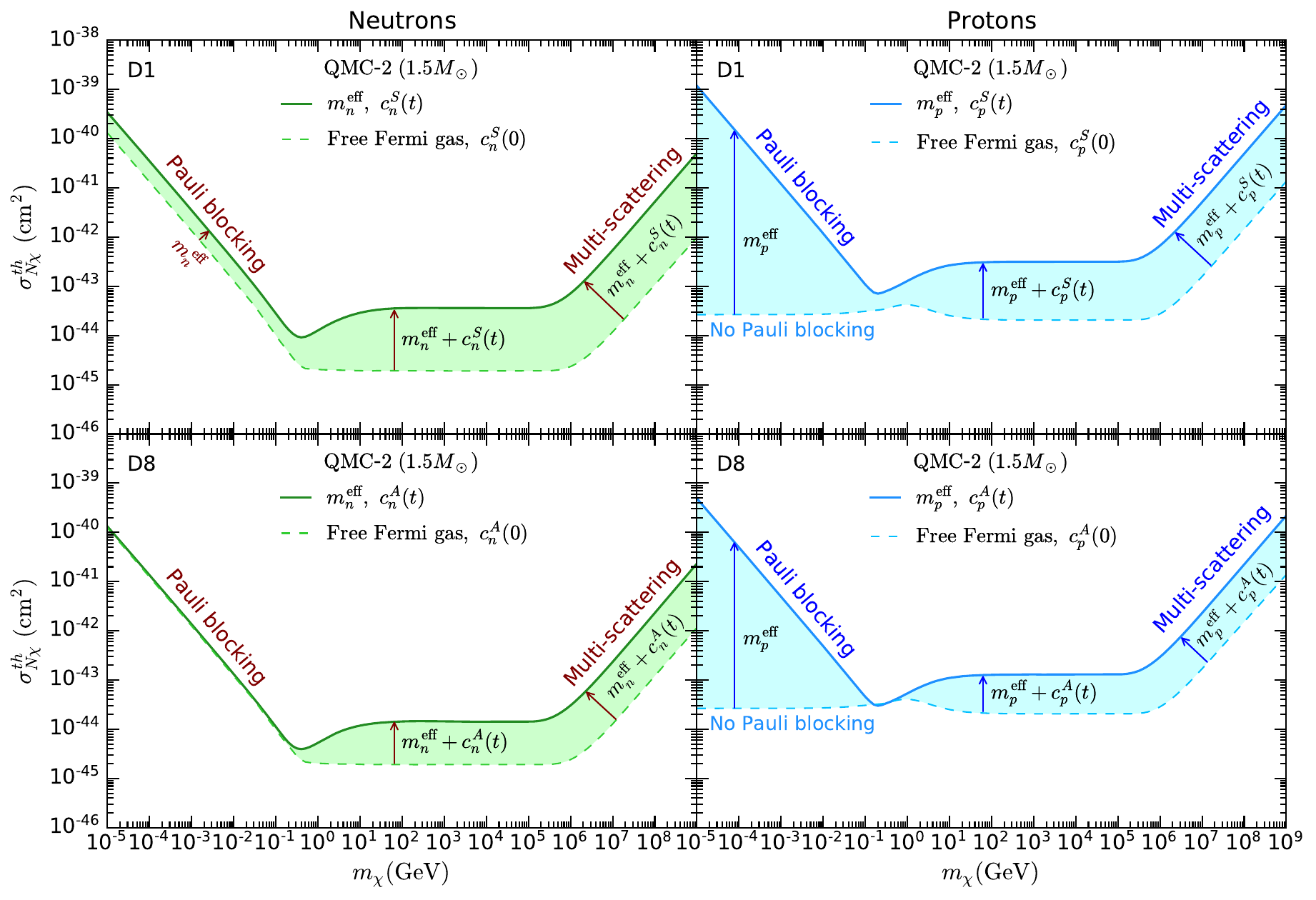}
\caption{Threshold cross section for  neutron (left) and proton (right) targets with scalar-scalar (D1, top) and axial-axial (D8, bottom) interactions with DM in a QMC-2 ($1.5\Msun$) NS. 
The solid lines represent the result obtained using the interacting baryon framework ($\mbeff$) and including the dependence of the hadronic matrix elements on the momentum transfer, $c_N^{S,A}(t)$. The dashed lines correspond to the free Fermi gas approximation and nucleon couplings at zero momentum transfer. }
    \label{fig:sigmathcomp}
\end{figure}

The threshold cross section $\sigmathi$ for a given target is defined  as the cross section for which the capture rate in the optically thin regime $C(m_\chi,\sigma(m_\chi,\Lambda))$ reaches the geometric limit  $C_{geom}$~\cite{Bell:2020jou}. Above this natural threshold the capture rate saturates, and hence $\sigma = \sigmathi$ cannot be distinguished from $\sigma\ge\sigmathi$.
For the intermediate DM mass range where NS capture is most efficient, i.e., where neither Pauli blocking nor multiscattering are relevant, a value of $\sigmathn = \textrm{few}\times10^{-45}\cm^2$ is commonly assumed in the literature.  However, as we shall see the combined effects of nucleon interactions and momentum dependence of the hadronic matrix elements imply that this value cannot be reached.

In Fig.~\ref{fig:sigmathcomp}, we show the impact of accounting for both strong interactions and nucleon structure, on the threshold cross section for both neutrons (left) and protons (right)  and operators D1 (top) and D8 (bottom) in the NS benchmark QMC-2 ($1.5\Msun$). The dashed lines represent the results obtained in the free Fermi gas approximation, which yields $\sigmathn\simeq2\times10^{-45}\cm^2$ in the $1\GeV\lesssim m_\chi\lesssim 4\times 10^5\GeV$ range and $\sigmathp\simeq2\times10^{-44}\cm^2$ in a similar DM mass range for both operators. Below $m_\chi\sim1\GeV$ the NS sensitivity is hampered by Pauli blocking, while multiple scattering comes into play above $m_\chi\sim4\times10^5\GeV$. 
When the corrections for nucleon structure and strong interaction are introduced (solid lines), the region of constant sensitivity is reduced to $100\GeV\lesssim m_\chi\lesssim  10^5\GeV$ and the value of the threshold cross section is no longer the same for D1 and D8. 
This is because of the dependence of the nucleon couplings on the nucleon effective mass in the case of scalar operators, which induces a larger suppression in the capture rate for D1 (see section~\ref{sec:capresnucleons}). 
In this DM mass regime, we find $\sigmathn\simeq3.6\times10^{-44}\cm^2$ for D1 and $\sigmathn\simeq1.4\times10^{-44}\cm^2$  for D8. 
For protons, the DM sensitivity is also lowered by a similar amount. 
In the multi-scattering regime, for both targets, we observe an even more pronounced effect, with $\sigmathn$ reaching a factor of $\sim52 \,(20)$ larger than in the ideal Fermi gas approach for D1 (D8), and $\sigmathp$  a factor of $\sim 35 \,(16)$.

In the Pauli blocking regime for DM-neutron scattering, only D1 (and other scalar and pseudoscalar operators) are noticeably affected by the introduction of effective masses, which leads to a threshold cross section which is larger by a factor of $\sim2.5$  in Fig.~\ref{fig:sigmathcomp}. For protons, the  threshold cross section remains essentially constant in the free Fermi gas approach, and does not increase as it does for neutron targets. This is because of the apparent absence of Pauli blocking for protons in some regions of the star where $\kinFp=0$. However, in the interacting baryon framework, protons are always degenerate, mildly towards the NS surface (see Fig.~\ref{fig:muradprofs}). The correct approach thus yields a $\sigmathp$ several orders of magnitude larger in the very light DM mass regime than that of the free Fermi gas approach (see cyan shaded regions).  This effect is particularly striking at the lowest DM masses.
We therefore conclude that the free Fermi gas approximation leads to highly erroneous conclusions regarding the NS sensitivity to scattering on proton targets, in the $m_\chi\lesssim0.1\GeV$ mass range.

\begin{figure}[t]
    \centering 
\includegraphics[width=0.6\textwidth]{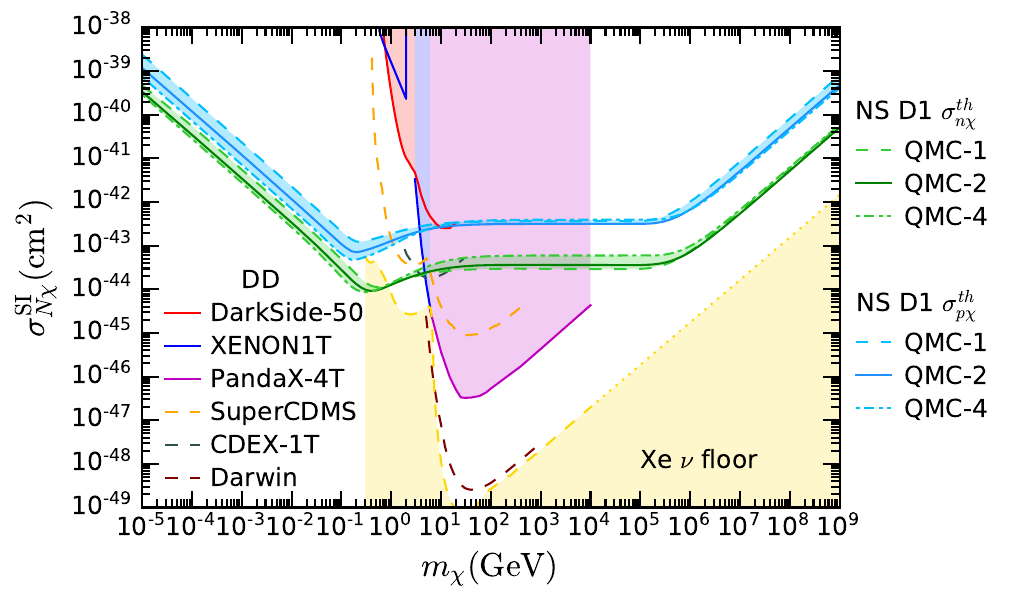}
\includegraphics[width=\textwidth]{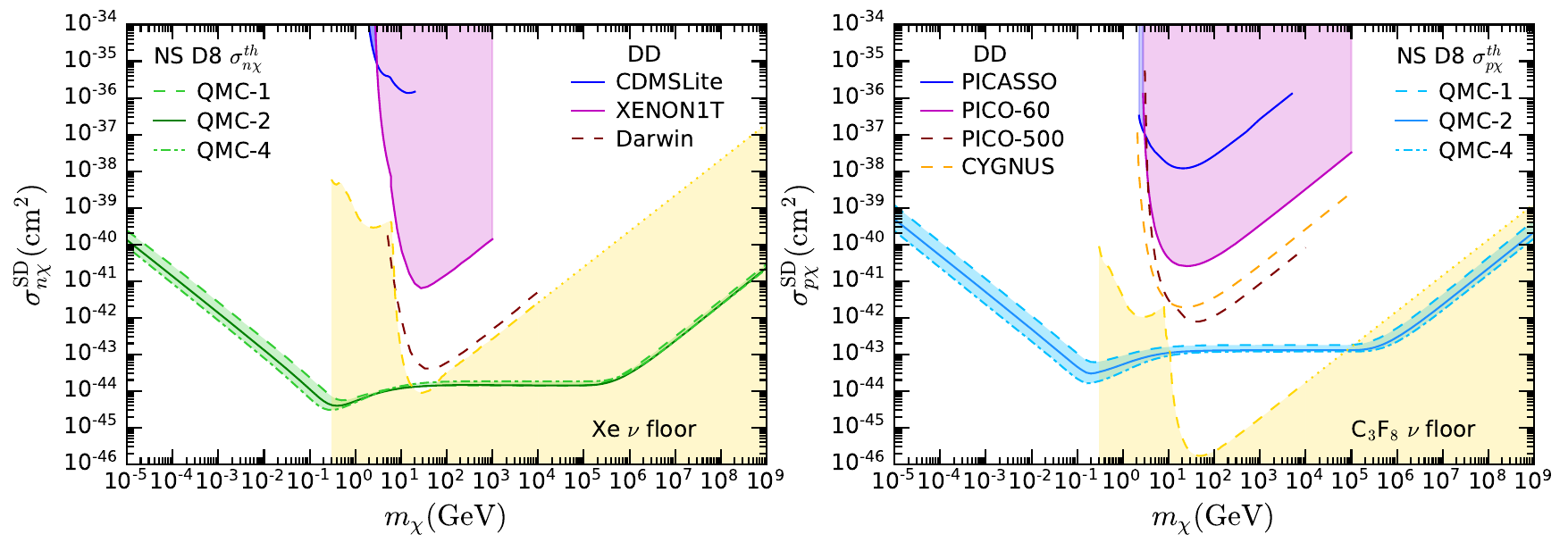}
    \caption{ DM-nucleon threshold cross section for operators D1 (top) and D8 (bottom) for the QMC EoS family. The solid line represents $\sigmath$, computed assuming the NS configuration QMC-2. 
    We also show for comparison the leading spin-independent (SI) and spin-dependent (SD) DD limits from CDMSLite~\cite{Agnese:2017jvy}, DarkSide-50~\cite{Agnes:2018ves},  Xenon1T~\cite{Aprile:2019dbj,Aprile:2019xxb,Aprile:2019jmx,Aprile:2020thb}, PandaX-4T~\cite{PandaX-4T:2021bab}, PICASSO~\cite{Behnke:2016lsk} and PICO-60~\cite{Amole:2019fdf},  projected sensitivities from SuperCDMS SNOLAB Ge/Si~\cite{Agnese:2016cpb}, CDEX-1T~\cite{Yue:2016epq}, CYGNUS  $10\m^3$~\cite{Vahsen:2020pzb},  PICO-500~\cite{VasquezJauregui:2017} and Darwin~\cite{Aalbers:2016jon}, as well as the neutrino coherent scattering background for xenon and $\mathrm{C_3F_8}$ bubble chamber detectors~\cite{Ruppin:2014bra}.  
    } 
    \label{fig:sigmath}
\end{figure}

To compare the NS sensitivity with the reach of direct detection experiments, 
we show in Fig.~\ref{fig:sigmath} the threshold cross section for neutron and proton targets for operators  D1 (top) and D8 (bottom). The solid green (light blue) lines correspond to $\sigmathn$ ($\sigmathp$) calculated using the NS QMC-2 ($1.5 M_\odot$),  while the shaded bands 
indicate the variation with 
the NS model along the QMC family, from  QMC-1 ($1M_\odot$, dashed lines)  to QMC-4 ($1.9 M_\odot$, dash-dotted). 
The background for direct detection experiments arising from the coherent scattering of solar and atmospheric neutrinos (the``neutrino floor") is shown as a shaded yellow region. 

The variation of the threshold cross section associated with the choice of NS configuration is relatively small, because of a number of competing effects. 
In the DM mass region affected by Pauli blocking, increasing the NS mass decreases the threshold cross section. This behaviour is reversed in the intermediate DM mass regime for neutron targets. This is because of the suppression stemming from $t$-dependent neutron form factors and strong interactions, which is stronger in massive NSs and less stringent in lighter NS configurations~\cite{Bell:2020obw}, thereby reducing the ratio of the capture rate in QMC-4 to that in QMC-1 
for a given $\Lambda$. In the multiple scattering regime, this ratio increases again, especially for operators whose form factors do not depend on $\mbeff$ such as D8, and  QMC-1 gives  the upper limit of the lower band for these cases, while in the others the above mentioned behaviour is maintained. 
For D8, these effects considerably reduce the uncertainty in $\sigmathn$ associated with the NS EoS in the intermediate and large DM mass range and only in the latter for D1. 
For protons the tendency of less massive NSs to give rise to a larger $\sigmathp$ is never reversed for D8, while D1 shows a similar behaviour to that of $\sigmathn$. In the proton case, the same considerations hold regarding the effect of the form factors and strong interactions in different NSs, but there is an additional factor that also plays a role. Specifically, there is a substantial difference in the proton content among distinct NS configurations (see the middle left hand panel of Fig.~\ref{fig:NSradprofs1}). The proton number density decreases as the NS mass decreases, more rapidly so than that of neutrons. The interplay of these factors determines which NS configuration gives the largest $\sigmathp$. 
It is worth mentioning that the uncertainty in $\sigmathp$ due to the NS EoS is greatly reduced by the use of the interacting baryon framework, especially in the $100\GeV\lesssim m_\chi\lesssim2\times10^5\GeV$ region.

For scalar-scalar spin-independent interactions (D1), the 
PandaX-4T~\cite{PandaX-4T:2021bab} 
(magenta) direct detection experiment currently constrains cross sections lower than $\sigmath$ in the $\sim10\GeV$ to $\sim10\TeV$ region.  In the rest of the parameter space, especially in the low mass region, the projected NS sensitivity surpasses any present or future DD experiment; sensitivity below the neutrino floor can be reached for a significant mass range. 
In the case of spin-dependent interactions (D8) with either neutrons or protons, the projected NS sensitivity greatly surpasses the reach of all current and future DD experiments, for all DM masses. Moreover, for SD scattering off neutrons, the NS sensitivity is well below the neutrino floor for all DD targets in most of the parameter space.

\section{Discussion}
\label{sec:discussion}

Neutron stars (NSs), the densest stars known, are the sole compact stellar object capable of accelerating infalling dark matter (DM) to (quasi-)relativistic speeds. 
As such, the capture of DM in these stellar remnants can potentially provide a very sensitive indirect probe of DM interactions with standard matter, complimentary to Earth-based searches. 
As is well known, the NS sensitivity is hampered in the sub-GeV DM mass regime by Pauli blocking, and also above the PeV scale, where a single collision provides insufficient energy loss for gravitational capture. 
When DM scatters off baryonic targets under the extreme conditions found in the NS interior, there are two additional factors, intrinsic to the physics of NSs, that also limit the NS reach. These are the facts that (i) the degenerate nucleons undergo strong interactions,  and (ii) given the large energy transfers expected in NSs, baryons cannot be treated as point-like particles as in direct detection experiments. 

In this paper, we extend the capture formalism to properly account for these effects. As in our previous work, our treatment also correctly incorporates the NS internal structure, general relativistic corrections, Pauli blocking and multiple scattering.  
We examine the impact of these corrections on the rate at which DM is captured via scattering on nucleon targets, using an effective field theory approach, and project the NS sensitivity to DM-nucleon interactions for a broad range of DM masses.

Once strong interactions are taken into account, the target mass is no longer the bare mass but is replaced with a radially dependent effective mass which reaches its minimum value at the centre of the NS. This effective mass is responsible for reducing the capture rate in the intermediate and large DM mass ranges; the size of this suppression is larger in heavier NSs. 
In this  framework, the calculation of the Fermi energy of the target is also modified. This correction is particularly important for DM-proton scattering. In the free Fermi gas approximation, the proton Fermi energy goes to zero in the outer regions of the core, with the exact region depending on the NS mass. As a result, treating protons as an ideal Fermi gas leads to the incorrect conclusion that their contribution to the capture rate is not affected by Pauli blocking and, in the light DM mass regime, can even surpass the capture rate due to scattering on neutrons.  
However, with a careful treatment of the strong interactions, we find that protons are (mildly) degenerate throughout most of the stellar interior and that the scattering of light DM with protons is indeed Pauli suppressed. 
Therefore, the capture rate in the sub-GeV mass regime is usually dominated by scattering on neutrons rather than protons;  with the sole exception being DM-proton pseudoscalar interactions.   

When correctly including the nucleon internal structure by using a momentum dependent nucleon coupling, collisions with large momentum transfers are highly suppressed.  This, in turn, implies another considerable suppression of the capture rate in the non-Pauli blocked regime. 
In addition, the suppression of collisions with large energy transfer results in a lower average DM energy loss per interaction. Therefore, there is a consequent reduction in the DM mass scale for which a single collision provides insufficient energy transfer for capture (the multi-scattering region).  Moreover, the suppression of the capture rate in the multi-scattering regime is greater than when the zero momentum transfer approximation is used for the nucleon couplings.  Again, these effects are all larger in heavier NSs.

The ultra-dense NS inner core may harbour exotic matter. 
Under beta equilibrium, strangeness-changing weak interactions allow hyperonic matter to appear as a stable constituent in massive NSs. 
We calculate the contribution of DM-hyperon scattering to the capture rate in the improved framework. We find that the hyperonic contribution is of similar size to that of neutrons for effective operators with pseudoscalar interactions. This enhancement of the capture rate is larger in operators  whose scattering amplitudes depend on higher powers of the energy transfer, for which the hyperonic contribution can surpass that of neutrons. 

Finally, we project the reach of NSs to DM-nucleon  scalar-scalar and axial-axial interactions, by estimating their  threshold cross section. Above this natural threshold, the capture rate saturates to its geometric limit.  
When nucleon structure and strong interactions are treated carefully, we find that the threshold cross section for DM-neutron scattering is no longer ${\cal O}(10^{-45}\cm^2)$, as usually assumed in the literature, and it now depends on the type of the interaction. Interactions that induce nucleon couplings which depend on the nucleon effective mass, such as the scalar-scalar operator, have larger $\sigmathn$, while those whose nucleon couplings depend only on the transferred momentum can reach $\sigmathn\sim\textrm{few}\times 10^{-45}\cm^2$ but only in a very narrow region of the parameter space. 
The figure of merit now exhibits three distinctive regions: (i) the low DM mass, Pauli-suppressed, regime (ii) the intermediate DM mass region, where suppression arising from nucleon strong interactions and momentum dependent form factors is in play, and where $\sigmathn$ is ${\cal O}(10^{-44}\cm^2)$, and (iii) the high DM mass, multiple-scattering regime, for which capture is also affected by the effects described. 
In the transition between the first two regions, the threshold cross section reaches its absolute minimum. 

Despite the reduced NS sensitivity, the reach of NSs is still able to outperform the leading and future direct detection experiments, in the $m_\chi\lesssim10\GeV$ region for spin-independent interactions and in the full DM mass range for spin-dependent (SD) interactions. For SD scattering off neutrons, NSs are sensitive to cross sections well below the neutrino floor. Of course, for some dark matter models, such as some regions of the  parameter space in models featuring neutralino-like DM, the DM-nucleon cross section will be sufficiently small that there will be no detectable effect in either direct detection experiments or neutron stars.

It is worth noting that the calculations presented in this paper require that one  solves the NS structure equations and determine the microscopic properties of the particle species in the NS, such as number densities and effective masses. To this end, we have assumed the QMC equation of state (EoS), an EoS that satisfies current observational constraints and  accounts for the presence of hyperons in the inner core. To assess the uncertainty due to the EoS, we have compared our results with those for a different type of EoS, the BSk24 functional and found little dependence on the EoS model. 
Note also that the corrections addressed in this paper  greatly reduce the uncertainty in the threshold cross section associated with the choice of the EoS.

\section{Conclusion}
\label{sec:conclusion}

We have extended the framework for dark matter (DM) capture in neutron stars (NSs) to account for the effect of hadronic structure, through the use of momentum dependent form factors, and to incorporate baryon interactions in the degenerate NS medium, rather than modelling the baryons as a free Fermi gas. These corrections decrease the rate at which DM is captured by scattering on nucleons.  Specifically, the threshold cross section, i.e., the cross section for which the capture rate saturates the geometric limit, is increased by at least $\sim$ one order of magnitude for $m_\chi\gtrsim10\GeV$. The exact size of the effect depends on the NS configuration and on the type of DM-nucleon interaction, being stronger for scalar and pseudoscalar EFT operators. Interestingly, incorporating these effects reduces the variation of the threshold cross section associated with different choices for the NS equation of state.

For proton targets, the use of the interacting baryon approach to obtain the correct Fermi energy is particularly important.  The standard free Fermi gas approach leads to the incorrect conclusion that protons are non-degenerate in the outer regions of the core, and hence not subject to Pauli blocking, greatly enhancing the capture rate due to DM-proton scattering, even surpassing that due to DM-neutron scattering.  However, with the use of the correct Fermi energy, obtained by incorporating nucleon strong interactions via effective masses, the Pauli suppression of the proton final states is recovered. 

Heavy NSs may contain exotic matter in the form of hyperons in the innermost regions of the NS core. We have found that scattering of DM with hyperons can boost the DM capture rate in the case of pseudoscalar DM-baryon interactions, with the hyperon contribution at least comparable to that of neutrons over the entire DM mass range.

Finally, we note that, despite somewhat reduced capture rates compared to the standard treatment in the literature, the projected NS sensitivity remains much greater than that for direct detection experiments for  both the spin-independent scattering of light (sub-GeV) DM, or for the spin-dependent scattering of DM of any mass.

\section*{Acknowledgements}
This work was supported by the Australian Research Council through the ARC Centre of Excellence for Dark Matter Particle Physics CE200100008 (NFB, SR and AWT) and the Discovery Project DP180100497 (AWT). MV was supported by an Australian Government Research Training Program Scholarship, FA by a  Melbourne Research Scholarship from the University of Melbourne and TM by a Beacon of Enlightenment PhD Scholarship from The University of Adelaide.

\appendix

\section{Hadronic matrix elements for scattering operators}
\label{sec:operators}

Ten dimension six effective operators for fermionic DM interacting with quarks can be constructed, without considering flavour violation (see Table~\ref{tab:operatorshe}). The coefficients for the squared matrix elements in the fourth column of Table~\ref{tab:operatorshe}  read, 
\begin{eqnarray}
c_{\cal B}^S &=& \frac{2 m_{\cal B}^2}{v^2}\left[\sum_{q=u,d,s}f_{T_q}^{(\cal B)}+\frac{2}{9}f_{T_G}^{(\cal B)}\right]^2,\\
c_{\cal B}^P &=& \frac{2 m_{\cal B}^2}{v^2}\left[\sum_{q=u,d,s}\left(1-3\frac{\overline{m}}{m_q}\right)\Delta_q^{(\cal B)}\right]^2,\\
c_{\cal B}^V &=& 9,\\
c_{\cal B}^A &=&  \left[\sum_{q=u,d,s}\Delta_q^{(\cal B)}\right]^2,\\
c_{\cal B}^T &=& \left[\sum_{q=u,d,s}\delta_q^{(\cal B)}\right]^2,
\end{eqnarray}
where  $v=246$ GeV is the vacuum expectation value of the SM higgs field, $\cal B$ is the baryonic species,  $\overline{m}\equiv(1/m_u+1/m_d+1/m_s)^{-1}$ and $f_{T_q}^{(\cal B)}$, $f_{T_G}^{(\cal B)}=1-\sum_{q=u,d,s} f_{T_q}^{(\cal B)}$, $\Delta_q^{(\cal B)}$ and $\delta_q^{(\cal B)}$ are the hadronic matrix elements, determined either experimentally or by lattice QCD simulations.

\subsection{Nucleons}
The values of the hadronic matrix elements for neutrons and protons used in this paper are listed in Table~\ref{tab:hadmatelem}. The values of $\Delta_q^{(p)}$ (and similarly for $\delta_q^{(p)}$) are obtained using isospin symmetry: 
\begin{equation}
\Delta_u^{N} =\Delta_d^{N^\star},\qquad \qquad
\Delta_s^{N} = \Delta_s^{N^\star},
\end{equation}
where $N^\star$ is the nucleon obtained interchanging $u\Longleftrightarrow d$ quarks.

\begin{table}[th]
    \centering
    \begin{tabular}{|c|c|c|c|c|}
    \hline
     $q$ & $f_{T_q}^{(n)}$ \cite{Belanger:2013oya} & $f_{T_q}^{(p)}$ \cite{Belanger:2013oya} & $\Delta^{(n)}_q$ & $\delta^{(n)}_q$ \cite{Belanger:2013oya} \\ 
     \hline
     $u$  & 0.0110 & 0.0153     & -0.319 \cite{QCDSF:2011aa} & -0.230 \\ \hline
     $d$ & 0.0273  & 0.0191     & 0.787 \cite{QCDSF:2011aa} & 0.840 \\ \hline
     $s$ & 0.0447  & 0.0447     & -0.040 \cite{Dienes:2013xya} & -0.046 \\ \hline
    \end{tabular}
    \caption{Hadronic matrix elements for neutrons and protons. }
    \label{tab:hadmatelem}
\end{table}

\subsection{Hyperons}\label{apx:hypff}

To calculate the $f_{T_q}^{(\cal B)}$ couplings for hyperons, we use the baryonic  sigma terms from ref.~\citep{Shanahan:2013cd}, listed in Table~\ref{tab:DeltaBar}, in the following way 
\begin{eqnarray}
f_{T_{u,d}}^{(\cal B)} &=& \frac{\sigma_{l \cal B}}{m_{\cal B}}\frac{m_{u,d}}{m_u+m_d},\\
f_{T_s}^{(\cal B)} &=& \frac{\sigma_{s}}{m_{\cal B}},
\end{eqnarray}
where the first relation assumes
\begin{equation}
    \frac{\sigma_u^{\cal B}}{m_u} = \frac{\sigma_d^{\cal B}}{m_d}.
\end{equation}
In addition, we assume
\begin{equation}
 \sigma_u^{\cal B} = \sigma_d^{{\cal B}^*}, \qquad \qquad
 \sigma_s^{\cal B} = \sigma_s^{{\cal B}^*}.
\end{equation}
For the dimension 6 operators where $c_q\propto m_q$, the nucleon couplings depend only on the following sum of the $f_{T_q}^{(\cal B)}$ values
\begin{eqnarray}
\sum_{q=u,d,s} f_{T_q}^{(\cal B)} = \frac{\sigma_{l \cal B}+\sigma_s}{m_{\cal B}},
\end{eqnarray}
and hence exact values for the individual $f_{T_q}^{(\cal B)}$ are unnecessary.
The axial vector \citep{Shanahan:2013apa,Alexandrou:2020sml} and tensor \citep{Zanotti:2017bte} couplings are listed in Table~\ref{tab:DeltaBar}. 

\begin{table}[th]
    \centering
    \begin{tabular}{|c|c|c|c|c|c|c|c|c|}
        \hline
         $\cal B$ & $\sigma_{l \cal B}$ (MeV) & $\sigma_s$ (MeV) & $\Delta_u$ & $\Delta_d$ & $\Delta_s$ & $\delta_u$ & $\delta_d$ & $\delta_s$ \\ \hline
         $\Lambda^0$ & $32\pm4$ & $176\pm19$ & 0 & 0 & 0.59 & 0 & 0 & 0.47 \\ \hline
         $\Xi^0$ & $13\pm 10$ & $334\pm21$ & -0.38 & 0 & 1.03 & -0.22 & 0 & 0.9  \\ \hline
         $\Xi^-$ & $13\pm 10$ & $334\pm21$ & 0 & -0.38 & 1.03 & 0 & -0.22 & 0.9 \\ \hline
    \end{tabular}
    \caption{Sigma commutators for scalar interactions for each hyperon $\cal B$ (first and second column). Spin matrix elements for axial vector~\citep{Shanahan:2013apa,Alexandrou:2020sml} and tensor interactions are given in the remaining columns. Tensor couplings for $\Xi$ are taken from ref.~\citep{Zanotti:2017bte}, while those for $\Lambda^0$ are our estimates.}
    \label{tab:DeltaBar}
\end{table}

\section{Deep Inelastic Scattering in Neutron Stars}
\label{sec:dis}

The kinematics of DIS of dark matter in NSs is not similar to any case that has previously been treated in the literature, at least not to our knowledge. In contrast to DIS of neutrinos and boosted DM, we are interested in much larger DM masses and relatively lower energies.  Specifically, $E_\chi/m_\chi=1/\sqrt{B}$, where $B$ is the time component of the Schwarzchild metric, which falls in the range $[\sim0.2,\sim0.75]$.  Therefore, we have  $E_\chi\lesssim 2m_\chi$. Following ref.~\cite{Agashe:2014yua}, we derive the DIS cross section in NSs for the operators in Table~\ref{tab:operatorshe}. 

The parton level differential cross section is given by
\begin{equation}
    \frac{d\hat{\sigma}}{d\hat{t}} = \frac{\hat{s}}{8\pi\hat{\gamma}^2}\frac{\hat{s}-(m_\chi^2+x^2m_n^2)}{\hat{s}^2-(m_\chi^2-x^2m_n^2)^2}|\overline{M}(\hat{s},\hat{t},m_i=x m_n)|^2,
\end{equation}
where $\hat{\gamma} = \gamma(\hat{s}, m_\chi, xm_n)$, $\hat{t}=-Q^2$ is the squared 4-momentum transfer,  $\hat{s} = (1-x)(m_\chi^2-x m_n^2) +xs$, $\Msq$ is defined at the parton level with couplings $g_i=g_q$, and $x$ is the fraction of the nucleon momentum ($P$) carried by the parton, $p= xP$. We define $y$, the fractional energy lost by the DM in the nucleon rest frame~\cite{Agashe:2014yua}
\begin{equation}
    y = \frac{2q\cdot P}{2k\cdot P} = \frac{-\hat{t}}{\hat{s}-m_\chi^2-x^2m_n^2}, 
\end{equation}
and obtain
\begin{equation}
    dQ^2 = (\hat{s}-m_\chi^2 - x^2m_n^2)dy = x(s - m_\chi^2 - m_n^2)dy. 
\end{equation}
We then use the parton distribution functions (PDFs), $f_i$, to obtain the nucleon level differential DIS cross section
\begin{equation}
    \frac{d^2\sigma}{dx\, dy} =(\hat{s} - m_\chi^2-x^2m_n^2) \frac{\hat{s}}{8\pi\hat{\gamma}^2}\frac{\hat{s}-(m_\chi^2+x^2m_n^2)}{\hat{s}^2-(m_\chi^2-x^2m_n^2)^2}\sum_i f_i(x, Q^2)|\overline{M}(\hat{s},\hat{t},x m_n)|^2.
\end{equation}
The integration bounds for the DIS cross section are generically $0<x<1$ and $0<y<y_{max}$, where $y_{max}$ is set by imposing $\cos\theta\leq 1$, and in general $y_{max}\neq 1$~\cite{Agashe:2014yua}.  The value of $y_{max}$ is set by
\begin{equation}
    y_{max} = \frac{-\hat{t}_{min}}{\hat{s} - m_\chi^2 - x^2 m_n^2}
     = \frac{(\hat{s} -m_\chi^2 -x^2m_n^2)^2 - 4 x^2 m_\chi^2 m_n^2}{\hat{s}(\hat{s} -m_\chi^2 - x^2m_n^2)}. 
\end{equation}

The capture rate requires integrating the differential cross section over $\hat{s}$ (see Eqs.~\ref{eq:capturefinalM2text} and \ref{eq:intrate}), which must be done at the parton level, i.e., before integrating over $x$. We perform this integration following  ref.~\cite{Bell:2020jou}. The differential capture rate will then scale as 
\begin{align}
    \int_0^1 dx\int_{\hat{s}_0-\delta\hat{s}}^{\hat{s}_0+\delta\hat{s}}d\hat{s}\int_0^{y_{max}}dy  \frac{d^2\sigma}{dx\, dy} \,\Theta(Q^2 -1\GeV^2), \label{eq:DIS}
\end{align}
where the step function enforces the momentum transfer to be above the $1\GeV$ threshold where the PDFs are reliable, and 
\begin{align}
    \hat{s}_0 & = m_\chi^2 + 2x E_n E_\chi,\\
    \delta\hat{s} & = 2xm_\chi \sqrt{E_n^2 -m_n^2}\sqrt{\frac{1-B(r)}{B(r)}},\\
    E_n & \simeq m_n + \mu_{F,n}. 
\end{align}
We numerically evaluate the DIS cross section using the MSTW2008 NLO PDFs~\cite{Martin:2009iq}. 

\begin{figure}[t]
    \centering
    \includegraphics[width=\textwidth]{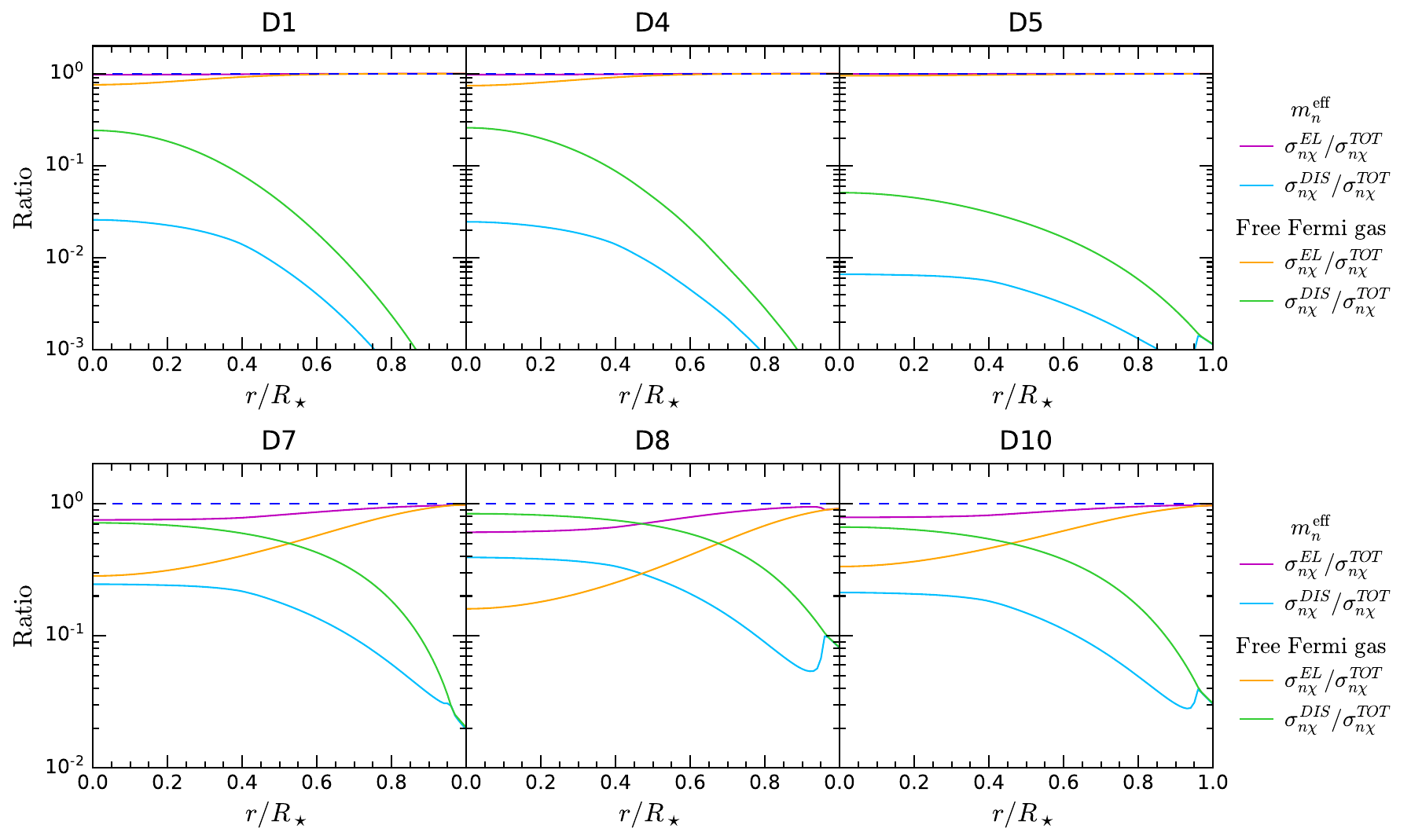}
    \caption{Ratios of the elastic (EL) and deep inelastic  scattering (DIS) cross sections  to the total cross section (TOT=EL+DIS) as a function of the NS radius, for neutron targets in the NS QMC-4 and six of the dimension-6 EFT operators.  We have taken $m_\chi=10^6\GeV$, and shown the comparison for both the interacting baryon and free Fermi gas approaches. Momentum dependent form factors have been included in the elastic calculations.  }  
    \label{fig:DISratio}
\end{figure}

In Fig.~\ref{fig:DISratio}, we show the ratios of the elastic (EL) and deep inelastic scattering (DIS) cross sections
to the total cross section (TOT=DIS+EL),  as a function of the radial coordinate $r$ for the NS  QMC-4, neutron targets and $m_\chi=10^6\GeV$. We consider two scenarios: the free Fermi gas approach and the interactive baryon approach characterised by $\mneff$. The radial coordinate determines the value of $B$, the Fermi energy of the target and $\mneff$. 
In both the $\mneff$ and free Fermi gas approaches, the DIS contribution (light blue and green lines, respectively) increases towards the centre of the star, where $B$ takes lower values and hence the DM kinetic energy is higher. The ratio $\sigma^{DIS}/\sigma^{TOT}$ is smaller in the $\mneff$ approach, compared to the free Fermi gas approach, due to a smaller neutron effective mass. In the correct interactive baryon approach, the ratio of the DIS contribution (light blue lines) to the total cross section is at most ${\cal O}(40\%)$ at the centre of the star for D8, ${\cal O}(20\%)$ for D7, D9-D10, and much lower for the remaining operators.  
As a result, for most operators the elastic cross section provides a very good approximation to the total cross section (compare magenta with dashed blue lines).
Note that these cross sections are weighted by $r^2dr$ in the capture rate calculation of Eq.~\ref{eq:capturefinalM2text} (i.e. weighted by volume) which further reduces the importance of the DIS contribution. 

It is worth noting that we have neglected the effect of Pauli blocking on the DIS cross section. In deep inelastic scattering, one must have a baryon in the final state and with a nucleon target this is almost always a nucleon. As shown in both theoretical calculations~\cite{Melnitchouk:1992gd} and direct experimental studies~\cite{BEBCWA59:1989ayi}, this nucleon has low momentum in the laboratory frame, typically 300 MeV or less. Such nucleons will be totally Pauli blocked in the core of a NS, and hence the deep inelastic cross section drastically reduced. 
As a result, the contribution of the deep inelastic process to the capture rate will have a negligible effect on our conclusions. 

We performed a similar calculation for hyperon targets and found that as for neutrons, the DIS contribution to the total cross section is more important for operators D7-D10 in the absence of Pauli blocking of the fragmented baryonic final states. For D8 and  $\Xi^-$ targets, the DIS cross section can even surpass the   elastic scattering contribution (including form factors) and reach $\sim60\%$ of the total cross section at the centre of the star.  $\Xi^-$ provides the largest hyperonic contribution to the capture rate, however, this (elastic scattering) contribution is already more than one order of magnitude lower than that of neutrons for D7-D10.  
Even if Pauli blocking does not suppress the DIS final states, the contribution of the deep inelastic process to the capture of DM is negligible, since it would enhance the capture rate by scattering on $\Xi^-$ at most by a factor of $\sim2$. 

\section{Capture Rate when  Pauli Blocking is negligible}
\label{sec:capratethinnondeg}

This procedure is similar to that outlined in ref.~\citep{Bell:2020jou} for the computation of the capture rate in the intermediate mass range.
The interaction rate in Eq.~\ref{eq:intrate} (free Fermi gas approximation) for matrix elements that depend on $t$,  $|\overline{M}|^2 = \bar{g}(s) t^n$,  can be written as 
\begin{equation}
\Omega^{-}(r) = \frac{\zeta(r)}{32\pi^3}\int dE_i ds  \frac{\gamma(s)}{s^2-[m_i^2-m_\chi^2]^2}  \frac{E_i}{m_\chi}\sqrt{\frac{B(r)}{1-B(r)}} \fFD(E_i,r)(1-\fFD(E_i^{'},r)) \frac{\bar{g}(s)}{n+1}\left(\frac{\gamma^2(s)}{s}\right)^n, 
\end{equation}
after performing the integral over $t$.  Since, we are working in the limit  $\kinFi\rightarrow0$, the integration interval for $s$ reduces to $[s_0-\delta s, s_0+\delta s]$, with $\delta s <\sqrt{E_i^2-m_i^2}$, and $E_i$ is constrained to the range $m_i<E_i<m_i+\kinFi$. The following simplified expressions are obtained 
\begin{eqnarray}
s_0 &=& m_i^2+m_\chi^2 + 2\frac{E_i m_\chi}{\sqrt{B(r)}},\\
\delta s &=& 2\sqrt{\frac{1-B(r)}{B(r)}}m_\chi\sqrt{E_i^2-m_i^2}, \\ 
\frac{\gamma(s)}{s^2-[m_i^2-m_\chi^2]^2} 
&\sim& \frac{\sqrt{1-B(r)}}{2\left(m_i^2+m_\chi^2+2m_i m_\chi/\sqrt{B(r)}\right)},\\
\frac{\gamma^2(s)}{s} &\rightarrow& \frac{4(1-B(r))m_\chi^2}{B(r)\left(1+\mu^2\right)+2\sqrt{B(r)}\mu}. 
\end{eqnarray}

In the limit $\kinFi\rightarrow0$, the integrand becomes a delta function. 
Estimating the integral over $s$, by keeping $\kinFi$ finite but small, leads to a factor of $\bar{g}(s_0) 2\delta s$. We then have
\begin{align}
\Omega^{-}(r) \sim \zeta(r)\frac{\bar{g}(s_0)}{16\pi^3} \frac{\sqrt{E_\chi^2-m_\chi^2}}{E_\chi\left(m_i^2+m_\chi^2+2m_i E_\chi\right)}  \frac{\left[\frac{4(1-B(r))m_\chi^2}{B(r)\left(1+\mu^2\right)+2\sqrt{B(r)}\mu}\right]^n}{n+1}\nn \\ \times \int_{m_i}^{m_i+\kinFi(r)} dE_i E_i\sqrt{E_i^2-m_i^2} 
\fFD(E_i,r). 
\end{align}
Performing the integral over $E_i$ gives a factor of  $\pi^2 n_{free}(r)$, thus
\begin{eqnarray}
\Omega^{-}(r) &\sim& \frac{n_i(r)}{16\pi} \frac{\sqrt{E_\chi^2-m_\chi^2}}{E_\chi\left(m_i^2+m_\chi^2+2m_i E_\chi\right)} \frac{\bar{g}(s_0)}{n+1} 
\left[\frac{4(1-B(r))m_\chi^2}{B(r)\left(1+\mu^2\right)+2\sqrt{B(r)}\mu}\right]^n.   
\end{eqnarray}
The capture rate is then 
\begin{eqnarray}
C &\sim& \frac{1}{4 \vstar} \frac{\rho_\chi}{m_\chi} \frac{1}{m_i^2+m_\chi^2+2m_i m_\chi/\sqrt{B}} {\rm Erf}\left(\sqrt{\frac{3}{2}}\frac{\vstar}{v_d}\right)\nn\\
&\times&\int_0^{\Rstar}  r^2 dr \, n_i(r)  \frac{1-B(r)}{B(r)} \frac{\bar{g}(s_0)}{n+1} \left[\frac{4(1-B(r))m_\chi^2}{B(r)\left(1+\mu^2\right)+2\sqrt{B(r)}\mu}\right]^n.   \label{eq:capratenondeg} 
\end{eqnarray}
This expression can be rewritten in terms of the cross section (note that in this case we do not need to average over $s$, as $s$ has a fixed value), 
\begin{eqnarray}
\sigma(r) &=& \int dt \frac{d\sigma}{dt} = \frac{1}{64\pi m_\chi^2 m_i^2} \frac{B(r)}{(1-B(r))} \bar{g}(s_0) \int dt \, t^n \\ 
&=& \frac{1}{16\pi \left(m_i^2+m_\chi^2+2m_i m_\chi/\sqrt{B}\right)}  \frac{\bar{g}(s_0)}{(n+1)} \left[\frac{4(1-B(r))m_\chi^2}{B(r)\left(1+\mu^2\right)+2\sqrt{B(r)}\mu}\right]^{n}, 
\end{eqnarray}
which leads to, 
\begin{eqnarray}
C &\sim& \frac{4\pi}{\vstar} \frac{\rho_\chi}{m_\chi}  {\rm Erf}\left(\sqrt{\frac{3}{2}}\frac{\vstar}{v_d}\right)\int_0^{\Rstar}  r^2 dr \, n_i(r)  \frac{1-B(r)}{B(r)} \sigma(r)\label{eq:csimplenomu}. 
\end{eqnarray}
Indeed Eq.~\ref{eq:csimplenomu}, is identical to the expression for the intermediate mass range given in ref.~\cite{Bell:2020jou}. The only difference is that we should keep  all terms in $m_i,m_\chi$ in the  cross section, otherwise the computation will fail for $m_\chi\sim m_i$.

\section{Uncertainties associated with the Equation of State}
\label{sec:uncereos}

\begin{figure*}[t]
    \centering
    \includegraphics[width=0.9\textwidth]{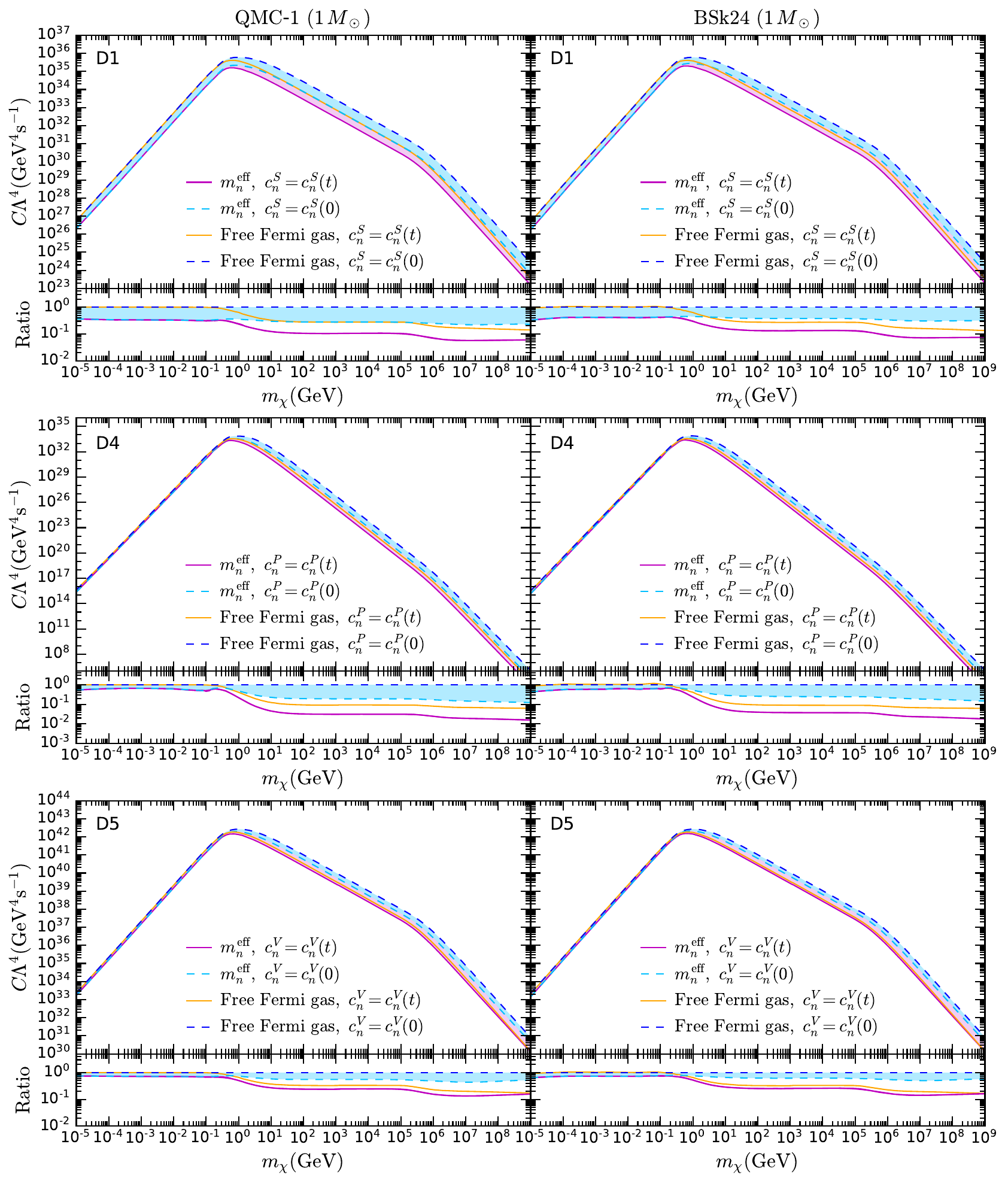}
    \caption{Capture rate in the optically thin limit for D1 (top row), D4 (middle row) and D5 (bottom row) operators as a function of the DM mass $m_\chi$ for neutron targets, using the free Fermi gas approach with constant neutron form factors (dashed blue) and form factors  that depend on the transferred momentum (orange), as well as the interacting baryon  approach 
    for constant neutron couplings (dashed light blue) and couplings that depend on $t$ (magenta), for $1\Msun$ NSs with EoS QMC-1 (left) and  BSk24 (right). In the lower panel of each plot, we show the ratios of the capture rates with respect to that for the free Fermi gas calculation with constant neutron couplings.}
    \label{fig:Cmdmcomp1}
\end{figure*}

\begin{figure*}[h]
    \centering
    \includegraphics[width=0.9\textwidth]{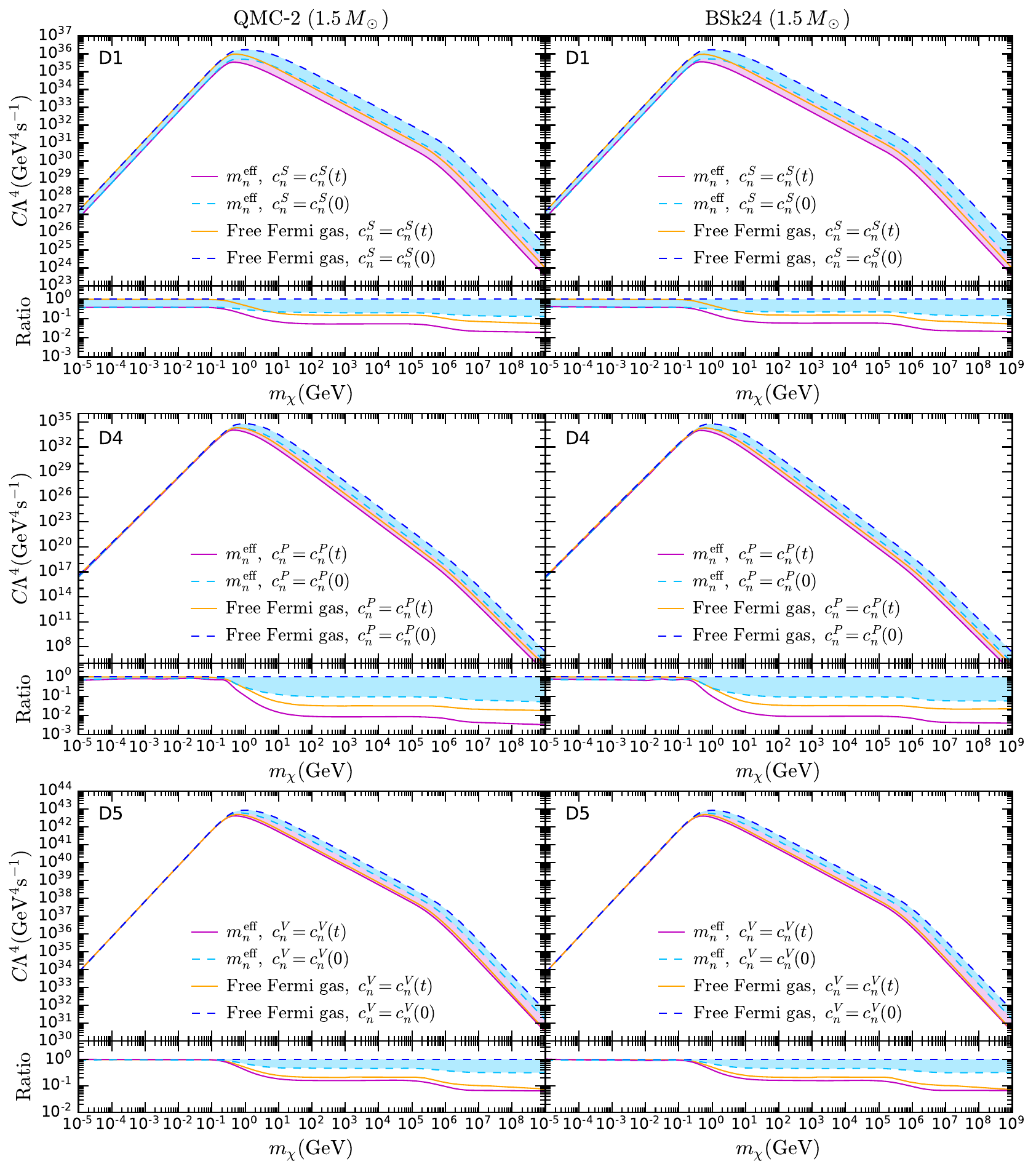}
    \caption{Capture rate in the optically thin limit for D1 (top row), D4 (middle row) and D5 (bottom row) operators as a function of the DM mass $m_\chi$ for neutron targets, using the free Fermi gas approach with constant neutron form factors (dashed blue) and form factors  that depend on the transferred momentum (orange), as well as the interacting baryon  approach 
    for constant neutron couplings (dashed light blue) and couplings that depend on $t$ (magenta), for $1.5\Msun$ NSs with EoS QMC-2 (left) and  BSk24 (right). In the lower panel of each plot, we show the ratios of the capture rates with respect to that for the free Fermi gas calculation with constant neutron couplings.}
    \label{fig:Cmdmcomp2}
\end{figure*}

For DM capture in NSs, the most significant source of uncertainty is the equation of state (EoS) of dense matter. This is an open problem in nuclear astrophysics, and there are several EoSs that satisfy current observational constraints. For NS masses below $1.6M_\odot$ there are many more models allowed. Of these models, we have chosen to compare our results with those for a different type of EoS, namely  the Brussels-Montreal functional BSk24~\cite{Goriely:2013,Pearson:2018tkr}. 
BSk24 is a unified Skyrme-type (non-relativistic) EoS, and assumes the NS core is made of $npe\mu$ matter. For this EoS,  the single particle spectrum is given by~\cite{Reddy:1997yr}
\begin{equation}
 E_i(p_i)= \frac{p^2_i}{2\mbeff(n_b)} +U_i(n_b),       
\end{equation}
and the nucleon effective masses $\mbeff(n_b)$ are calculated using Eq.~(A10) of ref.~\cite{Chamel:2009yx},  with parameters from Table~II of ref.~\cite{Goriely:2013}.  These $\mbeff(n_b)$ give rise to radial profiles very different to those of Fig.~\ref{fig:NSradprofs1}.

In Figs.~\ref{fig:Cmdmcomp1} and \ref{fig:Cmdmcomp2}, we compare our results for the QMC EoS 
with those for the 
BSk24 functional for NSs of the same mass, namely $\Mstar=1\Msun$ and $\Mstar=1.5\Msun$\footnote{These configurations of the BSk24 functional correspond to the benchmark models BSk24-1 and BSk24-2 in refs.~\cite{Bell:2020jou,Bell:2020lmm}.}, assuming neutron targets. Note that we do not compare heavier NS configurations, since BSk24 is a minimal EoS that does not consider the presence of exotic matter in the inner core. 
We see that the size of corrections due to momentum dependent form factors and strong interactions have very little model dependence on the EoS (compare left with right ratio panels).
This is true despite the distinct $\mbeff$ profiles obtained with QMC and BSk24.  
For the $1.5\Msun$ NS, the  absolute difference between the capture rates for the full calculation (magenta lines) is within $2-10\%$, depending on the type of interaction. 
Operators with neutron couplings that are a function of $\mneff$, such as D1 and D4, feature the largest differences. For the $1\Msun$ NS, this variation is greater, with the maximum difference being within $10-30\%$. For scattering on protons, the maximum variation in the capture rates between the QMC and BSk24 calculations is within the range $20-35\%$ for the $1\Msun$ NS, and $20-50\%$ for the $1.5\Msun$ NS. 
 
In terms of the threshold cross section, we find that the maximum absolute difference in $\sigmath$ for D1 and D8, due to the choice of EoS, is of similar size to that for the corresponding capture rate. Specifically, for D1 this variation does not exceed $30\%$ ($36\%$) for neutron 
(proton) targets, and for D8 is less than $19\%$ ($17\%$). Thus,  $\sigmathn$ and $\sigmathp$ are of the same order of magnitude for both EoSs and  their  exact values for BSk24 are very close to those given in section~\ref{sec:thxs}. 
In conclusion, these differences are very small compared to the overall size of the effects upon which we focus.


\label{Bibliography}

\lhead{\emph{Bibliography}} 

\bibliography{Bibliography} 

\providecommand{\href}[2]{#2}\begingroup\raggedright\begin{thebibliography}{100}

\bibitem{Gould:1987ir}
A.~Gould, ``{Resonant Enhancements in WIMP Capture by the Earth},''
\href{http://dx.doi.org/10.1086/165653}{{\em Astrophys. J.} {\bfseries 321}
  (1987) 571}.

\bibitem{Gould:1987ju}
A.~Gould, ``{{WIMP} Distribution in and Evaporation From the Sun},''
\href{http://dx.doi.org/10.1086/165652}{{\em Astrophys. J.} {\bfseries 321}
  (1987) 560}.

\bibitem{Press:1985ug}
W.~H. Press and D.~N. Spergel, ``{Capture by the sun of a galactic population
  of weakly interacting massive particles},''
  \href{http://dx.doi.org/10.1086/163485}{{\em Astrophys. J.} {\bfseries 296}
  (1985) 679--684}.

\bibitem{Griest:1986yu}
K.~Griest and D.~Seckel, ``{Cosmic Asymmetry, Neutrinos and the Sun},''
  \href{http://dx.doi.org/10.1016/0550-3213(87)90293-8}{{\em Nucl. Phys. B}
  {\bfseries 283} (1987) 681--705}. [Erratum: Nucl.Phys.B 296, 1034--1036
  (1988)].

\bibitem{Silk:1985ax}
J.~Silk, K.~A. Olive, and M.~Srednicki, ``{The Photino, the Sun and High-Energy
  Neutrinos},'' \href{http://dx.doi.org/10.1103/PhysRevLett.55.257}{{\em Phys.
  Rev. Lett.} {\bfseries 55} (1985) 257--259}.

\bibitem{Krauss:1985ks}
L.~M. Krauss, K.~Freese, W.~Press, and D.~Spergel, ``{Cold dark matter
  candidates and the solar neutrino problem},''
  \href{http://dx.doi.org/10.1086/163767}{{\em Astrophys. J.} {\bfseries 299}
  (1985) 1001}.

\bibitem{Jungman:1995df}
G.~Jungman, M.~Kamionkowski, and K.~Griest, ``{Supersymmetric dark matter},''
  \href{http://dx.doi.org/10.1016/0370-1573(95)00058-5}{{\em Phys. Rept.}
  {\bfseries 267} (1996) 195--373},
\href{http://arxiv.org/abs/hep-ph/9506380}{{\ttfamily arXiv:hep-ph/9506380
  [hep-ph]}}.

\bibitem{Busoni:2013kaa}
G.~Busoni, A.~De~Simone, and W.-C. Huang, ``{On the Minimum Dark Matter Mass
  Testable by Neutrinos from the Sun},''
  \href{http://dx.doi.org/10.1088/1475-7516/2013/07/010}{{\em JCAP} {\bfseries
  1307} (2013) 010},
\href{http://arxiv.org/abs/1305.1817}{{\ttfamily arXiv:1305.1817 [hep-ph]}}.

\bibitem{Garani:2017jcj}
R.~Garani and S.~Palomares-Ruiz, ``{Dark matter in the Sun: scattering off
  electrons vs nucleons},''
  \href{http://dx.doi.org/10.1088/1475-7516/2017/05/007}{{\em JCAP} {\bfseries
  1705} no.~05, (2017) 007},
\href{http://arxiv.org/abs/1702.02768}{{\ttfamily arXiv:1702.02768 [hep-ph]}}.

\bibitem{Busoni:2017mhe}
G.~Busoni, A.~De~Simone, P.~Scott, and A.~C. Vincent, ``{Evaporation and
  scattering of momentum- and velocity-dependent dark matter in the Sun},''
  \href{http://dx.doi.org/10.1088/1475-7516/2017/10/037}{{\em JCAP} {\bfseries
  1710} no.~10, (2017) 037},
\href{http://arxiv.org/abs/1703.07784}{{\ttfamily arXiv:1703.07784 [hep-ph]}}.

\bibitem{Ilie:2020nzp}
C.~Ilie, C.~Levy, J.~Pilawa, and S.~Zhang, ``{Constraining Dark Matter
  properties with the first generation of stars},''
  \href{http://arxiv.org/abs/2009.11474}{{\ttfamily arXiv:2009.11474
  [astro-ph.CO]}}.

\bibitem{Ilie:2020iup}
C.~Ilie, C.~Levy, J.~Pilawa, and S.~Zhang, ``{Probing below the neutrino floor
  with the first generation of stars},''
  \href{http://arxiv.org/abs/2009.11478}{{\ttfamily arXiv:2009.11478
  [astro-ph.CO]}}.

\bibitem{Ilie:2021iyh}
C.~Ilie and C.~Levy, ``{Multi-component multiscatter capture of Dark Matter},''
  \href{http://arxiv.org/abs/2105.09765}{{\ttfamily arXiv:2105.09765
  [astro-ph.CO]}}.

\bibitem{Gould:1989ez}
A.~Gould and G.~Raffelt, ``{Cosmion Energy Transfer in Stars: The Knudsen
  Limit},'' \href{http://dx.doi.org/10.1086/168569}{{\em Astrophys. J.}
  {\bfseries 352} (1990) 669}.

\bibitem{Gould:1989hm}
A.~Gould and G.~Raffelt, ``{Thermal Conduction by Massive Particles},''
  \href{http://dx.doi.org/10.1086/168568}{{\em Astrophys. J.} {\bfseries 352}
  (1990) 654}.

\bibitem{Vincent:2013lua}
A.~C. Vincent and P.~Scott, ``{Thermal conduction by dark matter with velocity
  and momentum-dependent cross-sections},''
  \href{http://dx.doi.org/10.1088/1475-7516/2014/04/019}{{\em JCAP} {\bfseries
  04} (2014) 019}, \href{http://arxiv.org/abs/1311.2074}{{\ttfamily
  arXiv:1311.2074 [astro-ph.CO]}}.

\bibitem{Geytenbeek:2016nfg}
B.~Geytenbeek, S.~Rao, P.~Scott, A.~Serenelli, A.~C. Vincent, M.~White, and
  A.~G. Williams, ``{Effect of electromagnetic dipole dark matter on energy
  transport in the solar interior},''
  \href{http://dx.doi.org/10.1088/1475-7516/2017/03/029}{{\em JCAP} {\bfseries
  03} (2017) 029}, \href{http://arxiv.org/abs/1610.06737}{{\ttfamily
  arXiv:1610.06737 [hep-ph]}}.

\bibitem{Tanaka:2011uf}
{\bfseries Super-Kamiokande} Collaboration, T.~Tanaka {\em et~al.}, ``{An
  Indirect Search for WIMPs in the Sun using 3109.6 days of upward-going muons
  in Super-Kamiokande},''
  \href{http://dx.doi.org/10.1088/0004-637X/742/2/78}{{\em Astrophys. J.}
  {\bfseries 742} (2011) 78},
\href{http://arxiv.org/abs/1108.3384}{{\ttfamily arXiv:1108.3384
  [astro-ph.HE]}}.

\bibitem{Choi:2015ara}
{\bfseries Super-Kamiokande} Collaboration, K.~Choi {\em et~al.}, ``{Search for
  neutrinos from annihilation of captured low-mass dark matter particles in the
  Sun by Super-Kamiokande},''
  \href{http://dx.doi.org/10.1103/PhysRevLett.114.141301}{{\em Phys. Rev.
  Lett.} {\bfseries 114} no.~14, (2015) 141301},
\href{http://arxiv.org/abs/1503.04858}{{\ttfamily arXiv:1503.04858 [hep-ex]}}.

\bibitem{Bell:2021esh}
N.~F. Bell, M.~J. Dolan, and S.~Robles, ``{Searching for Dark Matter in the Sun
  using Hyper-Kamiokande},''
  \href{http://dx.doi.org/10.1088/1475-7516/2021/11/004}{{\em JCAP} {\bfseries
  11} (2021) 004}, \href{http://arxiv.org/abs/2107.04216}{{\ttfamily
  arXiv:2107.04216 [hep-ph]}}.

\bibitem{Adrian-Martinez:2016gti}
{\bfseries ANTARES} Collaboration, S.~Adrian-Martinez {\em et~al.}, ``{Limits
  on Dark Matter Annihilation in the Sun using the ANTARES Neutrino
  Telescope},'' \href{http://dx.doi.org/10.1016/j.physletb.2016.05.019}{{\em
  Phys. Lett.} {\bfseries B759} (2016) 69--74},
\href{http://arxiv.org/abs/1603.02228}{{\ttfamily arXiv:1603.02228
  [astro-ph.HE]}}.

\bibitem{Adrian-Martinez:2016ujo}
{\bfseries ANTARES} Collaboration, S.~Adrián-Martínez {\em et~al.}, ``{A
  search for Secluded Dark Matter in the Sun with the ANTARES neutrino
  telescope},'' \href{http://dx.doi.org/10.1088/1475-7516/2016/05/016}{{\em
  JCAP} {\bfseries 1605} no.~05, (2016) 016},
\href{http://arxiv.org/abs/1602.07000}{{\ttfamily arXiv:1602.07000 [hep-ex]}}.

\bibitem{Aartsen:2016zhm}
{\bfseries IceCube} Collaboration, M.~G. Aartsen {\em et~al.}, ``{Search for
  annihilating dark matter in the Sun with 3 years of IceCube data},''
  \href{http://dx.doi.org/10.1140/epjc/s10052-019-6702-y,
  10.1140/epjc/s10052-017-4689-9}{{\em Eur. Phys. J.} {\bfseries C77} no.~3,
  (2017) 146}, \href{http://arxiv.org/abs/1612.05949}{{\ttfamily
  arXiv:1612.05949 [astro-ph.HE]}}.
[Erratum: Eur. Phys. J.C79,no.3,214(2019)].

\bibitem{Batell:2009zp}
B.~Batell, M.~Pospelov, A.~Ritz, and Y.~Shang, ``{Solar Gamma Rays Powered by
  Secluded Dark Matter},''
  \href{http://dx.doi.org/10.1103/PhysRevD.81.075004}{{\em Phys. Rev.}
  {\bfseries D81} (2010) 075004},
\href{http://arxiv.org/abs/0910.1567}{{\ttfamily arXiv:0910.1567 [hep-ph]}}.

\bibitem{Schuster:2009au}
P.~Schuster, N.~Toro, and I.~Yavin, ``{Terrestrial and Solar Limits on
  Long-Lived Particles in a Dark Sector},''
  \href{http://dx.doi.org/10.1103/PhysRevD.81.016002}{{\em Phys. Rev.}
  {\bfseries D81} (2010) 016002},
\href{http://arxiv.org/abs/0910.1602}{{\ttfamily arXiv:0910.1602 [hep-ph]}}.

\bibitem{Bell:2011sn}
N.~F. Bell and K.~Petraki, ``{Enhanced neutrino signals from dark matter
  annihilation in the Sun via metastable mediators},''
  \href{http://dx.doi.org/10.1088/1475-7516/2011/04/003}{{\em JCAP} {\bfseries
  1104} (2011) 003},
\href{http://arxiv.org/abs/1102.2958}{{\ttfamily arXiv:1102.2958 [hep-ph]}}.

\bibitem{Feng:2016ijc}
J.~L. Feng, J.~Smolinsky, and P.~Tanedo, ``{Detecting dark matter through dark
  photons from the Sun: Charged particle signatures},''
  \href{http://dx.doi.org/10.1103/PhysRevD.93.115036,
  10.1103/PhysRevD.96.099903}{{\em Phys. Rev.} {\bfseries D93} no.~11, (2016)
  115036}, \href{http://arxiv.org/abs/1602.01465}{{\ttfamily arXiv:1602.01465
  [hep-ph]}}.
[Erratum: Phys. Rev.D96,no.9,099903(2017)].

\bibitem{Leane:2017vag}
R.~K. Leane, K.~C.~Y. Ng, and J.~F. Beacom, ``{Powerful Solar Signatures of
  Long-Lived Dark Mediators},''
  \href{http://dx.doi.org/10.1103/PhysRevD.95.123016}{{\em Phys. Rev.}
  {\bfseries D95} no.~12, (2017) 123016},
\href{http://arxiv.org/abs/1703.04629}{{\ttfamily arXiv:1703.04629
  [astro-ph.HE]}}.

\bibitem{HAWC:2018szf}
{\bfseries HAWC} Collaboration, A.~Albert {\em et~al.}, ``{Constraints on
  Spin-Dependent Dark Matter Scattering with Long-Lived Mediators from TeV
  Observations of the Sun with HAWC},''
  \href{http://dx.doi.org/10.1103/PhysRevD.98.123012}{{\em Phys. Rev. D}
  {\bfseries 98} (2018) 123012},
  \href{http://arxiv.org/abs/1808.05624}{{\ttfamily arXiv:1808.05624
  [hep-ph]}}.

\bibitem{Bell:2021pyy}
N.~F. Bell, J.~B. Dent, and I.~W. Sanderson, ``{Solar Gamma Ray Constraints on
  Dark Matter Annihilation to Secluded Mediators},''
  \href{http://arxiv.org/abs/2103.16794}{{\ttfamily arXiv:2103.16794
  [hep-ph]}}.

\bibitem{McCullough:2010ai}
M.~McCullough and M.~Fairbairn, ``{Capture of Inelastic Dark Matter in White
  Dwarves},'' \href{http://dx.doi.org/10.1103/PhysRevD.81.083520}{{\em Phys.
  Rev. D} {\bfseries 81} (2010) 083520},
  \href{http://arxiv.org/abs/1001.2737}{{\ttfamily arXiv:1001.2737 [hep-ph]}}.

\bibitem{Hooper:2010es}
D.~Hooper, D.~Spolyar, A.~Vallinotto, and N.~Y. Gnedin, ``{Inelastic Dark
  Matter As An Efficient Fuel For Compact Stars},''
  \href{http://dx.doi.org/10.1103/PhysRevD.81.103531}{{\em Phys. Rev. D}
  {\bfseries 81} (2010) 103531},
  \href{http://arxiv.org/abs/1002.0005}{{\ttfamily arXiv:1002.0005 [hep-ph]}}.

\bibitem{Amaro-Seoane:2015uny}
P.~Amaro-Seoane, J.~Casanellas, R.~Sch\"odel, E.~Davidson, and J.~Cuadra,
  ``{Probing dark matter crests with white dwarfs and IMBHs},''
  \href{http://dx.doi.org/10.1093/mnras/stw433}{{\em Mon. Not. Roy. Astron.
  Soc.} {\bfseries 459} no.~1, (2016) 695--700},
  \href{http://arxiv.org/abs/1512.00456}{{\ttfamily arXiv:1512.00456
  [astro-ph.CO]}}.

\bibitem{Cermeno:2018qgu}
M.~Cerme\~no and M.~A. P\'erez-Garc\'\i{}a, ``{Gamma rays from dark mediators
  in white dwarfs},'' \href{http://dx.doi.org/10.1103/PhysRevD.98.063002}{{\em
  Phys. Rev. D} {\bfseries 98} no.~6, (2018) 063002},
  \href{http://arxiv.org/abs/1807.03318}{{\ttfamily arXiv:1807.03318
  [hep-ph]}}.

\bibitem{Dasgupta:2019juq}
B.~Dasgupta, A.~Gupta, and A.~Ray, ``{Dark matter capture in celestial objects:
  Improved treatment of multiple scattering and updated constraints from white
  dwarfs},'' \href{http://dx.doi.org/10.1088/1475-7516/2019/08/018}{{\em JCAP}
  {\bfseries 08} (2019) 018}, \href{http://arxiv.org/abs/1906.04204}{{\ttfamily
  arXiv:1906.04204 [hep-ph]}}.

\bibitem{Panotopoulos:2020kuo}
G.~Panotopoulos and I.~Lopes, ``{Constraints on light dark matter particles
  using white dwarf stars},''
  \href{http://dx.doi.org/10.1142/S0218271820500583}{{\em Int. J. Mod. Phys. D}
  {\bfseries 29} no.~08, (2020) 2050058},
  \href{http://arxiv.org/abs/2005.11563}{{\ttfamily arXiv:2005.11563
  [hep-ph]}}.

\bibitem{Ilie:2020vec}
C.~Ilie, J.~Pilawa, and S.~Zhang, ``{Comment on \textquotedblleft{}Multiscatter
  stellar capture of dark matter\textquotedblright{}},''
  \href{http://dx.doi.org/10.1103/PhysRevD.102.048301}{{\em Phys. Rev. D}
  {\bfseries 102} no.~4, (2020) 048301},
  \href{http://arxiv.org/abs/2005.05946}{{\ttfamily arXiv:2005.05946
  [astro-ph.CO]}}.

\bibitem{Bell:2021fye}
N.~F. Bell, G.~Busoni, M.~E. Ramirez-Quezada, S.~Robles, and M.~Virgato,
  ``{Improved Treatment of Dark Matter Capture in White Dwarfs},''
  \href{http://dx.doi.org/10.1088/1475-7516/2021/10/083}{{\em JCAP} {\bfseries
  10} (4, 2021) 083}, \href{http://arxiv.org/abs/2104.14367}{{\ttfamily
  arXiv:2104.14367 [hep-ph]}}.

\bibitem{Goldman:1989nd}
I.~Goldman and S.~Nussinov, ``{Weakly Interacting Massive Particles and Neutron
  Stars},''
\href{http://dx.doi.org/10.1103/PhysRevD.40.3221}{{\em Phys. Rev.} {\bfseries
  D40} (1989) 3221--3230}.

\bibitem{Kouvaris:2007ay}
C.~Kouvaris, ``{WIMP Annihilation and Cooling of Neutron Stars},''
  \href{http://dx.doi.org/10.1103/PhysRevD.77.023006}{{\em Phys. Rev. D}
  {\bfseries 77} (2008) 023006},
  \href{http://arxiv.org/abs/0708.2362}{{\ttfamily arXiv:0708.2362
  [astro-ph]}}.

\bibitem{Kouvaris:2010vv}
C.~Kouvaris and P.~Tinyakov, ``{Can Neutron stars constrain Dark Matter?},''
  \href{http://dx.doi.org/10.1103/PhysRevD.82.063531}{{\em Phys. Rev. D}
  {\bfseries 82} (2010) 063531},
  \href{http://arxiv.org/abs/1004.0586}{{\ttfamily arXiv:1004.0586
  [astro-ph.GA]}}.

\bibitem{deLavallaz:2010wp}
A.~de~Lavallaz and M.~Fairbairn, ``{Neutron Stars as Dark Matter Probes},''
  \href{http://dx.doi.org/10.1103/PhysRevD.81.123521}{{\em Phys. Rev. D}
  {\bfseries 81} (2010) 123521},
  \href{http://arxiv.org/abs/1004.0629}{{\ttfamily arXiv:1004.0629
  [astro-ph.GA]}}.

\bibitem{McDermott:2011jp}
S.~D. McDermott, H.-B. Yu, and K.~M. Zurek, ``{Constraints on Scalar Asymmetric
  Dark Matter from Black Hole Formation in Neutron Stars},''
  \href{http://dx.doi.org/10.1103/PhysRevD.85.023519}{{\em Phys. Rev.}
  {\bfseries D85} (2012) 023519},
\href{http://arxiv.org/abs/1103.5472}{{\ttfamily arXiv:1103.5472 [hep-ph]}}.

\bibitem{Bell:2013xk}
N.~F. Bell, A.~Melatos, and K.~Petraki, ``{Realistic neutron star constraints
  on bosonic asymmetric dark matter},''
  \href{http://dx.doi.org/10.1103/PhysRevD.87.123507}{{\em Phys. Rev.}
  {\bfseries D87} no.~12, (2013) 123507},
\href{http://arxiv.org/abs/1301.6811}{{\ttfamily arXiv:1301.6811 [hep-ph]}}.

\bibitem{Baryakhtar:2017dbj}
M.~Baryakhtar, J.~Bramante, S.~W. Li, T.~Linden, and N.~Raj, ``{Dark Kinetic
  Heating of Neutron Stars and An Infrared Window On WIMPs, SIMPs, and Pure
  Higgsinos},'' \href{http://dx.doi.org/10.1103/PhysRevLett.119.131801}{{\em
  Phys. Rev. Lett.} {\bfseries 119} no.~13, (2017) 131801},
\href{http://arxiv.org/abs/1704.01577}{{\ttfamily arXiv:1704.01577 [hep-ph]}}.

\bibitem{Raj:2017wrv}
N.~Raj, P.~Tanedo, and H.-B. Yu, ``{Neutron stars at the dark matter direct
  detection frontier},''
  \href{http://dx.doi.org/10.1103/PhysRevD.97.043006}{{\em Phys. Rev.}
  {\bfseries D97} no.~4, (2018) 043006},
\href{http://arxiv.org/abs/1707.09442}{{\ttfamily arXiv:1707.09442 [hep-ph]}}.

\bibitem{Bell:2018pkk}
N.~F. Bell, G.~Busoni, and S.~Robles, ``{Heating up Neutron Stars with
  Inelastic Dark Matter},''
  \href{http://dx.doi.org/10.1088/1475-7516/2018/09/018}{{\em JCAP} {\bfseries
  1809} no.~09, (2018) 018},
\href{http://arxiv.org/abs/1807.02840}{{\ttfamily arXiv:1807.02840 [hep-ph]}}.

\bibitem{Bell:2019pyc}
N.~F. Bell, G.~Busoni, and S.~Robles, ``{Capture of Leptophilic Dark Matter in
  Neutron Stars},'' \href{http://dx.doi.org/10.1088/1475-7516/2019/06/054}{{\em
  JCAP} {\bfseries 1906} no.~06, (2019) 054},
\href{http://arxiv.org/abs/1904.09803}{{\ttfamily arXiv:1904.09803 [hep-ph]}}.

\bibitem{Garani:2018kkd}
R.~Garani, Y.~Genolini, and T.~Hambye, ``{New Analysis of Neutron Star
  Constraints on Asymmetric Dark Matter},''
  \href{http://dx.doi.org/10.1088/1475-7516/2019/05/035}{{\em JCAP} {\bfseries
  05} (2019) 035}, \href{http://arxiv.org/abs/1812.08773}{{\ttfamily
  arXiv:1812.08773 [hep-ph]}}.

\bibitem{Acevedo:2019agu}
J.~F. Acevedo, J.~Bramante, R.~K. Leane, and N.~Raj, ``{Warming Nuclear Pasta
  with Dark Matter: Kinetic and Annihilation Heating of Neutron Star Crusts},''
  \href{http://dx.doi.org/10.1088/1475-7516/2020/03/038}{{\em JCAP} {\bfseries
  03} (2020) 038}, \href{http://arxiv.org/abs/1911.06334}{{\ttfamily
  arXiv:1911.06334 [hep-ph]}}.

\bibitem{Joglekar:2019vzy}
A.~Joglekar, N.~Raj, P.~Tanedo, and H.-B. Yu, ``{Relativistic capture of dark
  matter by electrons in neutron stars},''
  \href{http://dx.doi.org/10.1016/j.physletb.2020.135767}{{\em Phys. Lett.}
  {\bfseries B} (2020) 135767},
  \href{http://arxiv.org/abs/1911.13293}{{\ttfamily arXiv:1911.13293
  [hep-ph]}}.

\bibitem{Joglekar:2020liw}
A.~Joglekar, N.~Raj, P.~Tanedo, and H.-B. Yu, ``{Dark kinetic heating of
  neutron stars from contact interactions with relativistic targets},''
  \href{http://dx.doi.org/10.1103/PhysRevD.102.123002}{{\em Phys. Rev. D}
  {\bfseries 102} no.~12, (2020) 123002},
  \href{http://arxiv.org/abs/2004.09539}{{\ttfamily arXiv:2004.09539
  [hep-ph]}}.

\bibitem{Bell:2020jou}
N.~F. Bell, G.~Busoni, S.~Robles, and M.~Virgato, ``{Improved Treatment of Dark
  Matter Capture in Neutron Stars},''
  \href{http://dx.doi.org/10.1088/1475-7516/2020/09/028}{{\em JCAP} {\bfseries
  09} (2020) 028}, \href{http://arxiv.org/abs/2004.14888}{{\ttfamily
  arXiv:2004.14888 [hep-ph]}}.

\bibitem{Bell:2020lmm}
N.~F. Bell, G.~Busoni, S.~Robles, and M.~Virgato, ``{Improved Treatment of Dark
  Matter Capture in Neutron Stars II: Leptonic Targets},''
  \href{http://dx.doi.org/10.1088/1475-7516/2021/03/086}{{\em JCAP} {\bfseries
  03} (2021) 086}, \href{http://arxiv.org/abs/2010.13257}{{\ttfamily
  arXiv:2010.13257 [hep-ph]}}.

\bibitem{Bell:2020obw}
N.~F. Bell, G.~Busoni, T.~F. Motta, S.~Robles, A.~W. Thomas, and M.~Virgato,
  ``{Nucleon Structure and Strong Interactions in Dark Matter Capture in
  Neutron Stars},''
  \href{http://dx.doi.org/10.1103/PhysRevLett.127.111803}{{\em Phys. Rev.
  Lett.} {\bfseries 127} no.~11, (2021) 111803},
  \href{http://arxiv.org/abs/2012.08918}{{\ttfamily arXiv:2012.08918
  [hep-ph]}}.

\bibitem{Leane:2021ihh}
R.~K. Leane, T.~Linden, P.~Mukhopadhyay, and N.~Toro, ``{Celestial-Body Focused
  Dark Matter Annihilation Throughout the Galaxy},''
  \href{http://dx.doi.org/10.1103/PhysRevD.103.075030}{{\em Phys. Rev. D}
  {\bfseries 103} no.~7, (2021) 075030},
  \href{http://arxiv.org/abs/2101.12213}{{\ttfamily arXiv:2101.12213
  [astro-ph.HE]}}.

\bibitem{Bertoni:2013bsa}
B.~Bertoni, A.~E. Nelson, and S.~Reddy, ``{Dark Matter Thermalization in
  Neutron Stars},'' \href{http://dx.doi.org/10.1103/PhysRevD.88.123505}{{\em
  Phys. Rev.} {\bfseries D88} (2013) 123505},
\href{http://arxiv.org/abs/1309.1721}{{\ttfamily arXiv:1309.1721 [hep-ph]}}.

\bibitem{Garani:2020wge}
R.~Garani, A.~Gupta, and N.~Raj, ``{Observing the thermalization of dark matter
  in neutron stars},''
  \href{http://dx.doi.org/10.1103/PhysRevD.103.043019}{{\em Phys. Rev. D}
  {\bfseries 103} no.~4, (2021) 043019},
  \href{http://arxiv.org/abs/2009.10728}{{\ttfamily arXiv:2009.10728
  [hep-ph]}}.

\bibitem{Dasgupta:2020dik}
B.~Dasgupta, A.~Gupta, and A.~Ray, ``{Dark matter capture in celestial objects:
  light mediators, self-interactions, and complementarity with direct
  detection},'' \href{http://dx.doi.org/10.1088/1475-7516/2020/10/023}{{\em
  JCAP} {\bfseries 10} (2020) 023},
  \href{http://arxiv.org/abs/2006.10773}{{\ttfamily arXiv:2006.10773
  [hep-ph]}}.

\bibitem{Kouvaris:2010jy}
C.~Kouvaris and P.~Tinyakov, ``{Constraining Asymmetric Dark Matter through
  observations of compact stars},''
  \href{http://dx.doi.org/10.1103/PhysRevD.83.083512}{{\em Phys. Rev.}
  {\bfseries D83} (2011) 083512},
\href{http://arxiv.org/abs/1012.2039}{{\ttfamily arXiv:1012.2039
  [astro-ph.HE]}}.

\bibitem{Kouvaris:2011fi}
C.~Kouvaris and P.~Tinyakov, ``{Excluding Light Asymmetric Bosonic Dark
  Matter},'' \href{http://dx.doi.org/10.1103/PhysRevLett.107.091301}{{\em Phys.
  Rev. Lett.} {\bfseries 107} (2011) 091301},
\href{http://arxiv.org/abs/1104.0382}{{\ttfamily arXiv:1104.0382
  [astro-ph.CO]}}.

\bibitem{Capela:2013yf}
F.~Capela, M.~Pshirkov, and P.~Tinyakov, ``{Constraints on primordial black
  holes as dark matter candidates from capture by neutron stars},''
  \href{http://dx.doi.org/10.1103/PhysRevD.87.123524}{{\em Phys. Rev. D}
  {\bfseries 87} no.~12, (2013) 123524},
  \href{http://arxiv.org/abs/1301.4984}{{\ttfamily arXiv:1301.4984
  [astro-ph.CO]}}.

\bibitem{Guver:2012ba}
T.~Güver, A.~E. Erkoca, M.~Hall~Reno, and I.~Sarcevic, ``{On the capture of
  dark matter by neutron stars},''
  \href{http://dx.doi.org/10.1088/1475-7516/2014/05/013}{{\em JCAP} {\bfseries
  1405} (2014) 013},
\href{http://arxiv.org/abs/1201.2400}{{\ttfamily arXiv:1201.2400 [hep-ph]}}.

\bibitem{Bramante:2013nma}
J.~Bramante, K.~Fukushima, J.~Kumar, and E.~Stopnitzky, ``{Bounds on
  self-interacting fermion dark matter from observations of old neutron
  stars},'' \href{http://dx.doi.org/10.1103/PhysRevD.89.015010}{{\em Phys.
  Rev.} {\bfseries D89} no.~1, (2014) 015010},
\href{http://arxiv.org/abs/1310.3509}{{\ttfamily arXiv:1310.3509 [hep-ph]}}.

\bibitem{Ellis:2017jgp}
J.~Ellis, A.~Hektor, G.~Hütsi, K.~Kannike, L.~Marzola, M.~Raidal, and
  V.~Vaskonen, ``{Search for Dark Matter Effects on Gravitational Signals from
  Neutron Star Mergers},''
  \href{http://dx.doi.org/10.1016/j.physletb.2018.04.048}{{\em Phys. Lett.}
  {\bfseries B781} (2018) 607--610},
\href{http://arxiv.org/abs/1710.05540}{{\ttfamily arXiv:1710.05540
  [astro-ph.CO]}}.

\bibitem{Ellis:2018bkr}
J.~Ellis, G.~Hütsi, K.~Kannike, L.~Marzola, M.~Raidal, and V.~Vaskonen,
  ``{Dark Matter Effects On Neutron Star Properties},''
  \href{http://dx.doi.org/10.1103/PhysRevD.97.123007}{{\em Phys. Rev.}
  {\bfseries D97} no.~12, (2018) 123007},
\href{http://arxiv.org/abs/1804.01418}{{\ttfamily arXiv:1804.01418
  [astro-ph.CO]}}.

\bibitem{Bramante:2017ulk}
J.~Bramante, T.~Linden, and Y.-D. Tsai, ``{Searching for dark matter with
  neutron star mergers and quiet kilonovae},''
  \href{http://dx.doi.org/10.1103/PhysRevD.97.055016}{{\em Phys. Rev. D}
  {\bfseries 97} no.~5, (2018) 055016},
  \href{http://arxiv.org/abs/1706.00001}{{\ttfamily arXiv:1706.00001
  [hep-ph]}}.

\bibitem{Nelson:2018xtr}
A.~Nelson, S.~Reddy, and D.~Zhou, ``{Dark halos around neutron stars and
  gravitational waves},''
\href{http://arxiv.org/abs/1803.03266}{{\ttfamily arXiv:1803.03266 [hep-ph]}}.

\bibitem{Tolman:1939jz}
R.~C. Tolman, ``{Static solutions of Einstein's field equations for spheres of
  fluid},''
\href{http://dx.doi.org/10.1103/PhysRev.55.364}{{\em Phys. Rev.} {\bfseries 55}
  (1939) 364--373}.

\bibitem{Oppenheimer:1939ne}
J.~R. Oppenheimer and G.~M. Volkoff, ``{On Massive neutron cores},''
\href{http://dx.doi.org/10.1103/PhysRev.55.374}{{\em Phys. Rev.} {\bfseries 55}
  (1939) 374--381}.

\bibitem{Bramante:2017xlb}
J.~Bramante, A.~Delgado, and A.~Martin, ``{Multiscatter stellar capture of dark
  matter},'' \href{http://dx.doi.org/10.1103/PhysRevD.96.063002}{{\em Phys.
  Rev.} {\bfseries D96} no.~6, (2017) 063002},
\href{http://arxiv.org/abs/1703.04043}{{\ttfamily arXiv:1703.04043 [hep-ph]}}.

\bibitem{Ruester:2005fm}
S.~B. Ruester, M.~Hempel, and J.~Schaffner-Bielich, ``{The outer crust of
  non-accreting cold neutron stars},''
  \href{http://dx.doi.org/10.1103/PhysRevC.73.035804}{{\em Phys. Rev.}
  {\bfseries C73} (2006) 035804},
\href{http://arxiv.org/abs/astro-ph/0509325}{{\ttfamily arXiv:astro-ph/0509325
  [astro-ph]}}.

\bibitem{RocaMaza:2008ja}
X.~Roca-Maza and J.~Piekarewicz, ``{Impact of the symmetry energy on the outer
  crust of non-accreting neutron stars},''
  \href{http://dx.doi.org/10.1103/PhysRevC.78.025807}{{\em Phys. Rev.}
  {\bfseries C78} (2008) 025807},
\href{http://arxiv.org/abs/0805.2553}{{\ttfamily arXiv:0805.2553 [nucl-th]}}.

\bibitem{Pearson:2011}
J.~M. Pearson, S.~Goriely, and N.~Chamel, ``Properties of the outer crust of
  neutron stars from hartree-fock-bogoliubov mass models,''
  \href{http://dx.doi.org/10.1103/PhysRevC.83.065810}{{\em Phys. Rev. C}
  {\bfseries 83} (Jun, 2011) 065810}.

\bibitem{Kreim:2013rqa}
S.~Kreim, M.~Hempel, D.~Lunney, and J.~Schaffner-Bielich, ``{Nuclear Masses and
  Neutron Stars},'' \href{http://dx.doi.org/10.1016/j.ijms.2013.02.015}{{\em
  Int. J. Mass Spectr. Ion Process.} {\bfseries 349-350} (2013) 63--68},
\href{http://arxiv.org/abs/1303.1343}{{\ttfamily arXiv:1303.1343 [nucl-th]}}.

\bibitem{Chamel:2015oqa}
N.~Chamel, A.~F. Fantina, J.~L. Zdunik, and P.~Haensel, ``{Neutron drip
  transition in accreting and nonaccreting neutron star crusts},''
  \href{http://dx.doi.org/10.1103/PhysRevC.91.055803}{{\em Phys. Rev.}
  {\bfseries C91} no.~5, (2015) 055803},
\href{http://arxiv.org/abs/1504.04537}{{\ttfamily arXiv:1504.04537
  [astro-ph.HE]}}.

\bibitem{Antic:2020zuk}
S.~Anti\'c, J.~R. Stone, J.~C. Miller, K.~M.~L. Martinez, A.~W. Thomas, and
  P.~A.~M. Guichon, ``{Outer crust of a cold, nonaccreting neutron star within
  the quark-meson-coupling model},''
  \href{http://dx.doi.org/10.1103/PhysRevC.102.065801}{{\em Phys. Rev. C}
  {\bfseries 102} no.~6, (2020) 065801},
  \href{http://arxiv.org/abs/2006.16521}{{\ttfamily arXiv:2006.16521
  [nucl-th]}}.

\bibitem{Gudmundsson:1983}
E.~H. {Gudmundsson}, C.~J. {Pethick}, and R.~I. {Epstein}, ``{Structure of
  neutron star envelopes},'' \href{http://dx.doi.org/10.1086/161292}{{\em
  Astrophys. J.} {\bfseries 272} (Sep, 1983) 286--300}.

\bibitem{Potekhin:1997mn}
A.~Y. Potekhin, G.~Chabrier, and D.~G. Yakovlev, ``{Internal temperatures and
  cooling of neutron stars with accreted envelopes},'' {\em Astron. Astrophys.}
  {\bfseries 323} (1997) 415,
\href{http://arxiv.org/abs/astro-ph/9706148}{{\ttfamily arXiv:astro-ph/9706148
  [astro-ph]}}.

\bibitem{Chamel:2008Lr}
N.~{Chamel} and P.~{Haensel}, ``{Physics of Neutron Star Crusts},''
  \href{http://dx.doi.org/10.12942/lrr-2008-10}{{\em Living Reviews in
  Relativity} {\bfseries 11} no.~1, (Dec., 2008) 10},
  \href{http://arxiv.org/abs/0812.3955}{{\ttfamily arXiv:0812.3955
  [astro-ph]}}.

\bibitem{Haensel:2007yy}
P.~Haensel, A.~Y. Potekhin, and D.~G. Yakovlev, ``{Neutron stars 1: Equation of
  state and structure},''
\href{http://dx.doi.org/10.1007/978-0-387-47301-7}{{\em Astrophys. Space Sci.
  Libr.} {\bfseries 326} (2007) pp.1--619}.

\bibitem{Weber:2006ep}
F.~Weber, R.~Negreiros, P.~Rosenfield, and M.~Stejner, ``{Pulsars as
  Astrophysical Laboratories for Nuclear and Particle Physics},''
  \href{http://dx.doi.org/10.1016/j.ppnp.2006.12.008}{{\em Prog. Part. Nucl.
  Phys.} {\bfseries 59} (2007) 94--113},
\href{http://arxiv.org/abs/astro-ph/0612054}{{\ttfamily arXiv:astro-ph/0612054
  [astro-ph]}}.

\bibitem{Baym:2017whm}
G.~Baym, T.~Hatsuda, T.~Kojo, P.~D. Powell, Y.~Song, and T.~Takatsuka, ``{From
  hadrons to quarks in neutron stars: a review},''
  \href{http://dx.doi.org/10.1088/1361-6633/aaae14}{{\em Rept. Prog. Phys.}
  {\bfseries 81} no.~5, (2018) 056902},
\href{http://arxiv.org/abs/1707.04966}{{\ttfamily arXiv:1707.04966
  [astro-ph.HE]}}.

\bibitem{RikovskaStone:2006ta}
J.~Rikovska-Stone, P.~A.~M. Guichon, H.~H. Matevosyan, and A.~W. Thomas,
  ``{Cold uniform matter and neutron stars in the quark-mesons-coupling
  model},'' \href{http://dx.doi.org/10.1016/j.nuclphysa.2007.05.011}{{\em Nucl.
  Phys. A} {\bfseries 792} (2007) 341--369},
  \href{http://arxiv.org/abs/nucl-th/0611030}{{\ttfamily
  arXiv:nucl-th/0611030}}.

\bibitem{Whittenbury:2015ziz}
D.~L. Whittenbury, H.~H. Matevosyan, and A.~W. Thomas, ``{Hybrid stars using
  the quark-meson coupling and proper-time Nambu\textendash{}Jona-Lasinio
  models},'' \href{http://dx.doi.org/10.1103/PhysRevC.93.035807}{{\em Phys.
  Rev. C} {\bfseries 93} no.~3, (2016) 035807},
  \href{http://arxiv.org/abs/1511.08561}{{\ttfamily arXiv:1511.08561
  [nucl-th]}}.

\bibitem{Fortin:2014mya}
M.~Fortin, J.~Zdunik, P.~Haensel, and M.~Bejger, ``{Neutron stars with hyperon
  cores: stellar radii and equation of state near nuclear density},''
  \href{http://dx.doi.org/10.1051/0004-6361/201424800}{{\em Astron. Astrophys.}
  {\bfseries 576} (2015) A68}, \href{http://arxiv.org/abs/1408.3052}{{\ttfamily
  arXiv:1408.3052 [astro-ph.SR]}}.

\bibitem{Motta:2019tjc}
T.~F. Motta, A.~M. Kalaitzis, S.~Anti\'c, P.~A.~M. Guichon, J.~R. Stone, and
  A.~W. Thomas, ``{Isovector Effects in Neutron Stars, Radii and the GW170817
  Constraint},'' \href{http://dx.doi.org/10.3847/1538-4357/ab218e}{{\em
  Astrophys. J.} {\bfseries 878} no.~2, (2019) 159},
  \href{http://arxiv.org/abs/1904.03794}{{\ttfamily arXiv:1904.03794
  [nucl-th]}}.

\bibitem{Chodos:1974pn}
A.~Chodos, R.~L. Jaffe, K.~Johnson, and C.~B. Thorn, ``{Baryon Structure in the
  Bag Theory},'' \href{http://dx.doi.org/10.1103/PhysRevD.10.2599}{{\em Phys.
  Rev. D} {\bfseries 10} (1974) 2599}.

\bibitem{Guichon:1987jp}
P.~A.~M. Guichon, ``{A Possible Quark Mechanism for the Saturation of Nuclear
  Matter},'' \href{http://dx.doi.org/10.1016/0370-2693(88)90762-9}{{\em Phys.
  Lett. B} {\bfseries 200} (1988) 235--240}.

\bibitem{Guichon:1995ue}
P.~A.~M. Guichon, K.~Saito, E.~N. Rodionov, and A.~W. Thomas, ``{The Role of
  nucleon structure in finite nuclei},''
  \href{http://dx.doi.org/10.1016/0375-9474(96)00033-4}{{\em Nucl. Phys. A}
  {\bfseries 601} (1996) 349--379},
  \href{http://arxiv.org/abs/nucl-th/9509034}{{\ttfamily
  arXiv:nucl-th/9509034}}.

\bibitem{Guichon:2018uew}
P.~A.~M. Guichon, J.~R. Stone, and A.~W. Thomas,
  ``{Quark\textendash{}Meson-Coupling (QMC) model for finite nuclei, nuclear
  matter and beyond},''
  \href{http://dx.doi.org/10.1016/j.ppnp.2018.01.008}{{\em Prog. Part. Nucl.
  Phys.} {\bfseries 100} (2018) 262--297},
  \href{http://arxiv.org/abs/1802.08368}{{\ttfamily arXiv:1802.08368
  [nucl-th]}}.

\bibitem{Guichon:2004xg}
P.~A.~M. Guichon and A.~W. Thomas, ``{Quark structure and nuclear effective
  forces},'' \href{http://dx.doi.org/10.1103/PhysRevLett.93.132502}{{\em Phys.
  Rev. Lett.} {\bfseries 93} (2004) 132502},
  \href{http://arxiv.org/abs/nucl-th/0402064}{{\ttfamily
  arXiv:nucl-th/0402064}}.

\bibitem{Thomas:2021kio}
A.~W. Thomas, ``{Role of Quarks in Nuclear Structure},''
  \href{http://dx.doi.org/10.1093/acrefore/9780190871994.013.1}{{\em Oxford
  Encyclopedia of Science} (07, 2020) },
  \href{http://arxiv.org/abs/2105.12327}{{\ttfamily arXiv:2105.12327
  [nucl-th]}}.

\bibitem{Ozel:2016oaf}
F.~Ozel and P.~Freire, ``{Masses, Radii, and Equation of State of Neutron
  Stars},'' \href{http://dx.doi.org/10.1146/annurev-astro-081915-023322}{{\em
  Ann. Rev. Astron. Astrophys.} {\bfseries 54} (2016) 401},
\href{http://arxiv.org/abs/1603.02698}{{\ttfamily arXiv:1603.02698
  [astro-ph.HE]}}.

\bibitem{Antoniadis:2016hxz}
J.~Antoniadis, T.~M. Tauris, F.~Ozel, E.~Barr, D.~J. Champion, and P.~C.~C.
  Freire, ``{The millisecond pulsar mass distribution: Evidence for bimodality
  and constraints on the maximum neutron star mass},''
  \href{http://arxiv.org/abs/1605.01665}{{\ttfamily arXiv:1605.01665
  [astro-ph.HE]}}.

\bibitem{Thomas:2001kw}
A.~W. Thomas and W.~Weise, \href{http://dx.doi.org/10.1002/352760314X}{{\em
  {The Structure of the Nucleon}}}.
\newblock Wiley, Germany, 2001.

\bibitem{Zanotti:2017bte}
{\bfseries QCDSF/UKQCD} Collaboration, J.~Zanotti, J.~Bickerton, R.~Horsley,
  Y.~Nakamura, P.~Rakow, G.~Schierholz, P.~Shanahan, and R.~Young,
  ``{Transverse spin densities of octet baryons using Lattice QCD},''
  \href{http://dx.doi.org/10.22323/1.256.0163}{{\em PoS} {\bfseries
  LATTICE2016} (2017) 163}.

\bibitem{Alarcon:2017ivh}
J.~M. Alarc\'on and C.~Weiss, ``{Nucleon form factors in dispersively improved
  chiral effective field theory: Scalar form factor},''
  \href{http://dx.doi.org/10.1103/PhysRevC.96.055206}{{\em Phys. Rev. C}
  {\bfseries 96} no.~5, (2017) 055206},
  \href{http://arxiv.org/abs/1707.07682}{{\ttfamily arXiv:1707.07682
  [hep-ph]}}.

\bibitem{Goodman:2010ku}
J.~Goodman, M.~Ibe, A.~Rajaraman, W.~Shepherd, T.~M.~P. Tait, and H.-B. Yu,
  ``{Constraints on Dark Matter from Colliders},''
  \href{http://dx.doi.org/10.1103/PhysRevD.82.116010}{{\em Phys. Rev.}
  {\bfseries D82} (2010) 116010},
\href{http://arxiv.org/abs/1008.1783}{{\ttfamily arXiv:1008.1783 [hep-ph]}}.

\bibitem{BEBCWA59:1989ayi}
{\bfseries BEBC WA59} Collaboration, E.~Matsinos {\em et~al.}, ``{Backward
  Particle Production in Neutrino Neon Interactions},''
  \href{http://dx.doi.org/10.1007/BF01548585}{{\em Z. Phys. C} {\bfseries 44}
  (1989) 79}.

\bibitem{Melnitchouk:1992gd}
W.~Melnitchouk, A.~W. Thomas, and N.~N. Nikolaev, ``{Proton production bias in
  neutrino - hydrogen interactions},''
  \href{http://dx.doi.org/10.1007/BF01288472}{{\em Z. Phys. A} {\bfseries 342}
  (1992) 215--221}.

\bibitem{Motta:2019ywl}
T.~F. Motta, A.~W. Thomas, and P.~A.~M. Guichon, ``{Do Delta Baryons Play a
  Role in Neutron Stars?},''
  \href{http://dx.doi.org/10.1016/j.physletb.2020.135266}{{\em Phys. Lett. B}
  {\bfseries 802} (2020) 135266},
  \href{http://arxiv.org/abs/1906.05459}{{\ttfamily arXiv:1906.05459
  [nucl-th]}}.

\bibitem{Hahn:2004fe}
T.~Hahn, ``{CUBA: A Library for multidimensional numerical integration},''
  \href{http://dx.doi.org/10.1016/j.cpc.2005.01.010}{{\em Comput. Phys.
  Commun.} {\bfseries 168} (2005) 78--95},
  \href{http://arxiv.org/abs/hep-ph/0404043}{{\ttfamily arXiv:hep-ph/0404043}}.

\bibitem{Hahn:2014fua}
T.~Hahn, ``{Concurrent Cuba},''
  \href{http://dx.doi.org/10.1088/1742-6596/608/1/012066}{{\em J. Phys. Conf.
  Ser.} {\bfseries 608} no.~1, (2015) 012066},
  \href{http://arxiv.org/abs/1408.6373}{{\ttfamily arXiv:1408.6373
  [physics.comp-ph]}}.

\bibitem{Mathematica}
W.~R. Inc., ``Mathematica, {V}ersion 12.1.''
\newblock \url{https://www.wolfram.com/mathematica}. Champaign, IL, 2020.

\bibitem{Goriely:2013}
S.~{Goriely}, N.~{Chamel}, and J.~M. {Pearson}, ``{Further explorations of
  Skyrme-Hartree-Fock-Bogoliubov mass formulas. XIII. The 2012 atomic mass
  evaluation and the symmetry coefficient},''
  \href{http://dx.doi.org/10.1103/PhysRevC.88.024308}{{\em \prc} {\bfseries 88}
  no.~2, (Aug., 2013) 024308}.

\bibitem{Pearson:2018tkr}
J.~M. Pearson, N.~Chamel, A.~Y. Potekhin, A.~F. Fantina, C.~Ducoin, A.~K.
  Dutta, and S.~Goriely, ``{Unified equations of state for cold non-accreting
  neutron stars with Brussels–Montreal functionals – I. Role of symmetry
  energy},'' \href{http://dx.doi.org/10.1093/mnras/sty2413,
  10.1093/mnras/stz800}{{\em Mon. Not. Roy. Astron. Soc.} {\bfseries 481}
  no.~3, (2018) 2994--3026}, \href{http://arxiv.org/abs/1903.04981}{{\ttfamily
  arXiv:1903.04981 [astro-ph.HE]}}.
[erratum: Mon. Not. Roy. Astron. Soc.486,no.1,768(2019)].

\bibitem{Agnese:2017jvy}
{\bfseries SuperCDMS} Collaboration, R.~Agnese {\em et~al.}, ``{Low-mass dark
  matter search with CDMSlite},''
  \href{http://dx.doi.org/10.1103/PhysRevD.97.022002}{{\em Phys. Rev. D}
  {\bfseries 97} no.~2, (2018) 022002},
  \href{http://arxiv.org/abs/1707.01632}{{\ttfamily arXiv:1707.01632
  [astro-ph.CO]}}.

\bibitem{Agnes:2018ves}
{\bfseries DarkSide} Collaboration, P.~Agnes {\em et~al.}, ``{Low-Mass Dark
  Matter Search with the DarkSide-50 Experiment},''
  \href{http://dx.doi.org/10.1103/PhysRevLett.121.081307}{{\em Phys. Rev.
  Lett.} {\bfseries 121} no.~8, (2018) 081307},
  \href{http://arxiv.org/abs/1802.06994}{{\ttfamily arXiv:1802.06994
  [astro-ph.HE]}}.

\bibitem{Aprile:2019dbj}
{\bfseries XENON} Collaboration, E.~Aprile {\em et~al.}, ``{Constraining the
  spin-dependent WIMP-nucleon cross sections with XENON1T},''
  \href{http://dx.doi.org/10.1103/PhysRevLett.122.141301}{{\em Phys. Rev.
  Lett.} {\bfseries 122} no.~14, (2019) 141301},
\href{http://arxiv.org/abs/1902.03234}{{\ttfamily arXiv:1902.03234
  [astro-ph.CO]}}.

\bibitem{Aprile:2019xxb}
{\bfseries XENON} Collaboration, E.~Aprile {\em et~al.}, ``{Light Dark Matter
  Search with Ionization Signals in XENON1T},''
  \href{http://dx.doi.org/10.1103/PhysRevLett.123.251801}{{\em Phys. Rev.
  Lett.} {\bfseries 123} no.~25, (2019) 251801},
\href{http://arxiv.org/abs/1907.11485}{{\ttfamily arXiv:1907.11485 [hep-ex]}}.

\bibitem{Aprile:2019jmx}
{\bfseries XENON} Collaboration, E.~Aprile {\em et~al.}, ``{Search for Light
  Dark Matter Interactions Enhanced by the Migdal Effect or Bremsstrahlung in
  XENON1T},'' \href{http://dx.doi.org/10.1103/PhysRevLett.123.241803}{{\em
  Phys. Rev. Lett.} {\bfseries 123} no.~24, (2019) 241803},
  \href{http://arxiv.org/abs/1907.12771}{{\ttfamily arXiv:1907.12771
  [hep-ex]}}.

\bibitem{Aprile:2020thb}
{\bfseries XENON} Collaboration, E.~Aprile {\em et~al.}, ``{Search for Coherent
  Elastic Scattering of Solar $^8$B Neutrinos in the XENON1T Dark Matter
  Experiment},'' \href{http://dx.doi.org/10.1103/PhysRevLett.126.091301}{{\em
  Phys. Rev. Lett.} {\bfseries 126} (2021) 091301},
  \href{http://arxiv.org/abs/2012.02846}{{\ttfamily arXiv:2012.02846
  [hep-ex]}}.

\bibitem{PandaX-4T:2021bab}
{\bfseries PandaX-4T} Collaboration, Y.~Meng {\em et~al.}, ``{Dark Matter
  Search Results from the PandaX-4T Commissioning Run},''
  \href{http://dx.doi.org/10.1103/PhysRevLett.127.261802}{{\em Phys. Rev.
  Lett.} {\bfseries 127} no.~26, (2021) 261802},
  \href{http://arxiv.org/abs/2107.13438}{{\ttfamily arXiv:2107.13438
  [hep-ex]}}.

\bibitem{Behnke:2016lsk}
E.~Behnke {\em et~al.}, ``{Final Results of the PICASSO Dark Matter Search
  Experiment},''
  \href{http://dx.doi.org/10.1016/j.astropartphys.2017.02.005}{{\em Astropart.
  Phys.} {\bfseries 90} (2017) 85--92},
  \href{http://arxiv.org/abs/1611.01499}{{\ttfamily arXiv:1611.01499
  [hep-ex]}}.

\bibitem{Amole:2019fdf}
{\bfseries PICO} Collaboration, C.~Amole {\em et~al.}, ``{Dark Matter Search
  Results from the Complete Exposure of the PICO-60 C$_3$F$_8$ Bubble
  Chamber},'' \href{http://dx.doi.org/10.1103/PhysRevD.100.022001}{{\em Phys.
  Rev. D} {\bfseries 100} no.~2, (2019) 022001},
  \href{http://arxiv.org/abs/1902.04031}{{\ttfamily arXiv:1902.04031
  [astro-ph.CO]}}.

\bibitem{Agnese:2016cpb}
{\bfseries SuperCDMS} Collaboration, R.~Agnese {\em et~al.}, ``{Projected
  Sensitivity of the SuperCDMS SNOLAB experiment},''
  \href{http://dx.doi.org/10.1103/PhysRevD.95.082002}{{\em Phys. Rev. D}
  {\bfseries 95} no.~8, (2017) 082002},
  \href{http://arxiv.org/abs/1610.00006}{{\ttfamily arXiv:1610.00006
  [physics.ins-det]}}.

\bibitem{Yue:2016epq}
Q.~Yue, K.~Kang, J.~Li, and H.~T. Wong, ``{The CDEX Dark Matter Program at the
  China Jinping Underground Laboratory},''
  \href{http://dx.doi.org/10.1088/1742-6596/718/4/042066}{{\em J. Phys. Conf.
  Ser.} {\bfseries 718} no.~4, (2016) 042066},
  \href{http://arxiv.org/abs/1602.02462}{{\ttfamily arXiv:1602.02462
  [physics.ins-det]}}.

\bibitem{Vahsen:2020pzb}
S.~E. Vahsen {\em et~al.}, ``{CYGNUS: Feasibility of a nuclear recoil
  observatory with directional sensitivity to dark matter and neutrinos},''
  \href{http://arxiv.org/abs/2008.12587}{{\ttfamily arXiv:2008.12587
  [physics.ins-det]}}.

\bibitem{VasquezJauregui:2017}
{\bfseries PICO} Collaboration, E.~V\'azquez~J\'auregui, ``{PICO-500L:
  Simulations for a 500L Bubble Chamber for Dark Matter Search},'' in {\em {XV
  International Conference on Topics in Astroparticle and Underground Physics,
  TAUP2017 Sudbury ON, Canada; July 24-28}}.
\newblock
  \url{https://indico.cern.ch/event/606690/contributions/2591726/attachments/1498457/2332757/Eric_Vazquez_Jauregui_TAUP_2017.pdf}.

\bibitem{Aalbers:2016jon}
{\bfseries DARWIN} Collaboration, J.~Aalbers {\em et~al.}, ``{DARWIN: towards
  the ultimate dark matter detector},''
  \href{http://dx.doi.org/10.1088/1475-7516/2016/11/017}{{\em JCAP} {\bfseries
  1611} (2016) 017},
\href{http://arxiv.org/abs/1606.07001}{{\ttfamily arXiv:1606.07001
  [astro-ph.IM]}}.

\bibitem{Ruppin:2014bra}
F.~Ruppin, J.~Billard, E.~Figueroa-Feliciano, and L.~Strigari,
  ``{Complementarity of dark matter detectors in light of the neutrino
  background},'' \href{http://dx.doi.org/10.1103/PhysRevD.90.083510}{{\em Phys.
  Rev. D} {\bfseries 90} no.~8, (2014) 083510},
  \href{http://arxiv.org/abs/1408.3581}{{\ttfamily arXiv:1408.3581 [hep-ph]}}.

\bibitem{Belanger:2013oya}
G.~Belanger, F.~Boudjema, A.~Pukhov, and A.~Semenov, ``{micrOMEGAs3: A program
  for calculating dark matter observables},''
  \href{http://dx.doi.org/10.1016/j.cpc.2013.10.016}{{\em Comput. Phys.
  Commun.} {\bfseries 185} (2014) 960--985},
\href{http://arxiv.org/abs/1305.0237}{{\ttfamily arXiv:1305.0237 [hep-ph]}}.

\bibitem{QCDSF:2011aa}
{\bfseries QCDSF} Collaboration, G.~S. Bali {\em et~al.}, ``{Strangeness
  Contribution to the Proton Spin from Lattice QCD},''
  \href{http://dx.doi.org/10.1103/PhysRevLett.108.222001}{{\em Phys. Rev.
  Lett.} {\bfseries 108} (2012) 222001},
  \href{http://arxiv.org/abs/1112.3354}{{\ttfamily arXiv:1112.3354 [hep-lat]}}.

\bibitem{Dienes:2013xya}
K.~R. Dienes, J.~Kumar, B.~Thomas, and D.~Yaylali, ``{Overcoming Velocity
  Suppression in Dark-Matter Direct-Detection Experiments},''
  \href{http://dx.doi.org/10.1103/PhysRevD.90.015012}{{\em Phys. Rev. D}
  {\bfseries 90} no.~1, (2014) 015012},
  \href{http://arxiv.org/abs/1312.7772}{{\ttfamily arXiv:1312.7772 [hep-ph]}}.

\bibitem{Shanahan:2013cd}
P.~Shanahan, A.~Thomas, and R.~Young, ``{Scale setting, sigma terms and the
  Feynman-Hellman theorem},'' \href{http://dx.doi.org/10.22323/1.164.0165}{{\em
  PoS} {\bfseries LATTICE2012} (2012) 165},
  \href{http://arxiv.org/abs/1301.3231}{{\ttfamily arXiv:1301.3231 [hep-lat]}}.

\bibitem{Shanahan:2013apa}
P.~Shanahan, A.~Thomas, K.~Tsushima, R.~Young, and F.~Myhrer, ``{Octet Spin
  Fractions and the Proton Spin Problem},''
  \href{http://dx.doi.org/10.1103/PhysRevLett.110.202001}{{\em Phys. Rev.
  Lett.} {\bfseries 110} no.~20, (2013) 202001},
  \href{http://arxiv.org/abs/1302.6300}{{\ttfamily arXiv:1302.6300 [nucl-th]}}.

\bibitem{Alexandrou:2020sml}
C.~Alexandrou, S.~Bacchio, M.~Constantinou, J.~Finkenrath, K.~Hadjiyiannakou,
  K.~Jansen, G.~Koutsou, H.~Panagopoulos, and G.~Spanoudes, ``{Complete flavor
  decomposition of the spin and momentum fraction of the proton using lattice
  QCD simulations at physical pion mass},''
  \href{http://dx.doi.org/10.1103/PhysRevD.101.094513}{{\em Phys. Rev. D}
  {\bfseries 101} no.~9, (2020) 094513},
  \href{http://arxiv.org/abs/2003.08486}{{\ttfamily arXiv:2003.08486
  [hep-lat]}}.

\bibitem{Agashe:2014yua}
K.~Agashe, Y.~Cui, L.~Necib, and J.~Thaler, ``{(In)direct Detection of Boosted
  Dark Matter},'' \href{http://dx.doi.org/10.1088/1475-7516/2014/10/062}{{\em
  JCAP} {\bfseries 10} (2014) 062},
  \href{http://arxiv.org/abs/1405.7370}{{\ttfamily arXiv:1405.7370 [hep-ph]}}.

\bibitem{Martin:2009iq}
A.~D. Martin, W.~J. Stirling, R.~S. Thorne, and G.~Watt, ``{Parton
  distributions for the LHC},''
  \href{http://dx.doi.org/10.1140/epjc/s10052-009-1072-5}{{\em Eur. Phys. J. C}
  {\bfseries 63} (2009) 189--285},
  \href{http://arxiv.org/abs/0901.0002}{{\ttfamily arXiv:0901.0002 [hep-ph]}}.

\bibitem{Reddy:1997yr}
S.~Reddy, M.~Prakash, and J.~M. Lattimer, ``{Neutrino interactions in hot and
  dense matter},'' \href{http://dx.doi.org/10.1103/PhysRevD.58.013009}{{\em
  Phys. Rev.} {\bfseries D58} (1998) 013009},
\href{http://arxiv.org/abs/astro-ph/9710115}{{\ttfamily arXiv:astro-ph/9710115
  [astro-ph]}}.

\bibitem{Chamel:2009yx}
N.~Chamel, S.~Goriely, and J.~Pearson, ``{Further explorations of
  Skyrme-Hartree-Fock-Bogoliubov mass formulas. XI. Stabilizing neutron stars
  against a ferromagnetic collapse},''
  \href{http://dx.doi.org/10.1103/PhysRevC.80.065804}{{\em Phys. Rev. C}
  {\bfseries 80} (2009) 065804},
  \href{http://arxiv.org/abs/0911.3346}{{\ttfamily arXiv:0911.3346 [nucl-th]}}.

\end{thebibliography}\endgroup

 \includepdf[pages=-]{\supplementfilename}

\end{document}